\documentclass[UTF8,a4paper]{article}
\usepackage{amsmath}
\usepackage{bm}
\usepackage{graphicx}
\usepackage{subfigure}
\usepackage{ upgreek }
\usepackage{ stmaryrd }
\usepackage{ mathrsfs }
\usepackage{epstopdf}
\usepackage{caption}
\usepackage{ marvosym }
\usepackage{ tipa }
\usepackage{geometry}
\usepackage{ wasysym }
\usepackage{ulem}
\usepackage{indentfirst}
\usepackage{amsfonts,amssymb,dsfont}
\usepackage{accents}

\usepackage{setspace}
\usepackage{threeparttable}
\usepackage{amsmath,mathtools,amsthm}
\usepackage{array}
\usepackage{extarrows}
\usepackage{upgreek}

\makeatletter
\renewcommand*\env@matrix[1][\arraystretch]{%
	\edef\arraystretch{#1}%
	\hskip -\arraycolsep
	\let\@ifnextchar\new@ifnextchar
	\array{*\c@MaxMatrixCols c}}
\makeatother

\usepackage[pdfstartview=XYZ,
bookmarks=true,
colorlinks=true,
linkcolor=blue,
urlcolor=blue,
citecolor=blue,
pdftex,
bookmarks=true,
linktocpage=true, 
hyperindex=true
]{hyperref}
\usepackage{orcidlink}

\title{
Non-defective degeneracy in non-Hermitian bipartite system}
\author{Chen-Huan Wu
		\orcidlink{0000-0003-1020-5977} 
\thanks{chenhuanwu1@gmail.com},
\\
College of Physics and Electronic Engineering, Northwest Normal University, Lanzhou 730070, China
}

\begin{document}
	
	\maketitle

	\begin{abstract} 
Starting from a Hermitian operator
with two distinct eigenvalues,
we construct
a non-Hermitian bipartite system in Gaussian orthogonal ensemble according to random matrix theory,
where we introduce the off-diagonal fluctuations through random eigenkets
and realizing the bipartite configuration consisting of two $D\times D$ subsystems
(with $D$ the Hilbert space dimension).
As required by the global thermalization (chaos),
one of the two subsystems is full ranked,
while the other is rank deficient.
For the latter subsystem, there is a block with non-defective degeneracies 
containing the non-linear symmetries,
as well as the accumulation effect of the linear map in adjacent
eigenvectors.
The maximally mixed state made by the eigenvectors of this special region
exhibit not thermal ensmeble behavior (neither canonical or Gibbs),
and exhibit similar character with the corresponding reduced density,
which can be verified through the Loschmitch echo and variance of the imaginary spectrum.
This non-defective degeneracy region partly meets the Lemma in
 10.1103/PhysRevLett.122.220603
and theorem in 10.1103/PhysRevLett.120.150603.
The coexistence of strong entanglement and initial state fidelity in this region
make it possible to achieve a maximally mixed density which, however,
not be a thermal canonical ensemble
(with complete insensitivity to the environmental energy or temperature).
Outside this region,
the collection of eigenstates (reduced density) always exhibit 
restriction on the corresponding Hilbert space dimension,
and thus suppress the thermaliation.
There are abundant physics for those densities in Hermitian and non-Hermitian bases,
where we investigate seperately in this work.
		\\

	\end{abstract}
	\begin{small}

\section{Introduction}

Topology with robust geometric properties plays an important role in condensed matter physics, 
and for the materials of this class,
the quantum or thermal phases can be distinguished by open/close of spectral gap
and the presence/absence of underlying symmetry.
Such robust symmetries (conservations) can be found in a lot of aspects in Hermitian and non-Hermitian 
systems.
The real eigenvalues (local observable) are available in Hermitian system
or non-Hermitian system perserving the parity and time-reversal symmetry.
But there are also some researches reveal the possibility of
treating the complex eigenvalues as observable.
For non-Hermitian system, which
can be implemented in optics of topological materials\cite{Parto} and electrical circuit\cite{Helbig},
one of the most intriguing phenomena is the 
non-Hermitian skin effect where the eigenstates predominantly localize near the system
boundaries (edges) and sensitive to boundary
conditions.
This is in contrast to the extended Bloch wave of Hermitian Hamiltonian.
The electrical circuit analogues of non-Hermitian system\cite{Helbig} and the quantized quantum Hall effect in the materials
with quantized edge mode\cite{Yao,Wu,Kendirlik}
imply that the Hermitian/non-Hermitian physics can find a good correspondence in reality,
and these
also highlight the role of Hermitian to non-Hermitian transition
in the study of electronic transport as well as other condensed matter physics.
In non-Hermitian systems,
the remarkable role of system edges
reveals the importance of dimension (of the corresponding Hilbert space), and necessitates the concept of
generalized Brillouin zone which includes the topological invariance\cite{Yao}.
Unlike the Bloch Hamiltonian where the edge (zero) modes
are determined by the topological number, the
 non-Bloch bulk-boundary correspondence as well as the topological invariance in non-Hermitian system can usually not be described by the touched edge modes
 due to the nondefectively degenerated (untouched)
 eigenstate.
While in exceptional point, both the eigenvalues and
eigenstates of (defective) Hamiltonian coalesce.
Further, our work
get rid of the nonadaptance of the Bloch theory of Hermitian system, including the concepts of Bravais
vectors and lattice constant.
It is more efficient to find the effective dimension for each cluster/subcluster in a non-Hermitian system, and the confined channels with lowered dimension.

As a most intriguing phenomena,
the quantum (thermal) phase transition happen at zero(finite) temperature can used to distinguishing
the integrable/nonintegrable regimes\cite{Chung,rig}.
Previous works has theoretically and experimentally reveals that the quenches
 changes an initially stationary state into nonintegrable final state
at long-time limit is a successfull thermalization.
Also, there are some exceptions\cite{rig,rig2}:
in thermodynamic limit, the thermalization occurs in nonintegrable
regime (where the final state is governed by an nonintegrable Hamiltonian) 
but not in integrable regime (where the final state is governed by an integrable Hamiltonian).
In this work,
our results are in consistent with these points of view.
We consider two observables:
the first one is the $D$-dimensional randomly generated Hermitian matrix $H_{D}$ that is already in the quantum chaotic regime,
whose ratio of variances (diagonal to off-diagonal) is closed to 2;
The second one is constructed through elaborately prepared eigenbasis (eigenkets),
where the corresponding eigenkets are selected from that of a real symmetric random matrix.
As a result, we can obtain the matrix ${\bf M}$ whose entries are complex and read
 \begin{equation} 
	\begin{aligned}
M_{jj'}
=\sum_{\alpha}
\lambda_{\alpha}\psi_{j}^{\alpha}(\psi_{j'}^{\alpha})^*,
	\end{aligned}
\end{equation}
 where $\psi_{j}^{\alpha}$ is a vector instead of a trivial inner product (between the eigenstates of $H_{D}$ and that of the Hermitian operator $\hat{M}$).
We note that, the past treatment for the generation of non-Hermiticity in ${\bf M}$ from the eigenkets of real symmetric random matrix is to consider the density operator (integrable eigenstates) of $\hat{M}$
as a summation of rank-1 matrices
$M_{jj'}
=\sum_{k,l}
\langle E_{j}|k\rangle \langle k|
(\lambda_{\alpha}{\bf I})
|l\rangle \langle l|E_{j'}\rangle
=\sum_{k}\lambda_{k}
\langle E_{j}|k\rangle \langle k|E_{j'}\rangle$,
where the measurement of $\hat{M}$ in a non-Hermitian basis has a higher degrees of freedom in selecting the eigenkets.
This treatment weaken the finite correlations between the original integrable eigenstates
which is robust since they need to be mutually orthogonal and uniformly distributed with Haar measure.
Although such correlations can be ignored in the limit of large dimension, i.e., ${\rm Rank}\hat{M}\rightarrow\infty$ due to the emergent statistically independence in the central limit,
but it is not the case for a case with less flavor number like the case we discuss in this article
that the Hermitian operator provides only two distinct finite eigenvalues.
As a result,
the resulting $\psi_{j}^{\alpha}$ (which is necessarily be a vector now) cannot be a real Gaussian
distributed variable, but we found that it is able to share most features of a real Gaussian
distributed variable (GOE here) even in the presence of finite correlation between two subsystems 
distincted by the eigenvalues of $\hat{M}$.
We found that there is a clear correspondence between such correlations and the deviation of  $\psi_{j}^{\alpha}$ from a real Gaussian variable,
and such correlation can be revealed in terms of the emergent difference between the trace classes of two
subsystems.
With the degrees-of-freedom of the original Hermitian operator increase,
there is level repulsion between
all the subsystems except the one in end of spectrum.
Correspondently, for these (bulk) subsystems
${\rm Rank}{\bf J}^{T}_{\alpha}{\bf J}_{\alpha}\rightarrow 1$
but there degenerated parts remain non-defective as long as the whole system is non-Hermitian
(i.e., full-ranked).

We prove that, a collection of eigenstates (mixed state)
can maintain the robust initial state revival during time evolution,
and results in nonintegrable final state which fail to be fully thermalized with respect to observable.
Thus the collection of eigenstates here corresponds to a complete dephasing of the 
coherence between two subsystems,
and the emergent level repulsion can be observed from the imaginary part of the spectrum
 which signals the entanglement.
A most significant difference between initial states correspond to restricted and unrestricted Hilbert spaces is that the restricted one has an evolutionary spectrum exhibit level repulsion
at arbitarily long-time, and the fidelity spectrum
exhibits
dip-ramp-plateau structure for at least one of the subsystem which means it is maximally chaotic,
while the unrestricted one 
(as well as the expectation of a local observable for at least one of the subsystem)

A dip-ramp-plateau structure in spectrum can be a realiable signal of maximal chaos,
where the corresponding cluster is a locally conserved quantity and contains both the integrable and nonintegrable eigenstates
and keeping the non-deficent chraracter down to the degenerated region.
The level repulsion for each maximal chaotic cluster (or subsystem) is most obvious in the spectrum bulk
and the remaining nonintegrable states only locates in the edge.
In this work,
the coherence between two subsystems counteract the nonintegrablities of two subsystems
and results in chaotic spectrum of the global system.
Such coherence mechanism is closely related to the eigenstate structure
(or off-diagonal parts) of the two subsystems.
We also prove that the nonintegrable final state here corresponds to the
initially restricted Hilbert space (such that a parent Hamiltonian is inavailable),
in which case a global chaos is possible only if we taking all the nonintegrable eigenstate into consideration until
the spectrum asymptotically degenerated.
This cannot be seem in bipartite configuration by requires a large-$N$ limit.
Hall measurement provide a testbed for such experimentally realization\cite{w},
where in Hermitian case
the Hall plateaus are under quantum confinement and each localized edge state
is topologically protected,
while im non-Hermitian case
the emergent integrable eigenstates with well-defined eigenvalues as a result of solvable large-$N$ limit  
(like the number of carrier trajectory from source to drain)
and can be observed by the emergent Hall plateaus with shorter width
and follow Gaussian distribution which can be treated as the quasiparticle (and also qusi-one-dimensional) bands induced by the non-Hermitian effect.
For the latter case,
the narrowed channels (ir carrier densities) are independent of the edge geometry but symmetry
with respect to the bulk center.

Similarly, for the rank-deficient subsystem
there is a non-defectively degenerated block whose dimension is $(\frac{D}{2}-1)$
which we call nonlocal symmetry (NLS) block.
Within this block,
a biorthogonal set of eigenvector can be found,
and the subspaces
each which formed by the corresponding projector as the outer product between the left and right eigenvectors,
are pseudo-maximal-chaotic through the pseudo-edge-bulk correspondence
(the "pseudo" here indicates that the virtue edges for these subspaces are not a constant,
but equally spaced along the block diagonal),
and there superposition results in the NLS block whose diagonal elements are the same
and equal to $\frac{D-2}{D}$.
Such process removing the localization induced by the inhomogeneously distributed diagonal elements,
and results in the extensive NLS block.
Such extensive NLS block is necessary for the maximal chaos for the rank-deficient subsystem
in terms of the diagonal ETH.
Our model can be further extended to be multi-partite,
where each rank-deficient subsystem (still pseudo-maximal chaotic) contains a thermal state,
and there is still one full-rank subsystem which is maximal chaos.

Furthermore,
similar to the generalized Bloch phase factor proposed in Refs.\cite{Yao2},
we introduce the effective temperature (in terms of thermal ETH)
for the restricted or unrestricted Hilbert spaces.
For the initial state with restricted Hilbert space (initial thermal/dephased state),
the maximal chaos can be achieved at finite energy density (instead of the thermodynamic limit),
whose corresponding non-Bloch-like invariance can be described by an effective winding number,
and
the diagonal ensemble can be solved
for the whole final state (the case of final nonintegrability)
While for the initial state with unrestricted Hilbert space,
a approximated dephased state is available 
by the collection of configurations 
(that keeping the "virtue" initial state fidelity)
where the emergent conservation
generate effective cutoff during the long-time relaxation,
which helps to estimation the distance
for the current coherented state aways from the nonintegrable one.
Both cases are perfectly verified numerically.
Thus, our results show that the property of initial states can be revealed by
rewriting them as the effective phase factors and the long-time relaxation can be
predicted by the cumulative effect of the conspired effective phases.

\section{Bipartite system}

We conisder a thermalized bipartite Hamiltonian ${\bf M}$ consist of two subsystems,
and its projector operator reads
 \begin{equation} 
	\begin{aligned}
		&
\mathcal{P}_{i}=\sum_{j}
(\langle E_{j}|{\bf J})^{T}
(\langle E_{j}|{\bf J})
=\sum_{j}
{\bf J}^{T}|E_{j}\rangle
\langle E_{j}|{\bf J}.
	\end{aligned}
\end{equation}
The elements read
 \begin{equation} 
	\begin{aligned}
		\label{8221}
M_{jj'}
=\langle E_{j}|\hat{M}|E_{j'}\rangle
=\sum_{\alpha}
\langle E_{j}|{\bf J}^{T}_{\alpha}
(\lambda_{\alpha}{\bf I})
{\bf J}_{\alpha}|E_{j'}\rangle
=\sum_{\alpha}
\lambda_{\alpha}\psi_{j}^{\alpha}(\psi_{j'}^{\alpha})^*,
	\end{aligned}
\end{equation}
where we denote 
$\langle E_{j}|{\bf J}^{T}_{\alpha}
=\psi_{j}^{\alpha}$ is an eigenvectors of $\mathcal{H}_{i}^{\alpha}$.
${\bf I}$ is the identity matrix.
Thus for each nondegenerated eigenstate of ${\bf M}$,
there is acoresponding $\mathcal{H}_{i}^{\alpha}$
whose trace is $\le D$.
To make sure the elements $M_{jj'}$ satisfy the
 classical analog for the Gaussian variables in large system size,
 the matrices ${\bf J}_{\alpha}$ are formulated under the orthogonal
 and normalization conditions:
 $\psi_{j}^{\alpha}\psi_{j'}^{\alpha *}=\delta_{jj'},\ 
 \overline{\psi_{j}^{\alpha}\psi_{j'}^{\alpha *}}=\delta_{jj'}\frac{1}{D},\ 
 \sum_{j=1}^{D}|\psi_{j}^{\alpha}|^2=1\ (\forall \alpha)$.
 
 To realize the eigenstate thermalization which follows the prediction of ETH,
 the Hermitian local observable $\hat{M}$ in the non-Hermitian basis provided by $H_{j}$ follows the classical analogy.
 For $\hat{M}$ with $\alpha_{m}$ nonzero distinct eigenvalues
 (each one corresponds to a unique integrable eigenstate),
 at least $\alpha_{m}$ eigenfunction $\psi_{j}^{\alpha}$
 (i.e., $D\ge \alpha_{m}$) is needed for each eigenvalue (labelled by $\alpha$) of $\hat{M}$.
 That means there will be at least $\alpha_{m}^2$ eigenfunctions over all $\alpha$'s.
Each eigenfunction $\psi_{j}^{\alpha}$ contains $D$ elements,
where we only consider the case that $D(\ge \alpha_{m})$ 
and $\alpha_{m}$ are even integers.
For the case $D=\alpha_{m}$,
each eigenfunction caontains only the elements contributing to the non-fluctuation part,
which plays the key role in the above-mentioned classical analogy and cannot be seem from an integrable eigenstate,
if $D>\alpha_{m}$,
there will be additionally $(D-\alpha_{m})=n\alpha_{m}\ 
(n\in \mathbb{Z})$ elements contributing the fluctuation part,
 which can be seem from an integrable eigenstate which cause large eigenvalue functuation in an integrable system by the high
 barrier that prohibiting the quantum transitions.
 Note that the vanishing transitions here reflect the negligibly 
 correlation (or level repulsion) between quantum states of different symmetry sectors
 (usually labelled by good quantum numbers).
 
For $D=\alpha_{m}$, we arrange the eigenfunctions labelled by $j$'s to make the resulting eigenbasis satisfy
$
 		|\psi_{j}^{\alpha}|^2+|\psi_{j'}^{\alpha}|^2
 	=|\psi_{j}^{\alpha}|^2+|\psi_{j}^{\alpha'}|^2
 	=|\psi_{j}^{\alpha'}|^2+|\psi_{j'}^{\alpha'}|^2
 	=|\psi_{j'}^{\alpha}|^2+|\psi_{j'}^{\alpha'}|^2$.
which can be realized by the proper parameters in the nonfluctuation part.
For $D-\alpha_{m}=n\alpha_{m}$, 
there will be $D=(n+1)\alpha_{m}$ eigenfunctions,
which are statistically mutually independent,
for each $\alpha$.
Now these $D$ eigenfunctions (called the eigenbasis of $\alpha$) can be viewed as a collection of $(n+1)$ replicas
of the variant of the corresponding eigenbasis at $D=\alpha_{m}$,
i.e.,
in the new eigenbasis which including the fluctuation part,
the elements contributing to the nonfluctuation part locate
in the $1\cdots \alpha_{m}$,
$(\alpha_{m}+1)\cdots 2\alpha_{m}$, $\cdots$,
 $(D-\alpha_{m}+1) \cdots (n+1)\alpha_{m}$,
 respectively,
 and we then classify the $\psi_{j}^{\alpha}$'s
 into these $(n+1)$ parts labelled by $k=1,\cdots,(n+1)$.
Now the eigenbasis satisfies
$|\psi_{kq}^{\alpha}|^2+|\psi_{kq'}^{\alpha}|^2
 		=|\psi_{kq}^{\alpha}|^2+|\psi_{kq}^{\alpha'}|^2
 		=|\psi_{kq}^{\alpha'}|^2+|\psi_{kq'}^{\alpha'}|^2
 		=|\psi_{kq'}^{\alpha}|^2+|\psi_{kq'}^{\alpha'}|^2,\ 
 		|\psi_{kq}^{\alpha}|^2+|\psi_{k'q}^{\alpha}|^2
 	=|\psi_{kq'}^{\alpha'}|^2+|\psi_{k'q'}^{\alpha'}|^2$.

For bipartite system, 
the averages over random eigenkets satisfy
$\overline{\psi_{j}^{\alpha_{1}}\psi_{j}^{\alpha_{1} *}}=
\overline{\psi_{j}^{\alpha_{2}}\psi_{j}^{\alpha_{2} *}}=\frac{1}{D},
\overline{(\psi_{j}^{\alpha_{1}}\psi_{j}^{\alpha_{1} *})^2}=
\overline{(\psi_{j}^{\alpha_{2}}\psi_{j}^{\alpha_{2} *})^2}=\frac{3}{D^2},
\overline{\psi_{j}^{\alpha_{1}}\psi_{j}^{\alpha_{1} *}
	\psi_{j}^{\alpha_{2}}\psi_{j}^{\alpha_{2} *}}=
\overline{\psi_{j}^{\alpha_{1}}\psi_{j}^{\alpha_{1} *}}\ 
\overline{\psi_{j}^{\alpha_{2}}\psi_{j}^{\alpha_{2} *}}
=\frac{1}{D^2}\approx
(\overline{\psi_{j}^{\alpha_{1}}\psi_{j}^{\alpha_{2}}
	\psi_{j}^{\alpha_{1}}\psi_{j}^{\alpha_{2}}})^2$,
$\overline{\psi_{j}^{\alpha_{1}}\psi_{j'}^{\alpha_{1}*}
	\psi_{j}^{\alpha_{2}}\psi_{j'}^{\alpha_{2} *}}=
\overline{\psi_{j}^{\alpha_{1}}\psi_{j'}^{\alpha_{1} *}}\ 
\overline{\psi_{j}^{\alpha_{2}}\psi_{j'}^{\alpha_{2} *}}=0$,
$\overline{\psi_{j}^{\alpha_{1}}\psi_{j'}^{\alpha_{1} *}}
=\overline{\psi_{j}^{\alpha_{2}}\psi_{j'}^{\alpha_{2} *}}=
\overline{\psi_{j}^{\alpha_{1}}\psi_{j}^{\alpha_{2} *}}=0$.
However,
the off-diagonal parts of the two subsystems ${\bf \Psi}_{\alpha_{1}}$ and ${\bf \Psi}_{\alpha_{2}}$, which provide the non-Hermiticity, 
inevitably lead to:
$\overline{\psi_{j}^{\alpha_{1}}\psi_{j'}^{\alpha_{2} *}}\neq 0$,
$\overline{\psi_{j}^{\alpha_{1}}\psi_{j'}^{\alpha_{1} *}}=
\overline{\psi_{j}^{\alpha_{2}}\psi_{j'}^{\alpha_{2} *}}=0,
\overline{(\psi_{j}^{\alpha_{1}}\psi_{j'}^{\alpha_{1} *})^2}\neq 
\overline{(\psi_{j}^{\alpha_{2}}\psi_{j'}^{\alpha_{2} *})^2}$,
$\overline{
(\psi_{j}^{\alpha_{1}}\psi_{j'}^{\alpha_{1} *})^2+
(\psi_{j}^{\alpha_{2}}\psi_{j'}^{\alpha_{2} *})^2}=\frac{2}{D^2}$,
$\overline{(\psi_{j}^{\alpha_{1}}\psi_{j'}^{\alpha_{1} *})^2 (\psi_{j}^{\alpha_{2}}\psi_{j'}^{\alpha_{2} *})^2}\neq 
\overline{(\psi_{j}^{\alpha_{1}}\psi_{j'}^{\alpha_{1} *})^2}\ 
\overline{(\psi_{j}^{\alpha_{2}}\psi_{j'}^{\alpha_{2} *})^2}$
where the inequility is due to the finite correlation between the off-diagonal fourth moments of two subsystems. Further, we have
$\overline{\psi_{j}^{\alpha_{1}}\psi_{j}^{\alpha_{1}}
	\psi_{j'}^{\alpha_{1}}\psi_{j'}^{\alpha_{1}}}=
\overline{\psi_{j}^{\alpha_{2}}\psi_{j}^{\alpha_{2}}
	\psi_{j'}^{\alpha_{2}}\psi_{j'}^{\alpha_{2}}}=\frac{1}{D^2}+o(\frac{1}{D})$,
where the error term decays with increasing system size,
and evenly spread among all the subsystems,
thus the decaying of error term here only contributes to the enhanced localizations in diagonal ensemble.
In large system size limit,
a collection of such localized states will be the discrete integrable states without the level repulsion between them.
As a directe evidence,
if we replace the intermediate matrices for each subsystem 
by the rank-1 matrices whose degenerated regions are defective
(in large system size limit, the difference between defective rank-1 matrices and the
non-defective rank-1 matrices only exhibited in the off-diagonal part),
i.e., the matrices whose entries are
$
\langle E_{j}|\mathcal{J}_{\alpha_{1}}\rangle
(\lambda_{\alpha}{\bf I})
\langle\mathcal{J}_{\alpha_{1}}|E_{j'}\rangle
$ and $
\langle E_{j}|\mathcal{J}_{\alpha_{2}}\rangle
(\lambda_{\alpha}{\bf I})
\langle\mathcal{J}_{\alpha_{2}}|E_{j'}\rangle
$, respectively,
which are similar matrices,
there off-diagonal variance will equal to the 
$\overline{\psi_{j}^{\alpha_{1}}\psi_{j}^{\alpha_{1} *}
	\psi_{j'}^{\alpha_{1}}\psi_{j'}^{\alpha_{1} *}}$
(or $\overline{\psi_{j}^{\alpha_{2}}\psi_{j}^{\alpha_{1} *}
	\psi_{j'}^{\alpha_{1}}\psi_{j'}^{\alpha_{2} *}}$),
as a direct result of 
$|\mathcal{J}_{\alpha}\rangle
=(|\psi_{1}^{\alpha}|,|\psi_{2}^{\alpha}|,\cdots,|\psi_{D}^{\alpha}|)^{T}$.

Thus the absence of non-defective degeneracies 
originates from the symmetry shape of the intermediate matrices ${\bf J}^{T}_{\alpha}{\bf J}_{\alpha}$
where the actural intermediate matrices (if available) should be weakly asymmetry.
Such approximation results in the exact diagonalizability and leads to another exactly diagonalizable
matrix, $\psi_{j}^{\alpha *}\psi_{j'}^{\alpha}=\psi_{j'}^{\alpha *}\psi_{j}^{\alpha}$,
which in the mean time, artifically eliminate the inter-subsystems correlation (which should be exists for a bipartite system) through $\overline{\psi_{j}^{\alpha_{1}}\psi_{j'}^{\alpha_{2} *}}=
0$ or $\overline{(\psi_{j}^{\alpha_{1}}\psi_{j'}^{\alpha_{1} *})^2
	(\psi_{j}^{\alpha_{1}}\psi_{j'}^{\alpha_{1} *})^2
}=\overline{(\psi_{j}^{\alpha_{1}}\psi_{j'}^{\alpha_{1} *})^2}\ 
\overline{(\psi_{j}^{\alpha_{2}}\psi_{j'}^{\alpha_{2} *})^{2}}$.

Thus base on a random matrix in quantum chaotic regime
whose eigenvectors $|E_{j}\rangle$ provides the ingredients of such inavailability of an exact solution of ${\bf J}^{\alpha}$ implies that 
$\langle E_{j}|{\bf J}^{T}_{\alpha}=\psi_{j}^{\alpha}$ is indeed a Wigner-type expansion
where the expansion coefficients necessarily
carries the statistical properties of the chaotic systems\cite{Wang}
as a result of the coexistence of perturbed and unperturbed states.
In other word,
${\bf J}^{T}_{\alpha}$ cannot be solved as a effective tensor here, but a functional type response
whose exact form should contains infinite terms to exhausting the responses in all possible orders.
In this sence, the unperturbed basis of the original Hermitian operator plays a less important role,
compares to that of the intermediate term when the ingredient vectors $\psi_{j}^{\alpha}$ set in.
But note that this statement is only valid in the case of few-subsystem.
Such requirement for the expansion coefficients to be capable for different orders
can be mitigated with the increasing number of subsystems.

Bipartite character within each subsystem supporting the diagonal-type conservation and off-diagonal type inter-subsystems correlation.

The conservation within each subsystem only happen in diagonal channel,
while the off-diagonal parts between different subsystems inevitably correlated.
While the level repulsions (in log scale) within the non-defective degenerate region for all the bulk subsystems
will be more obvious with increasing number of subsystems,
and meanwhile the non-defective degenerate region behaves more like an extensive summation
that coupled with the nondegenerated region.
Correspondenly, the enhanced level repulsion (in log scale) within each subsystem also leads to the mutually independence between bulk subsystems.

Thus the entanglement between the degenerate and nondegenerate parts
of the rank-deficient subsystem will be enhanced with the increasing state number of degenerated bulk.
The Hermicity may induced by increasing number of subsystems,
like the case each rank-one subsystem loss the (exponentially decayed)
off-diagonal-type decoration
and results in defective degeneration with many-body localization.
Different to the case of nondegenerated thermalized spectrum,
here the non-Hermicity does not reflected by the finite and distinct excited states in the spectrum,
but through the orthogonal and self-normalized eigenvectors within the degenerated region.

\subsection{Biorthogonality}

In such a bipartite system,
there will be a non-local symmetry sector with dimension $D_{NLS}=\frac{D}{2}-1$, i.e., an $D_{NLS}$-dimensional eigenspace spanned by the $D_{NLS}$ mutually
orthogonal eigenvectors of
${\bf \Psi}_{\alpha_{1}}=
\pmb{\mathscr{S}}_{\alpha}\pmb{\mathscr{S}}_{\alpha}^{T}$
that corresponding to zero eigenvalue.
In this article we choose ${\bf \Psi}_{\alpha_{1}}$ to be rank-deficent,
which means there must be a full-rank subsystem ${\bf \Psi}_{\alpha_{2}}$.
For ${\bf \Psi}_{\alpha_{1}}$,
there will be a $\frac{D}{2}\times\frac{D}{2}$ block filled by the real entrices (determined by the Hilbert space dimension) equal to $\frac{1}{D}$,
which is rank-one and thus supporting only one nonzero eigenvalue.
However, as we mentioned above,
${\bf \Psi}_{\alpha_{1}}$ still has $D$ linear independent eigenvectors as its exactly diagonalizable,
while 
$\pmb{\uppsi}^{T}_{\alpha_{1}} \pmb{\lambda}_{\alpha_{1}}
[\pmb{\uppsi}_{\alpha_{1}}^{T}]^{-1}$ and 
$\pmb{\uppsi}^{-1}_{\alpha_{1}}\pmb{\lambda}_{\alpha_{1}}
\pmb{\uppsi}_{\alpha_{1}}$ (see Eq.(\ref{9141}))
contain the block whose real part of the entries are the same with that of ${\bf \Psi}_{\alpha_{1}}$ 
but the distribution of imaginary parts (despite vanishingly small $\sim O(10^{-10})$) are random and asymmetry.
We call this block as random imaginary part block (RIB).
Consequently, there are at most $(\frac{D}{2}+2)$ orthogonal
eigenvectors for $\pmb{\uppsi}^{T}_{\alpha_{1}} \pmb{\lambda}_{\alpha_{1}}
[\pmb{\uppsi}_{\alpha_{1}}^{T}]^{-1}$ and 
$\pmb{\uppsi}^{-1}_{\alpha_{1}}\pmb{\lambda}_{\alpha_{1}}
\pmb{\uppsi}_{\alpha_{1}}$.
Thus the minor random imaginary part in the RIB within a rank-deficient subsystem matrix
will cause the reducing of nullspace's dimension from $(\frac{D}{2}-1)$ to 1
(geometric multiplicity of eigenvalue zero).
In this case,
there is a linear independent (full ranked) basis of eigenvectors for ${\bf \Psi}_{\alpha_{1}}$,
which are, however, not mutually orthogonal.
As a result,
$\pmb{\uppsi}_{\alpha_{1}}\pmb{\uppsi}_{\alpha_{1}}^{T}$
is a diagonal matrix,
but there are $(D-D_{NLS})$ diagonal entries smaller than one,
and $D_{NLS}$ diagonal entries equal to one.
That means the nullspace of ${\bf \Psi}_{\alpha_{1}}$
is spanned by $D_{NLS}$ non-orthogonal eigenvectors
and each one of them form an orthogonal basis by itself.

For a bipartite non-Hermitian system
constructed in above way,
the NLS block (locates in the right-bottom corner of the rank-deficient subsystem ${\bf \Psi}_{\alpha_{1}}$) is of rank 1 and has the only nonzero eigenvalue equals to $(\frac{1}{2}-\frac{1}{D})$ (independent of $D$).
Now due to the self-normalization and the mutual orthogonality
(non-defective degeneracies by non-Hermiticity),
the $(D-j+1)$-th eigenvectors of NLS block reads
$
|{\bf \Psi}_{\alpha_{1};D-(j-1)}\rangle=
\left(0,0,\cdots,0,
\frac{j}{\sqrt{j (1 + j)}},
\frac{-1}{\sqrt{j(1 + j)}},
\cdots,
\frac{-1}{\sqrt{j(1 + j)}}\right)^{T}$ ($j\le D_{NLS}$),
such that
\begin{equation} 
	\begin{aligned}		
\label{1225}
&|{\bf \Psi}_{\alpha_{1};D}\rangle=		\left(\overbrace{ 0,\cdots,0,\sqrt{\frac{1}{2}},-\sqrt{\frac{1}{2}}	}^{D}
\right)^{T},\ \ 
|{\bf \Psi}_{\alpha_{1};D-1}\rangle=	\left(0,\cdots,0,\frac{\sqrt{6}}{3},-\sqrt{\frac{1}{6}},-\sqrt{\frac{1}{6}}\right)^{T},\\&
	\cdots,\\&
		|{\bf \Psi}_{\alpha_{1};D-(D_{NLS}-1)}\rangle=	
		\left(0,\cdots,0,
		\overbrace{
\sqrt{\frac{D_{NLS}}{D_{NLS}+1}},
-\sqrt{\frac{1}{(D_{NLS}+1)D_{NLS}}},
		\cdots,
			-\sqrt{\frac{1}{(D_{NLS})D_{NLS}+1}}}^{D_{NLS}+1}
			\right)^{T}.
	\end{aligned}
\end{equation}
Only the eigenvectors within the NLS block are Gaussian, i.e.,
the elements therein have zero mean and variance equals to $\frac{1}{D}$.
As such NLS block only exist within the rank-deficient subsystem,
the eigenvetctors therein bring many marvellous results to the whole system.
We shown in Fig.\ref{Vd} the square of the nonzero elements of eigenvectors within NLS block
up to $D_{NLS}=30$,
i.e., the nonzero diagonal elements of the pure state densities spanned by the non-defectively degenerated eigenvectors
therein $|{\bf \Psi}_{\alpha_{1};i}\rangle\langle{\bf \Psi}_{\alpha_{1};i}|$ ($i\in NLS$).
From the eigenvector structure within NLS block,
we can see that, despite each eigenvector contains the identically distributed elements (zero mean
and variance $1/D$) and thus be fully chaotic due to the reflecting boundary,
there is a strict constriction imposed by the 
$V_{d}(|{\bf \Psi}_{\alpha_{1};D-(D_{NLS}-1)}\rangle\langle{\bf \Psi}_{\alpha_{1};D-(D_{NLS}-1)}|)$
whose largest element $\frac{D_{NLS}}{D_{NLS}+1}$ is just the diagonal elements
of the superposed densities of the NLS block, i.e.,
$\frac{D_{NLS}}{D_{NLS}+1}=\sum_{j=1}^{D_{NLS}}\frac{1}{j(j+1)}$, or recursively,
$\frac{j}{j+1}=\frac{j-1}{j}+\sum_{j'=1}^{j-1}\frac{1}{j'(j'+1)}$.
The latter expression indicates an extensive conservation,
which providing an effective short-range entanglement and avoid the formation of long-range order
(ergodic-breaking).
Due to such extensive conservations form a 
subspace whose energy surface (block dimension) is determined by the $D$ (or $D_{NLS}$).
Such short-range (nearest-neighbor) correlation-saved ergodicity has also been studied in terms of different systems\cite{Halder}, including the so-called
 forward scattering approximation (FSA)\cite{Turner}.
Thus it is less random for the 
 $(\frac{D}{2}-1)$ nonunique eigenvectors that span the nulleigenspace ${\bf \Psi}_{\alpha_{1}}$.

\begin{figure}
	\centering
	\includegraphics[width=0.6\linewidth]{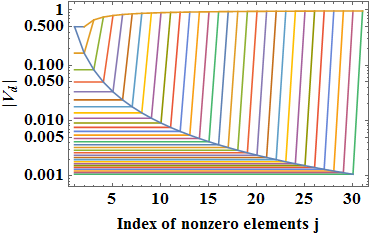}
	\captionsetup{font=small} 
\caption{Structure of
	$V_{d}(|{\bf \Psi}_{\alpha_{1};i}\rangle\langle {\bf \Psi}_{\alpha_{1};i}|)$,
	the vector consist of the diagonal elements of densities $|{\bf \Psi}_{\alpha_{1};i}\rangle\langle {\bf \Psi}^{*}_{\alpha_{1};i}|$
for $i=D-(D_{NLS}-1),\cdots,D$. Here we set $D_{NLS}=30$
(for $D=62$).
The upper curve indicates the largest element
(reflecting boundary) for each vector,
$\frac{j}{j+1}$.
The lower curve indicates the other nonzero elements for each vector, $\frac{1}{j(j+1)}$.
	}
\label{Vd}
\end{figure}

\begin{figure}
	\centering
	\includegraphics[width=1\linewidth]{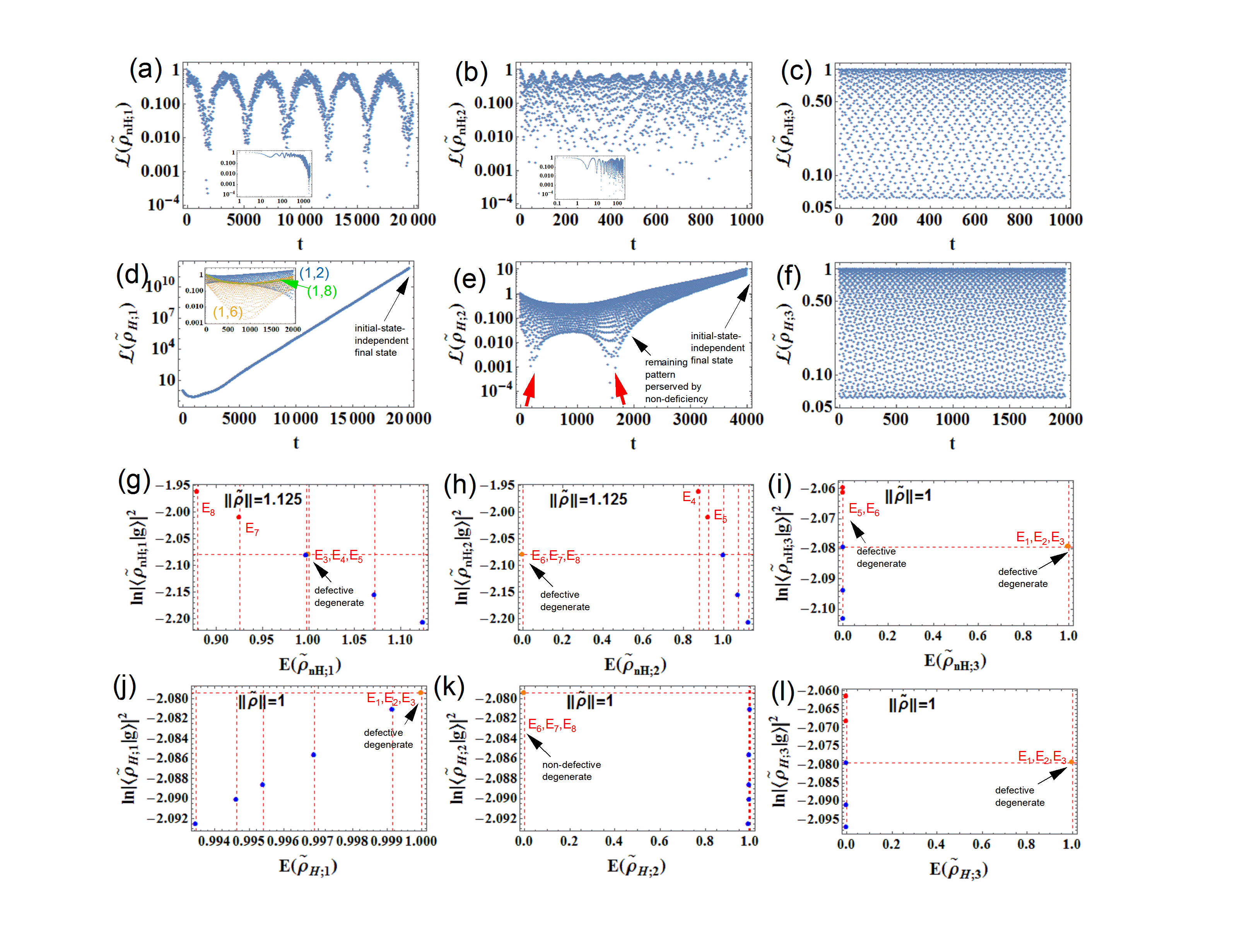}
	\captionsetup{font=small} 
\caption{
The Loschmidt echo spectrum for the mixed densities in non-Hermitian ((a)-(c)) and Hermitian basis
((d)-(f)), and the corresponding overlap bebtween eigenstates and the choosen ground state (g)-(l). 
(e) The regular fringes between the upper and lower edges are related to the thermal states
which are mutually independent.
The two red arrows indicate the 
persistive entanglement between thermal states before equilibrium,
and such dynamics of highly entangled thermal states is also convinced by the
non-defective degeneracies (k).
For the case that the boundary-dependence of the evolving quantum states
are dominated by the initial state fidelity, there are flat boundaries, as shown in (c,f).
From (d)-(e), despite $\tilde{\rho}_{H;1}$ and $\tilde{\rho}_{H;2}$ of the unrestricted Hilbert space,
$\mathcal{C}(\tilde{\rho}_{H;1},\tilde{\rho_{H;1}}')\neq 1$,
$\mathcal{C}(\tilde{\rho}_{H;2},\tilde{\rho_{H;2}}')\neq 1$,
imply finite correlation between two subsystems initially
which contribute to the localization effect,
as a result, there is persisting initial state character as indicated by the red arrows
which means the initial state information die out after a finite time.
As shown in (j)-(k),
their corresponding overlap exhibit only one nonthermal state,
which can be regarded as the low-entangled state (and coherent with the collection of rest).
Another numerical reults which supports our above discussion is:
$\sum_{i=1}^{D}|\langle\tilde{\rho}_{nH;1;i}|g \rangle|^2=
\sum_{i=1}^{D}|\langle\tilde{\rho}_{nH;2;i}|g \rangle|^2=
\sum_{i=1}^{D}|\langle\tilde{\rho}_{nH;3;i}|g\rangle|^2=
\sum_{i=1}^{D}|\langle\tilde{\rho}_{H;1;i}|g\rangle|^2=0.9989$
for $\tilde{\rho}_{res}$,
$\sum_{i=1}^{D}|\langle\tilde{\rho}_{H;2;i}|g\rangle|^2=0.986343,
\sum_{i=1}^{D}|\langle\tilde{\rho}_{H;3;i}|g\rangle|^2=0.994908$
for $\tilde{\rho}_{unres}$,
which imply the extensive and localized characters, respectively. 
The horizon and vertical red dashed lines indicate the lowest non-thermal eigenstates
(i.e., the $j$-th state such that ${\rm ln}|\langle \rho_{j}|g\rangle|^2
:=\sum_{i}{\rm ln}(\frac{1}{D}
|\langle \rho_{j}|{\bf \Psi}_{\alpha_{1};i}\rangle|^2)=-{\rm ln}D$ which $\approx -2.07944$ 
here for $D=8$).
For ${\bf \Psi}_{\alpha_{1}}$, the eigenvalues are all bounded under $\frac{1}{D}=0.125$
due to the existence of NLS block.
Comparing (a) and (b),
$\mathcal{C}(\tilde{\rho}_{nH;1})=8/7.99997\approx 1$,
$\mathcal{C}(\tilde{\rho}_{nH;2})=5/2.016\approx 2.48$,
and consequently there is much weaker coherence initially between $\tilde{\rho}_{nH;1}$ and $\tilde{\rho}_{nH;1}'$,
which can be seen from the insets where there is a well-performed (ill-performed)
dip-ramp-plateau
evolution at initial stage for $\tilde{\rho}_{nH;2}$ ($\tilde{\rho}_{nH;1}$).
While for Hermitian basis ((d)-(f)),
there is not such coherene-prevented thermalization expect the $\tilde{\rho}_{H;3}$
which is of the degenerated ETH case\cite{Anza}.
Note that the reduced (steady) state $\tilde{\rho}_{H;3}$ has an asymptotic approaching to
its maximally mixed counterpart (the thermal canonical state).
Inset of (d) show the $\mathcal{L}(\tilde{\rho}_{H;(a,b)})$
for $(a,b)=(1,2),(1,6),(1,8)$.
	}
\label{echooverlap}
\end{figure}

\begin{figure}
	\centering
	\includegraphics[width=0.6\linewidth]{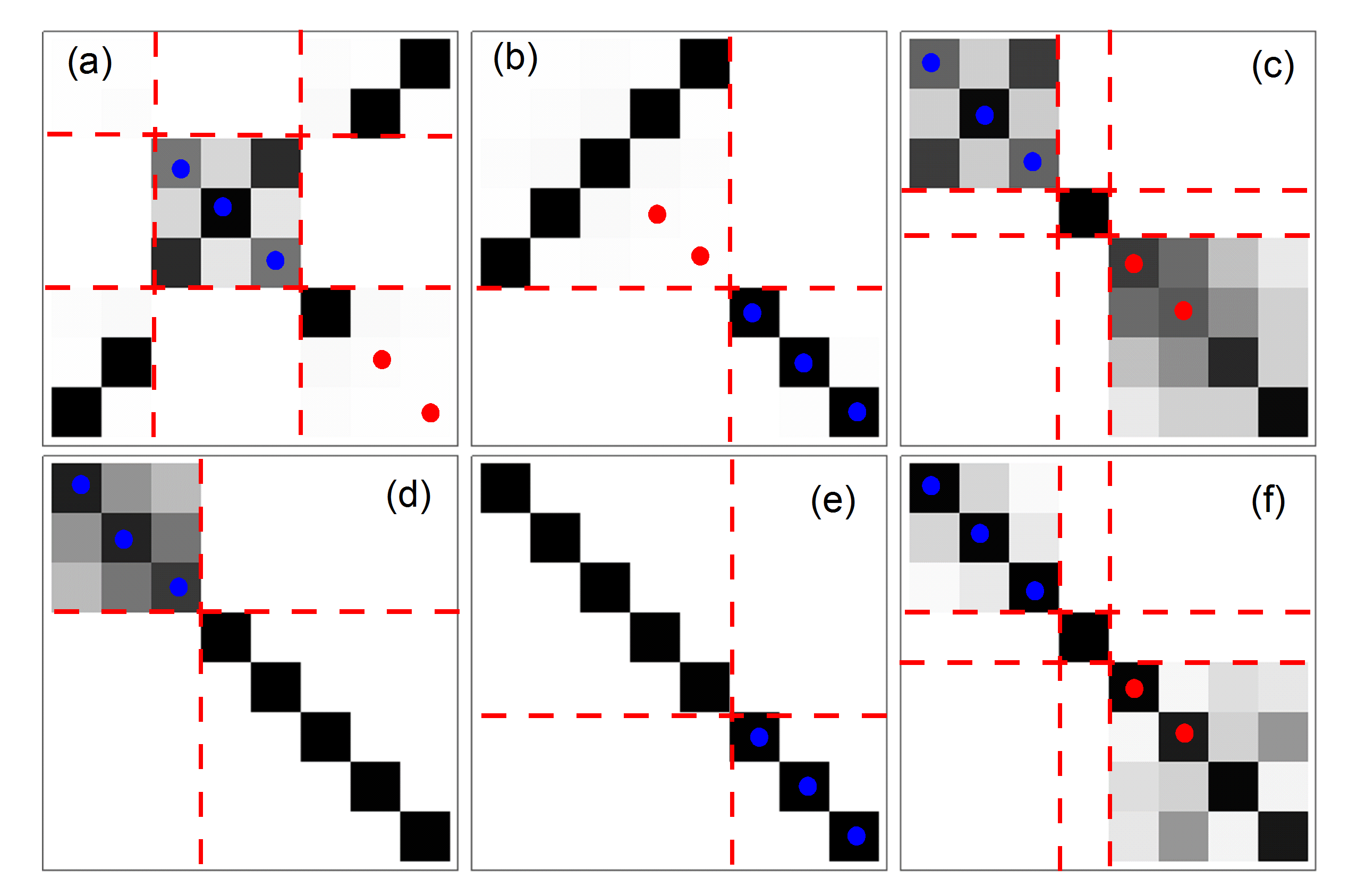}
	\captionsetup{font=small} 
\caption{
The nodes indicate the nonthermal eigenstates that have overlap with ground state
$>\frac{1}{D}$ (red) or $=\frac{1}{D}$ (blue).
The maximal chaos is possible as long as all the nontrivial (nonthermal) eigenstates are degenerated
in energy, no matter be the defective or non-defective types.
Large overlap between highly-entangled state with the low-entangled state is only possible
for a collection of thermal states where the boundary is localized instead of extensive.
 The maximal chaos is impossible when there are at least two nontrivial (nonthermal) eigenstates are non-degenerated in energy, which corresponds to the case of multi-edge as we discuss in Appendix.
	}
\label{echooverlap2}
\end{figure}

\begin{figure}
	\centering
	\includegraphics[width=0.8\linewidth]{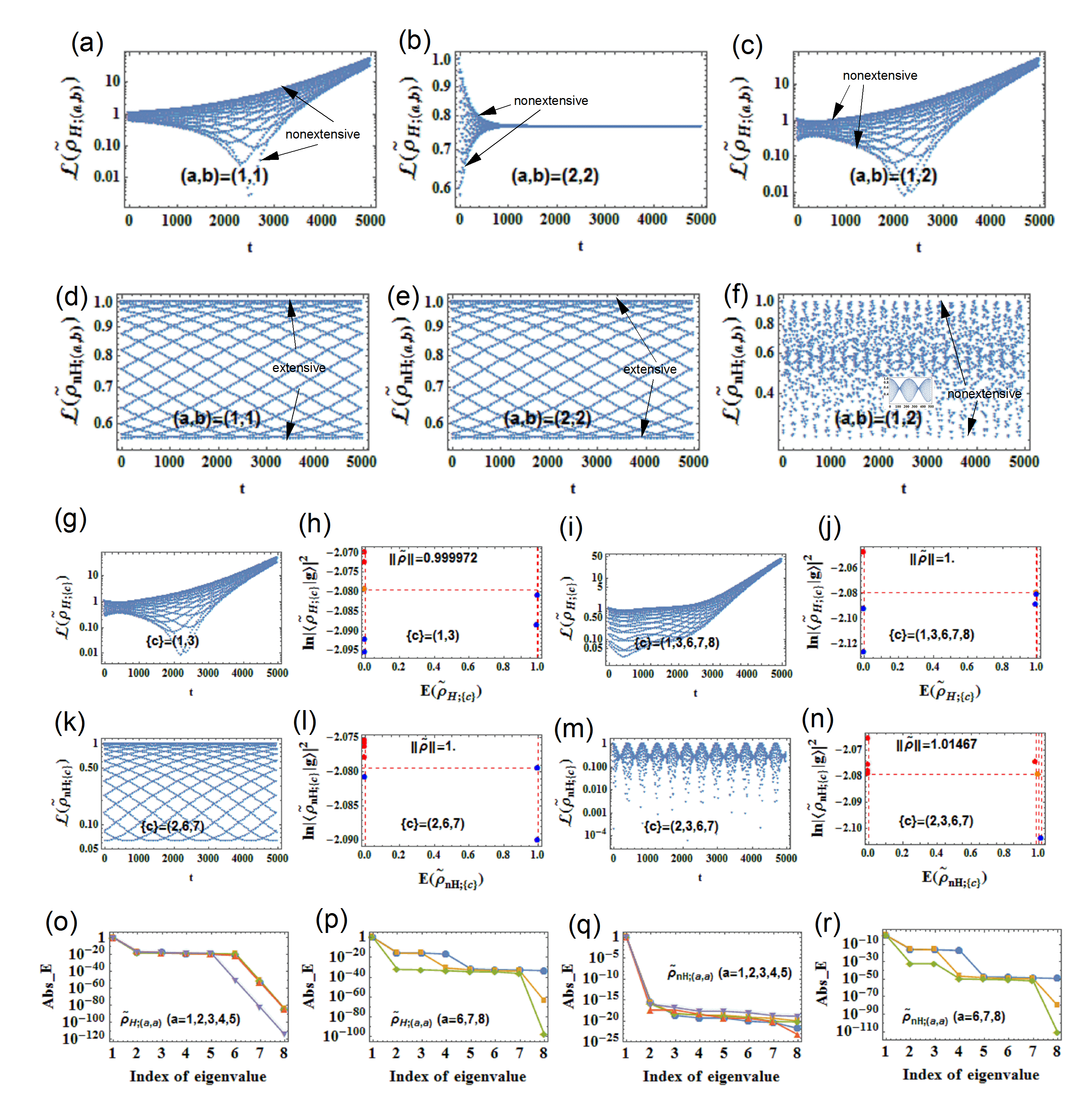}
\captionsetup{font=small} 
\caption{(a)-(c) ((d)-(f)) shows that
direct superposition between subsystems will not (will) breaks the thermal character
for the Hermitian (non-Hermitian) basis.
For Hermitian basis ((a)-(c),(g)-(j)),
mutually-independent thermal states form regular spectrum where each one of them 
depends on the boundaries.
We can see that different sum configurations in Hermitian basis only modifies the boundary of the
enveloped regularly-patterned spectrum and this is similar to the macroscopic observable
mentioned in Ref.\cite{Anza},
such that each thermal state has the same overlap with the mutually commuting (but actually emtangled) 
subspaces partitioned by the macroscopic observable.
Thus each thermal state in arbitary subsystem ($\tilde{\rho}_{H;(a,a)},\ a=1,\cdots,5$)
only response to the spectrum boundaries of the subspace
it belongs to. 
Specifically,
for non-thermal states in $\tilde{\rho}_{H;(a,a)},\ a=6,\cdots,8$)
there correlation to spectrum boundaries of all subsystems
are compatible and even dominated by the coherence within the subsystem it belongs to
(as a typical character of the non-defective degenerated region).
As shown in (g)-(j),
despite the subspaces of NLS have not edge-response,
they will not obstruct the global thermalization but only
increase the number of entangled quantum states within each subspace that has finite response to the edge.
While for the non-Hermitian basis ((d)-(f),(k)-(n)),
any superpositions (direct sum) involving more than one subsystem outside the NLS block, i.e.,
$\tilde{\rho}_{H;(a,a)},\ (a=1,\cdots,5)$,
the regularly-patterned spectrum is absent,
and the resulting spectrum exhibits chaos and initial state recurrence.
(d) to (f) show the superpositions of pure state outside the NLS region
resulst in
the emergence of chaotic (inregular) spectrum and indicates that there are some
off-diagonal elements not being exponentially decayed due to some certain conservations.
Indeed 
any subspaces within NLS 
can be regarded as infinite periodicity just like the full coherence set $\mathcal{Q}_{nH;1}$ 
(Fig.\ref{EO6}(f))
and thus
will not coherent with any other subspaces (outside or inside NLS).
and only subspaces with finite and magnitude-closed (comparable) periods can be coherent.
For non-Hermitian basis,
the quantum states indistinguishable through the edge-correspondence
(i.e., the non-thermal states)
can be seem along the horizon direction with constant echo,
and such quantum states oscillate in a fixed period determined by their own initial state property.
Coherence significantly modifies such period.
	}
\label{echooverlap3}
\end{figure}

\section{Initial state recurrence and a comparasion between mixed and reduced states}

For convinience in discussion,
hereafter we use the following notations for the maximally mixed states
(initial densisties)
\begin{equation} 
	\begin{aligned}		
&
\tilde{\rho}_{nH;1}:=\sum_{i=1}^{D}
|{\bf \Psi}_{\alpha_{1};i}\rangle\langle{\bf \Psi}_{\alpha_{1};i}^{*}|,\ 
\tilde{\rho}_{nH;2}:=\sum_{i\notin NLS}
|{\bf \Psi}_{\alpha_{1};i}\rangle\langle{\bf \Psi}_{\alpha_{1};i}^{*}|,\ 
\tilde{\rho}_{nH;3}:=\sum_{i\in NLS}
|{\bf \Psi}_{\alpha_{1};i}\rangle\langle{\bf \Psi}_{\alpha_{1};i}^{*}|,\\
&
\tilde{\rho}_{H;1}:=\sum_{i=1}^{D}
|{\bf \Psi}_{\alpha_{1};i}\rangle\langle{\bf \Psi}_{\alpha_{1};i}|,\ 
\tilde{\rho}_{H;2}:=\sum_{i\notin NLS}
|{\bf \Psi}_{\alpha_{1};i}\rangle\langle{\bf \Psi}_{\alpha_{1};i}|,\ 
\tilde{\rho}_{H;3}:=\sum_{i\in NLS}
|{\bf \Psi}_{\alpha_{1};i}\rangle\langle{\bf \Psi}_{\alpha_{1};i}|,\\
&
\tilde{\rho}_{nH;(a,b)}:=\sum_{i=a}^{b}
|{\bf \Psi}_{\alpha_{1};i}\rangle\langle{\bf \Psi}_{\alpha_{1};i}^{*}|,\
\tilde{\rho}_{H;(a,b)}:=\sum_{i=a}^{b}
|{\bf \Psi}_{\alpha_{1};i}\rangle\langle{\bf \Psi}_{\alpha_{1};i}|.
\end{aligned}	
\end{equation} 
where the last line refers to the densities in non-Hermitian/Hermitian basis with sum configuration
defined by $(a,b)$ with the range ${\rm dim}[a,\cdots,b]$.
Thus we have $\tilde{\rho}_{nH/H;2}=\tilde{\rho}_{nH/H;(1,\frac{D}{2}+1)}$,
$\tilde{\rho}_{nH/H;3}=\tilde{\rho}_{nH/H;(\frac{D}{2}+2,D)}$.
The above-mentioned eigenvector structure make sure that, for the NLS block, despite each eigenvector contains elements with variance $1/D$
(this result is shared by the Hermitian operator $H_{D}$),
it will not generate a symmetry sector that violating the ETH.
The key reason for keeping the ergodicity here is homogenuously distributed diagonal elements
in the block with nonzero entries in
$\tilde{\rho}_{nH;3}$ or $\tilde{\rho}_{H;3}$.
Also, we define the counterpart of the above mixed densities
as $\mathcal{Q}_{nH/H;\{c\}}$, which is the rank-1 reduced density.
For example, 
$\mathcal{Q}_{nH;1}=\left(
\sum_{i=1}^{D}
|{\bf \Psi}_{\alpha_{1};i}\right)
\left(\sum_{i=1}^{D}
\rangle\langle{\bf \Psi}_{\alpha_{1};i}^{*}|\right)$.

Here we present an important result ahead.
The densities with restricted Hilbert space ($\tilde{\rho}_{res}$), 
which cannot be thermalized,
including
$\tilde{\rho}_{nH;\{c\}}\ (\forall \{c\})$ and $\tilde{\rho}_{H;\{c\}}\ (\{c\}\in NLS)$.
While the densities $\tilde{\rho}_{H;\{c\}}$
with $\{c\}$ involving at least one element outside the NLS,
are of the unrestricted Hilbert space ($\tilde{\rho}_{unres}$), 
which can be thermalized.

In terms of the Loschmidt echo 
$\mathcal{L}(\tilde{\rho})=
|\langle g|e^{i\tilde{\rho}t} |g\rangle |^2$
with 
$\langle g|=\sum_{i=1}^{D}\langle{\bf \Psi}_{\alpha_{1};i}|$
defined as the ground state,
it is directly to view the many-body fidelity of the initial state $\tilde{\rho}$.
For a initial state governed by a restricted Hilbert space, $\tilde{\rho}_{res}$
it cannot be fully thermalized with respect to a local observable.
The Loschmidt echo for $\tilde{\rho}_{res}$ 
($\tilde{\rho}_{nH;3}$) and $\tilde{\rho}_{unres}$
exhibit large and vanishing recurrence, respectively.
For $\tilde{\rho}_{nH;2}
(\in \tilde{\rho}_{res})$,
the initial state information never be completely wiped out until the long-time limit,
despite the recurrences appear randomly.
For $\tilde{\rho}_{nH;2}$ and $\tilde{\rho}_{nH;3}$, there are
correlated initial states between two subsystems $\alpha_{1}$ and $\alpha_{2}$.
while $\tilde{\rho}_{nH;1}$ is initially delocalized.

In perspective of this,
we introduce the ratio 
$\mathcal{C}(\rho_{\alpha_{1}},\rho_{\alpha_{2}})=\frac{\frac{1}{D}{\rm Tr}[\rho_{\alpha_{1}}]  \frac{1}{D}{\rm Tr}[\rho_{\alpha_{2}}]}
{\frac{1}{D}V_{d}(\rho_{\alpha_{1}})\cdot V_{d}^{*}(\rho_{\alpha_{2}})}$ to
verified by the degree of locality of initial density,
where $V_{d}(\cdot)$ denotes the row vector whose elements are the diagonal elements of density $\cdot$.
For $\mathcal{C}=1$,
there is no coherence between the two initial densities (totally dephased);
For $\mathcal{C}\neq 1$,
there is coherence between the initial densities
whose restritive effect guarantees the coherence revival.
For $\tilde{\rho}_{nH;1}
(\in \tilde{\rho}_{res})$, it has a longer periodicity for the revival 
and each recurrence exhibits larger difference compares to others.

Specifically,
only the Loschmidt echos for
 $\tilde{\rho}_{nH;3}$ and  $\tilde{\rho}_{H;3}$
strictly follow the periodic pattern,
and in the mean time,
$\mathcal{L}(\tilde{\rho})$ turns back to 1 in a fixed periodicity,
as shown in Figs.\ref{echooverlap}(c),(f).
One of the reason is strong localization in initial state ($\mathcal{C}$ deviates away from 1).
Another reason is due to the non-defective degeneracies in NLS block,
where the ergodicity is not due to the exponentially large Hilbert space
(which allows the randomness of high-energy nonintegrable 
eigenstate such that it has very small overlap with the low-entangled state
like the selected ground state which cannot be thermalized)
but the special eigenvector struture (as shown in Eq.(\ref{1225}))
which avoid the localization by allowing the superposition of pure states within NLS block
be equivalent to a statistical mixture corresponding to a certain symmetry
(and thus generates a highly-entangled state, e.g.,
the elements of $\sum_{i\in NLS}|{\bf \Psi}_{\alpha_{1};i}\rangle$
have zero mean and variance equals to $\frac{D_{NLS}}{D}$).
Mathematically, this is allowed by the sparse matrix structure of teh pure states,
whose blocks with nonzero entries are ordered with acsending dimensions.

Mechanically, this is of
the same case with the Lemma proposed in Refs.\cite{Choi,Turner}:
For the target denisty $\tilde{\rho}$ within the exponential part of Loschmidt echo be extensive in energy levels,
the ground state can be decomposed into
the superposition of eigenstates, and at least one of the eigenstates has a overlap with ground state whose absolute square $\ge \frac{1}{D}$.
Here we further developing this lemma in our case,
with the choosen energy-dense (with large energy density) 
ground state $|g\rangle$.
For those mixed densities that cannot be thermalized ($\tilde{\rho}_{res}$)
and of some certain thermal ensemble,
the amount of nonthermal (and also nondegenerated) eigenstates of the density always larger than
$(||\tilde{\rho}||_{2}+1)$, and the echo
$\mathcal{L}(\tilde{\rho})$ is bounded from above by 1 and exhibit recurrence which last until the long-time limit
(in fact, such nonthermal states).
By nonthermal state,
we mean those eigenstates $\langle \tilde{\rho}$ that satisfy
$|\langle \tilde{\rho}|g\rangle|^2\ge \frac{1}{D}$ (relatively large overlap).
While for those mixed densities that can be thermalized ($\tilde{\rho}_{unres}$),
the amount of nonthermal (and also nondegenerated) eigenstates of the density always equal to
or smaller than
$(||\tilde{\rho}||_{2}+1)$, in which case the echo
$\mathcal{L}(\tilde{\rho})$ is not bounded under 1 
and the recurrences vanish in long-time limit
where the spectrum's upper and lower edges merge into a single curve with finite slope
(Fig.\ref{echooverlap}(d)),
during which process the entangling evolved thermal states approaching the thermal equilibrium
(despite not a static echo).
Before equilibrium,
the mutually independent thermal states
can be seem from the regular fringes enveloped by the upper and lower edges
before they merge.
But these rule only valid for some certain sets of $\{c\}$,
i.e., $\{c\}$ contains at least two (one) elements outside (within) the NLS,
or $\{c\}$ contains at least four elements outside the NLS.

While for $\tilde{\rho}_{nH/H;\{c\}}\ (\{c\}\in NLS)$,
there amount of nonthermal eigenstates is always $2=(||\tilde{\rho}||_{2}+1)$
for both the non-Hermitian and Hermitian bases,
where
the echo $\mathcal{L}(\tilde{\rho})$ exhibit fixed pattern (strictly periodic revival)
with the spectrum bounded by the upper and lower horizon boundaries
(roughly parallel but uncorrelated to each other).

As shown in Fig.\ref{echooverlap}(c),(f),
such regular fringers generate fixed pattern of spectrum corresponds to the undisordered
thermals states which can be evidenced by the large overlap with the cluster-decomposable
ground state.
Different to refs.\cite{Choi,Turner},
the corresponding many-body Hamiltonian within the exponential factor does not have extensive energies
(whose norm of the order of $D$), instead,
we have 
$||\tilde{\rho}_{nH;1}||_{2}=||\tilde{\rho}_{nH;2}||_{2}=\frac{D+1}{D}$,
$||\tilde{\rho}_{H;1}||_{2}=
||\tilde{\rho}_{H;2}||_{2}=
||\tilde{\rho}_{H;3}||_{2}=||\tilde{\rho}_{nH;3}||_{2}=1$.

As a low-entangled state, the ground state
$| g\rangle$ satisfies
\begin{equation} 
	\begin{aligned}		
\tilde{\rho}_{nH/H;3}|g\rangle
=\sum_{i\in NLS}|{\bf \Psi}_{\alpha_{1};i}\rangle
=\sum_{i\in NLS}(-1)^{a}|\tilde{\rho}_{nH/H;(i,i);1}\rangle,
	\end{aligned}
\end{equation}
where
$|\tilde{\rho}_{nH/H;(i,i);1}\rangle$ denotes the first eigenvector of the pure state 
$\tilde{\rho}_{nH/H;(i,i)}$,
and we have	
$\tilde{\rho}_{nH/H;(i,i)}|g\rangle=|\tilde{\rho}_{nH/H;(i,i);1}\rangle
=|{\bf \Psi}_{\alpha_{1};i}\rangle$ for $i\in NLS$ 
($|\tilde{\rho}_{H;(i,i);1}\rangle
=|{\bf \Psi}_{\alpha_{1};i}\rangle
\neq |\tilde{\rho}_{nH;(i,i);1}\rangle$ for $i\notin NLS$) 
and properly choosen $a=\pm 1$.
This can be further extended to
\begin{equation} 
	\begin{aligned}		
\tilde{\rho}_{nH/H;\{i\}}|g\rangle
=\sum_{\{i\}}|{\bf \Psi}_{\alpha_{1};i}\rangle
=\sum_{\{i\}}(-1)^{a}|\tilde{\rho}_{nH/H;(i,i);1}\rangle,
	\end{aligned}
\end{equation}
for subgroup $\{i\}\in NLS$ which means
the state
$\tilde{\rho}_{nH/H;\{i\}}|g\rangle$
can (and only can) be decomposed into the eigenvectors therein.
Difference between the NLS block and the else part of the rank-deficient subsystem
can be directly seem from the the overlap
$\frac{1}{D} \langle{\bf \Psi}_{\alpha_{1};i}|g\rangle$,
which equals to $\frac{1}{D}$ for $i\in NLS$ and $\lesssim \frac{1}{D}$ for $i \notin NLS$.
Such large overlap between the eigenvectors within NLS block, which are of 
highly-entangled eigenstates, with the low-entangled ground state,
is due to the coalesced eigenvalue (or reduced algebraic multiplicity)
such that the highly-entangled states are not random vectors but be mutually orthogonal
(thus be an orthonormal set)
due to the restricted Hilbert space.
This is different to the completely non-ergodic case like the Hermitian operator,
where $\langle E_{j}|(\frac{1}{D}\sum_{i}|E_{i}\rangle)=\delta_{ij}\frac{1}{D}$,
where the orthonormal set covers the whole non-degenerated spectrum.
There are special cases, e.g., for projectors $\mathcal{P}_{D}$ and $\mathcal{P}'_{D}$,
their eigenvectors form orthonormal set
but covers the whole eigenvalue-coalesced spectrum for the form projector,
and covers both the eigenvalue-coalesced/non-coalesced spectrum for the latter projector.
Such difference reflects the degrees of many-body localization
(or that of the restriction on Hilbert space) can be modified directly by changing the matrix rank.

For densities $\tilde{\rho}_{H;1}$ and $\tilde{\rho}_{H;2}$
which are of the unrestricted Hilbert space,
without exhibit any initial state recurrence, and the initial state informations are being wiped out quickly. their corresponding nontrivial state has ${\rm num}|\tilde{\rho}\rangle=
||\tilde{\rho}||=1$. This single nonthermal state 
Correspondently,
only the densities $\tilde{\rho}_{nH;3}$ and $\tilde{\rho}_{1H;3}$,
whose Loschmidt echo follows fixed pattern (without randomness) and the number of nontrivial eigenstate
satisfies  ${\rm num}[\langle \tilde{\rho}|]
=(||\tilde{\rho}||_{2}+1)=2$.

Next we define the following notation for the overlap with ground state:
$|\langle \rho|g\rangle|^2
:=\sum_{j}|\langle \rho|{\bf \Psi}_{\alpha_{1};j}\rangle|^2$,
which is necessary to preserve the symmetries
by filter the crossover components in off-diagonal channels.
For arbitary densities $\rho$ with random summation configuration,
its $j$-th eigenstate can be decomposed into a linear combination of certain set of eigenstates 
$|{\bf \Psi}_{\alpha_{1};i}\rangle$,
as far as $|\langle \rho_{j}|g\rangle|^2=1$.
Specifically,
for $\tilde{\rho}_{nH/H;(a,b)}$ with sum configuration $a,b\in NLS$,
$|\langle\tilde{\rho}_{nH;(a,b);i}|g\rangle|^2
=1$ for $i=1,\cdots,{\rm dim}[a,\cdots,b]$.
And correspondently,
the $i$-th eigenvector (for $i=1,\cdots,{\rm dim}[a,\cdots,b]$)
$\langle \tilde{\rho}_{nH/H;(a,b);i}|$ can be decomposed into
a linear combination of $\langle {\bf \Psi}_{\alpha_{1};j}|$'s
with $j=a,\cdots,b$:
$\sum_{j=a}^{b}
|\langle \tilde{\rho}_{nH/H;(a,b);i}|
{\bf \Psi}_{\alpha_{1};j}\rangle|^2
=1$.

The emergent non-thermal states can be observed by considering the overlap between certain densities 
(pure states or superposition of pure states):
$|\langle \tilde{\rho}_{nH/H;(a,b);i}|g\rangle|^2=
\sum_{j=1}^{D}|\langle \tilde{\rho}_{nH/H;(a,b);i}
|{\bf \Psi}_{\alpha_{1};j}\rangle|^2=
\sum_{j=a}^{b}|\langle \tilde{\rho}_{nH/H;(a,b);i}
|{\bf \Psi}_{\alpha_{1};j}\rangle|^2=
1$ for $a,b\in NLS$ and $\forall i\in \{1,\cdots,{\rm dim}[a,\cdots,b]\}$.
\footnote{Thus this again shows the discrete quantum Fourier transformation
between two mutually unbiased bases
such that $\langle \tilde{\rho}_{nH/H;(a,b);i}|g\rangle$ behaves like a exponential phase factor.}
Typically, we have
$|\langle \tilde{\rho}_{nH/H;3;i}|g\rangle|^2=
\sum_{j=1}^{D}|\langle \tilde{\rho}_{nH/H;3;i}
|{\bf \Psi}_{\alpha_{1};j}\rangle|^2=
\sum_{j\in NLS}|\langle \tilde{\rho}_{nH/H;3;i}
|{\bf \Psi}_{\alpha_{1};j}\rangle|^2=
1$ for $i=1,\cdots,D_{NLS}$,
and
$\sum_{j=1}^{D}
|\langle \tilde{\rho}_{nH/H;2;i}|
{\bf \Psi}_{\alpha_{1};j}\rangle|^2
=\sum_{j\in NLS}
|\langle \tilde{\rho}_{nH/H;2;i}|
{\bf \Psi}_{\alpha_{1};j}\rangle|^2
=1$ for $i\in NLS$.

Thus the non-defective degeneracy or the forward mapping in NLS blcok
generates such nontrivial result that
the unbiased superpositions of pure states within NLS block exhibits nonthermal character
(in opposite to the thermal case where the above-mentioned overlap is bounded
by $\frac{1}{D}$ which refers to the maximally mixed state\cite{Anza}).
Usually such initial density that of a restricted Hilbert space cannot be written in terms of spectral decomposition (diagonal ensemble).
Further, there is an additional conservation for the non-Hermitian basis compares to Hermitian basis,
$\sum_{j=1}^{D}\frac{1}{D}
\sum_{i=1}^{D}
|\langle\tilde{\rho}_{nH;(a,b);j}|{\bf \Psi}_{\alpha_{1};i}|^2=1$ for
arbitary configurations $(a,b)$,
while
$\sum_{j=1}^{D}\frac{1}{D}
\sum_{i=1}^{D}
|\langle\tilde{\rho}_{H;(a,b);j}|{\bf \Psi}_{\alpha_{1};i}|^2=1$ only for
 $a,b\in NLS$.
This difference is due to the initial state fidelity for the non-Hermitian basis
which efficiently break
the ergodicity on whole system (or energy surface),
and according to the Kolmogorov-Arnold-Moser theorem, this can be understood by the emergent
invariant tori in the phase space due to the local crimps of the energy surface\cite{Halder}
That is also why the latter case allows the global chaos for the whole system.
In Ref.\cite{Turner},
such nonthermal (nonergodic) states can be generated by a nonreciprocal "forward scattering
approximation" Hamiltonian 
whose existence is guaranteed by the parent Hamiltonian for the restricted Hilbert space.
Such conservation contributed by off-diagonal elements
obviously different to that described by the Bloch Chern number.

Any superposition of eigenstates of the NLS block
$\sum_{j\in NLS}\lambda_{j}|{\bf \Psi}_{\alpha_{1};j}\rangle$ is always an eigenstate of
$\tilde{\rho}_{nH/H;(a,b)}$
for $a,b\in NLS$ and $\{j\} \subset [a,b]$,
and 
\begin{equation} 
	\begin{aligned}	
&
\sum_{\{j\}}\lambda_{j}|{\bf \Psi}_{\alpha_{1};j}\rangle
=
\tilde{\rho}_{nH/H;(a,b)}^{\gamma}
\sum_{\{j\}}\lambda_{j}|{\bf \Psi}_{\alpha_{1};j}\rangle,\\
&
\sum_{\{j\}}\lambda_{j}|{\bf \Psi}_{\alpha_{1};j}\rangle
=\left(
\sum_{i=1}^{D}
|\tilde{\rho}_{nH/H;(a,b);i}^{*}\rangle\langle \tilde{\rho}_{nH/H;(a,b);i}|\right)^{\gamma}
\sum_{\{j\}}\lambda_{j}|{\bf \Psi}_{\alpha_{1};j}\rangle,
\end{aligned}	
\end{equation} 
such that an arbitary superpositions of the eigenstates of $\{j\}$
can be a decomposed into the superposition of the $1,\cdots,{\rm Tr}[\tilde{\rho}_{nH/H;(a,b)}]$-th
eigenvectors of
$\tilde{\rho}_{nH/H;(a,b)}$, or all eigenvectors ($i,\cdots,D$) of
$\sum_{i=1}^{D}
|\tilde{\rho}_{nH/H;(a,b);i}^{*}\rangle\langle \tilde{\rho}_{nH/H;(a,b);i}|$
(all with eigenvalue 1).
Such special symmetry is possible only for the rank-deficient subsystem,
 and the above equation is valid for all positive integer power $\gamma$,
which indicates a dimension-independent Krylov-subspace\cite{van}
and meanwhile well explain the regular evolution pattern
as shownw in Fig.\ref{echooverlap}(c),(f).
We further verify that
$\tilde{\rho}_{nH/H;(a,b)}\sum_{\{i\}}|{\bf \Psi}_{\alpha_{1};i}\rangle
=\sum_{\{j\}=[a,b]\cap \{i\}}|{\bf \Psi}_{\alpha_{1};j}\rangle,
\ a,b,i\in NLS$,

Specifically, for Hermitian basis,
$|{\bf \Psi}_{\alpha_{1};j}\rangle$ is always an eigenstate $\tilde{\rho}_{H;(a,b)}$
for $j\in [a,b]$.
While for
$\sum_{\j\}}|{\bf \Psi}_{\alpha_{1};j}\rangle$ is always the main eigenstate (principal component)
of
$\sum_{\{j\};\{j'\}}|{\bf \Psi}_{\alpha_{1};j}\rangle\langle {\bf \Psi}_{\alpha_{1};j'}|$
or
$\sum_{\{j\};\{j'\}}|{\bf \Psi}_{\alpha_{1};j}\rangle\langle {\bf \Psi}_{\alpha_{1};j'}^{*}|$
as long as ${\rm dim}[\{j\}\cap \{j'\}]\ge 1$.
As shown in Fig.\ref{echooverlap}(a)-(b),
the initial states of unrestricted Hilbert space
(by unbiased basis)
are $\sum_{\{i\{ \notin NLS}|{\bf \Psi}_{\alpha_{1};i}\rangle\langle {\bf \Psi}_{\alpha_{1};i}|$
and $\sum_{\forall \{i\}}|{\bf \Psi}_{\alpha_{2};i}\rangle\langle {\bf \Psi}_{\alpha_{2};i}|$,
unlike that of $\tilde{\rho}_{nH;3}$ which play an enssential role
in connecting two subsystems through off-diagonal fluctuations.

Typically, $\tilde{\rho}_{H;1}$ 
is the maximally mixed state (with all diagonal elements $\lesssim 1$
and the off-diagonal elements have zero mean and zero variance
such that the density is close to the identity operator; consistent with Refs.\cite{Popescu})
where
all eigenstates outside the NLS block are thermal states
and are forced to be equally overlapped with the choose ground state
($\lesssim\frac{1}{D}$)
and becomes indistinguishable due to the complete suppression on coherence.
Such environment-dependent overlap with initial state also referred
as priori probability in Refs.\cite{Popescu,Anza}.
While for $\tilde{\rho}_{H;3}$,
due to the existence of mutually-unbiased orthonormal basis
with strong entanglement,
it has special diagonal entries distribution
(see Fig) and the off-diagonal entries have nonzero mean 
(equal to$-D_{NLS}/(D^2-D)$) and nonzero variance.
Obviously from Fig.\ref{echooverlap} we see that $\tilde{\rho}_{H;3}$,
together with those in non-Hermitian basis,
cannot be thermalized (away from its initial state),
and thus should not be called thermal (canonical) ensemble states.

Due to the forward-scattering structure of the eigenvectors within NLS block,
these mutually-independent but (sub)system-size-dependent nontrivial nonthermal states
(with respect to ${\bf M}$)
describe a one-dimensional Hilbert subspace no matter how large $D_{NLS}(=\frac{D}{2}-1)$ is.
While $D_{NLS}(D_{NLS}-1)>2$ is satisfied as far as $D_{NLS}\ge 2$,
such that only the collection of orthonormal eigenvectors withi NLS block satisfies
\begin{equation} 
	\begin{aligned}	
\bigg|\sum_{i\in NLS}\langle{\bf \Psi}_{\alpha_{1};i}| {\bf \Psi}_{\alpha_{1};j}\rangle \bigg|^2
=\frac{ \sum_{i\in NLS}\langle{\bf \Psi}_{\alpha_{1};i}|\tilde{\rho}_{nH/H;3}
\sum_{i\in NLS}|{\bf \Psi}_{\alpha_{1};i}\rangle}
{D_{NLS}}
=
\langle{\bf \Psi}_{\alpha_{1};i}|{\bf \Psi}^{*}_{\alpha_{1};i}\rangle\ (\forall i)
=1,
\end{aligned}	
\end{equation} 
where $j\in NLS$.
This relation again reflects a (sub)system-size-dependence for the degenerated eigenstates within NLS block, and consistent with the theorem 1 of Ref.\cite{Anza}
where our case should corresponds to the zero spin such that the effective dimension of highly degenerated
Hilbert subspace always reduced to 1 to avoid the many-body localization in the subsystem ${\bf \Psi}_{\alpha_{1}}$.
That means there is only one valid pure state as a collection of highly-entangled 
eigenstates with NLS for such a bipartite system.
Thus in conclusion,
the non-defective degenerated eigenstates within NLS block can be treated as thermal state with respect to the rank-defective subsystem ${\bf \Psi}_{\alpha_{1}}$ in the maximally mixed limit
(i.e., proximity to canonical ensemble), which are subsystem-size dependent and thus unthermalized with respect to the global system; while part of (other that the one closest to the NLS block) the rest eigenstates
thermal with respect to the global system.

The difference between the reduced states (unitary invariant
under Haar measurement; i.e., product of collections of eigenvectors)
and the maximally mixed density in terms of ensemble (thermalized)
is  significant.
An example is teh above-mentioned restriction (on Hilbert space) emergent from rank reducing from 
$\mathcal{P}_{D}$ to $\mathcal{P}'_{D}$.
Here,
the pure state superposition (mixed state) within the NLS block $\tilde{\rho}_{nH/H;3}$
is different to the reduced state
$\mathcal{Q}_{H;3}$
whose bulk part of the spectrum is broadened by the reduced rank
and as a result, raise the thermalization and strong entanglement within the degenerated bulk.

While it is proved\cite{Shirai,Popescu} that each pure state (fully thermalized equilibrium steady state)
of $\mathcal{Q}_{H;3}$ becomes locally indistinguishable from that of the maximally mixed state $\tilde{\rho}_{H;3}$ (the restricted one).
In the mean time,
$\bigg|\langle{\bf \Psi}_{\alpha_{1};i}| {\bf \Psi}_{\alpha_{1};i}\rangle \bigg|^2
=
\langle{\bf \Psi}_{\alpha_{1};i}|\tilde{\rho}_{H;1}|{\bf \Psi}_{\alpha_{1};i}\rangle
\lesssim 1$ for $i\notin NLS$,
is a significant property for the macroobservables that support the ETH\cite{Anza}.
In fact, the $\tilde{\rho}_{H;1}$ has a weaker homogenization on the diagonal elements
compares to $\tilde{\rho}_{nH/H;3}$
(see Fig.\ref{EO6}(e)-(h)),
which is consistent with the maximally mixed density $\tilde{\rho}_{nH/H;3}$
with highest entanglement between the eigenvectors therein
(while $\tilde{\rho}_{H;1}$ is only almost be maximally mixed)
despite the exact homogenization only happen in the block with dimensional up to 
$(D_{NLS}+1)$ as discussed below Eq.(\ref{1225}).

Down to the degenerated levels where each eigenstates can be regarded as maximally thermalized ($N\rightarrow\infty$ limit) with respect to the corresponding subsystem
and be steady where we can nomore expecting any entanglement in both the non-Hermitian and Hermitian basis (see Fig.\ref{echooverlap}(c),(f))
\footnote{This can be further evidenced by the complete energy degeneracy (even down to log scale) for these pure states as well as the same off-diagonal elements $-\frac{2}{D}$ which vanish in large size},
and thus guarantee the ETH.

As shown in Fig.\ref{echooverlap3},
the boundaries of Loschmidt echo evolutionary spectrum provide a direct
evidence for whether the current system can be thermalized:
The system can be thermalized following ETH if the upper and lower boundaries
deviate from (and without revivals to) their initial values
in which case the spectrum must be regularly-patterned.
This class of systems is $\tilde{\rho}_{H;\{c\}}$ with arbitary sampled set $\{c\}\neq NLS$,
i.e., the system will be thermalized away from initial state as long as there including at least 
one subsystem $\tilde{\rho}_{H;\{c\}}$ ($\{c\}\notin NLS$) and meanwhile the rest part
$\tilde{\rho}_{H;\{c\}}$ ($\{c\}\in NLS$) will not cause any localization.
In such class of system, each subsystem therein can be treated as the commuting macroobservables\cite{Anza,Bocchieri}
with entangled spectrum boundaries,
which guarantees the global thermalization.
The system cannot be thermalized if the upper and lower boundaries do not
deviate from (or exhibit persistive recurrence to) their initial values.
This class of systems including $\tilde{\rho}_{nH;\{c\}}$ ($\forall \{c\}$) and $\tilde{\rho}_{H;3}$.
As shown in Fig.\ref{echooverlap3}(),
the spectrum keep regular as long as $\{c\}$ involving
zero or one subsystem outside the NLS region, i.e., $\tilde{\rho}_{nH;(a,a)}$ ($a\notin NLS$),
which means each quantum state entangled between subsystems
$\tilde{\rho}_{nH;(a,a)}$ ($a\notin NLS$) and $\tilde{\rho}_{nH;\{c\}}$ ($\{c\}\in NLS$)
and related to the boundaries of these subsystems.

The macroobservables $\tilde{\rho}_{H;(a,a)}\ (a\notin NLS)$ that support the global thermalization,
can be diagonalized by $|{\bf \Psi}_{\alpha_{i}}\rangle$, and
the spectrums for different subspaces have mutually commuting boundaries while each subspace contains more
non-defectively degenerated eigenstates
(such that the eigenvalues are seperated in log scale which is in consistent with the Haar measurement result of Ref.\cite{Anza})
and behave closer to the mutually unbiased bases in NLS region
(comparing (o) and (q) to (p) and (r), respectively).
In Fig.\ref{echooverlap3}(o),
the spectrum for $\tilde{\rho}_{H;(a,a)}\ (a=1,\cdots,4)$ are almost coalesced
which implies that each macroobservable here is nonextensive and there without coherence between them.
Thus the ETH guaranteed for Hermitian basis
is due to the indistingushibility
between the expectation under unitary invariant (Haar) measurement on the pure states (subspaces)
and that for the ensemble averaged state (of some local observable; like $\mathcal{P}_{D}'$).
Comparing (o) and (p) in Fig.\ref{echooverlap3},
we can see that although both the subspaces in Hermitian basis and non-Hermitian basis
are macroobservables,
those in Hermitian basis (o) exhibit much less "restriction" than that in (p):
Here the restriction refers to the character that
one of eigenstate (or eigenspace) is overwhelming among all,
such that the corresponding subspace is extensive.
While for Hermitian basis,
the subspaces outside the NLS region ($\tilde{\rho}_{H;(a,a)},\ a=1,\cdots,5$)
are the so-called typical states\cite{Reimann,Steinigeweg,Hamazaki}
due to the existence of non-defective degenerated eigenstates in the NLS block
which actually provide a infinitely degenerated nullspace.
This can also be evidence by the last row of Fig.\ref{EO4},
where arbitary combination (even with overlap) always results in a mesh-like spectrum
with nonextensive boundaries.
That means there could be exponential partition on the Hilbert space
by the macroobservables
(e.g., for a set of macroobservables with degrees-of-freedom 
$(D-D_{NLS})$ there could be a dimensionality up to $10^{D-D_{NLS}}$ for the Hilbert space\cite{Reimann}).
Since there will be much larger amount for the degrees-of-freedom
when consider the time-dependence of evolved quantum states,
it will be an enormous amount of dimensionality 
that capable for the mutually entangled (with level repulsion)
quantum states.

More importantly,
as indicated in Fig.\ref{echooverlap3}(a)-(c),
for Hermitian basis where each density is a local macroobservable,
the spectrum boundaries strictly correspond to nonextensive
sum (statistical mixture) of the mutually independent (but entangled) 
quantum states therein.
Such mixture of thermal states within each subspace can be visualized
by the spectrum with different time interval (narrower and wider time intervals
correspond to larger and lower frequencies or energies, respectively),  
where, as shown in Fig.\ref{EO4}(c,g,k,o),
each quantum state form a line-shape spectrum and participate the common
boundaries by their own varying period.
Thus a complete (with respect to the correspondig local macroobservable) set of
lines form a nonextensive mesh. 
In this perspective,
there should be infinite quantum states within each subspace.
The regularly-patterned mesh-like spectrums for the echo of pure state and their mixture
are due to the entanglement between quantum states (distinguished by the
contact moments to the boundaries).
While the coherence required by the locality for each subspace (local macroobservable)
can be seem in high-frequency limit (Fig.\ref{EO4}(d,h,l,p))
where the spectrum is a (dramatically oscillating) single curve
consist of the mesh-pattern quantum states (whose contacts to the boundary are nearly indistinguishable)
that are entangled to each other
and note that here all the existing mesh-pattern quantum states 
form a complete set of coherence
as evidenced by the line-shape spectrum (without any thermal relaxation)
which exhibits complementary effect as well as the strongest disentangling.
Here it is the oscillations (strongly variant echo during short time)
enlarging the "bulk" that is containable for the thermal states.
The mesh-pattern spectrum is available only when there is a large number of thermal states
or nonthermal states (in which case the coherence happen among the boundary).
While the asymptotically degenerated thermal states (indistinguishable)
results in the merged mesh (also, infinitely dense mesh) as indicated by arrows
in Figs.(\ref{EO5},\ref{EO6}),
and such merging will not be repeated during the evolution for a nonextensive boundary.

Similarly,
arbitary combination of subspaces in Hermitian basis (each one with mesh-pattern spectrum)
results in mesh-pattern spectrum (soly through entanglement) until
the combination is in (arbitary multiple of) full set such
that containing $\tilde{\rho}_{H;(a,a)}\ (a=1,\cdots,D)$.

While in the limit of short-time scale,
the combination of subspaces of incomplete set (full set)
where each one with a line-shape spectrum, results in
the oscillating line-shape spectrum (static line-shape spectrum; which can also be viewed as
infinitely dense mesh-like spectrum)
through entanglement (coherence).

Thus we can conclude that the complementary effect (coherence) between 
mesh-like spectrums form a line-shape spectrum that corresponds to a local observable.
This happen for complete set of mesh-like spectrums like 
all subspace in Hermitian basis form line-shape spectrum in Fig.\ref{echooverlap}(d),
and meanwhile, down to short-time (high-energy) limit all mesh-like spectrum
(among all the high-frequency region) coherent to form a line-shape spectrum
which is dramatically oscillating; 
While the entanglement between line-shape spectrums form mesh-like spectrum
that corresponds to a local macroobservable.

While in non-Hermitian basis, as shown in Fig.(\ref{EO6}),
each subspace  
with mesh-pattern spectrum can still be decomposed into the entangled mesh-pattern spectrums.
Arbitary subspaces outside the NLS (as an extensive local observable)
able to coherent to each other and form a chaotic (with large recurrence)
spectrum with many-body localization
which still able to decomposed into entangled mesh-pattern spectrums.
Each times of coherence create new repeating merging along the boundary with 
larger and larger periodicity,
and the preserving nonthermal states therein
allow the further coherence with other extensive local macroobservables.
Futher short-time detail exhibit line-shape spectrum (Fig.\ref{EO5}(d,h,l,p))
that containing the entangled mesh-pattern spectrums
in a complete set of coherence.

Different to Hermitian basis, the "complete set" for non-Hermitian basis
is inavailable by the diagonal part
(e.g., Fig.\ref{echooverlap}(a) as a combination of all subspaces in non-Hermitian basis
does not be a line-shape spectrum but still a broadened one with chaos).
Each time of coherently combination between 
 extensive subspaces results in more chaotic
spectrum,
where the coherence cancel the complementary degrees-of-freedom
in the extensive boundaries of the subspaces
and results in new degrees-of-freedom (another extensive boundary)
which is of higher frequency and compatible with the conservations formed
by the old degrees-of-freedom.

We show that only $\mathcal{Q}_{nH;1}$
exhibits complete many-body localization (a complete set of coherence)
and cannot be thermalized ($\in\rho_{res}$).
While $\mathcal{Q}_{H;1}$
becomes a quasi-equilibrium state (with drastic fluctuation) at long time
and preserve vanishingly small overlap with initial state, thus it can be thermalized
(with respect to some certain environment, and thus $\in\rho_{unres}$).
This is further evidence by their corresponding VIE spectrum,
where only for $\mathcal{Q}_{nH;1}$, there is level
repulsion for local observables
with two parallel upper-boundaries with vanishing slope in long time limit as
shown in Fig.\ref{EO6}(h),
during which process the off-diagonal part plays an essential role in subsystem conservation.

The Loschmidt echo provide a visualized way for the ETH diagnosis
and in consistent with our VIE results.
Thus for those densities with similar echo spectrum, they should share similar character in VIE spectrum too.
Specifically,
$\tilde{\rho}_{nH/H;3}$ and $\mathcal{Q}_{nH/H;3}$ perform similar echo
as well as the VIE spectrum.
This happen only in the NLS block with the mutually unbiased (and orthonormal) bases
where the subapces therein
will not obstruct the thermalization in Hermitian basis
(Fig.\ref{echooverlap3}(g)-(j)),
and also free from coherence with other subspaces in non-Hermitian basis (Fig.\ref{echooverlap3}(k)-(n)).
Thus this part of mixed/reduced densities meet the postulate proposed in Ref.\cite{Popescu},
that the pure states become (locally) indistinguishable for those from mixed density
and those from reduced density.
This indeed reflects the limit where the
maximal mixing of pure states (as well as maximal entropy) is achieved and each pure state be equiprobable.
Despite this equilibrium result quite like the occasion of thermodynamic limit,
as we discuss in above
it is indeed caused by its eigenvector structure,
which is, in other word, its initial state information.
For the former case,
it is the asymptotical degeneracy in large-$N$ limit case such
proximity between mixed (canonical) and reduced densities.
While in our case,
there can be an effective temperature (see Appendix),
reflecting the restriction-induced generalized thermal canonical\cite{Popescu,Anza} or Gibbs\cite{Hamazaki} ensembles.
And also, 
different to condition of exponentially large environmental density of states
or astronomically large time sclae to eliminate the off-diagonal contribution\cite{Anza},
the coexistence of entanglement (due to the linear map or
FSA) and initial state fidelity in NLS region
results in such indistingushability, 
even for the mixed density here not be a thermal canonical ensemble
(where a complete insensitivity to the environmental energy or temperature 
is achieved here\cite{Anza}).
Such indistingushability for the $\tilde{\rho}_{nH/H;3}$ and $\mathcal{Q}_{nH/H;3}$
is indeed caused by the weak coherence between
these two densities,
which is possible since the steady (quasi)equilibrium is achieved when including the off-diagonal component.

For maximally mixed states of macroobservables (in Hermitian basis),
the VIE spectrum without level repulsion is in consistent with Ref.\cite{Anza}
that a macroscopic system can only measure a
small number of observables,
e.g., the two observables ${\bf J}^{T}_{\alpha_{1}}{\bf J}_{\alpha_{1}}$
and ${\bf J}^{T}_{\alpha_{2}}{\bf J}_{\alpha_{2}}$ are indistingushable here
according to their expectation with respect to the density under unitary evolution.
While for reduced density there could be infinite number of local-type observables be measured
whose expectations constitute a continued distribution.

From the spectrums in log scale for the corresponding densities
(and compares to Fig.(\ref{echooverlap}(a)-(f))),
we also see that the broaded echo spectrum
at a single time through the mesh-like pattern
is directly related to the splitting of levels.
Also,
the existence of dominating single eigenspace (the first one)
for the reduced densities reveal the restriction on Hilbert space,
which is necesary for the initial state fidelity (recurrence)
as the dominating eigenspace overcomes the effect from trace class
and broaden the bulk that compatible with a large number of degenerated states.

It is hard to estimate if there is coherence between $\tilde{\rho}_{nH;2}$ and $\tilde{\rho}_{nH;3}$.
from their echo spectrum.
But this can be done by their long-time average with respect to a single local observable.
We found that
$e^{i\tilde{\rho}_{nH;1}t}
\rho_{0}e^{-i\tilde{\rho}_{nH;1}t}=\rho_{0}$ for $\rho_{0}=\tilde{\rho}_{nH;1},\tilde{\rho}_{nH;2},\tilde{\rho}_{nH;3}$,
which is a direct evidence that
$\tilde{\rho}_{nH;2}$ and $\tilde{\rho}_{nH;3}$ only entangled with each other
with their individual boundaries (periodicity) only determined by that of $\tilde{\rho}_{nH;3}$.
Also,
the coherence exist along the direction that orthogonal to the effect from $\tilde{\rho}_{nH;1}$ 
is not complete which is evidenced by the 
finite period of $\tilde{\rho}_{nH;1}$.
While a complete set of coherence should results in a inifinite period one 
without any broaden of spectrum can be seem,
like Fig.\ref{EO5}(j) for non-Hermitian basis 
and Fig.\ref{echooverlap}(d) for Hermitian basis. 
While for Hermitian basis,
$e^{i\tilde{\rho}_{H;1}t}
\rho_{0}e^{-i\tilde{\rho}_{H;1}t}=\rho_{0}$ for $\rho_{0}=\tilde{\rho}_{H;\{c\}}\ (\forall \{c\})$,
which is consistent with the fact that there is no coherence along the boundaries between arbitary
densities in Hermitian basis ($\tilde{\rho}_{H;\{c\}}$ and $\mathcal{Q}_{H;\{c\}}$),
or in other word,
there is always a definite (well-defined) eigenvalue for any one of them
originates from the expectation from some disorder.
In this case, the merging of mesh-pattern is caused directly by the disentangling (non-Hermitian skin)
effect from (environmental) disorder,
and for a complete coherence set, where the coherence must be and only can be happend
along the direction orthogonal to the entangled boundaries,
the echo becomes a single curve (Fig.\ref{echooverlap}(d)).
In fact for both the Hermitian and non-Hermitian bases,
strong enough coherence along such "allowed" direction enable to cause the merging 
(to a line-shape or infinitely dense mesh spectrum),
and one nature condition to realizing this is large enough number of thermal states
which will leads to asymptotic degeneracy and enforce the thermal equilibrium.
However, the subspaces of the NLS are the exceptions,
where it is the initial state fidelity prevents the happening of thermal equilibrium.
Such merging in incomplete set of coherence also indicated by the arrows in Fig.\ref{EO4}.
While the merging in Fig.\ref{EO5}(k) in non-Hermitian basis is of the difference case,
which is caused by the coherence along the boundary (i.e., the part of many-body localization)
instead of unique environmental disorder.
This is also why the mesh-mesh entanglement cannot be seem from the mesh-pattern echos in Hermitian basis
$\mathcal{L}(\tilde{\rho}_{H;\{c\}})$ and $\mathcal{L}(\mathcal{Q}_{H;\{c\}})$,
where the spectrum can only be decomposed into entangled line-shape spectrums
or one single mesh.
In fact,
for Hermitian basis
the arbitary sum configuration where only entanglement happen among the boundary
results in the spectrum that can either be a single mesh (narrower time interval) 
or multi curves (larger time interval)
due to the preclusion of nonthermal state by the virtue of commuting macroobservables
(with level repsulion between them).
For a clear illustration on this,
we shown in Fig.\ref{EO7} a comparasion of the spectrums of $\mathcal{Q}_{H;(a,b)=(2,6)}$ and 
$\tilde{\rho}_{nH;\{c\}=(3,5)}$.
Interestingly,
a thermalization process (dynamically) can be found 
for $\tilde{\rho}_{H;1},\tilde{\rho}_{H;2},\tilde{\rho}_{H;3}$ with respect to
the unitary evolution of $\tilde{\rho}_{nH;3}$, such that
$\lim_{T\rightarrow\infty}
\frac{1}{T}\int^{T}_{0} dt{\rm Tr}[e^{i\tilde{\rho}_{nH;3}t}
\tilde{\rho}_{H;\alpha}e^{-i\tilde{\rho}_{nH;3}t}\mathcal{O}]
={\rm Tr}[\tilde{\rho}_{H;\alpha}\mathcal{O}]+o(1)$ ($\alpha=1,2,3$).

\begin{figure}
	\centering
	\includegraphics[width=0.8\linewidth]{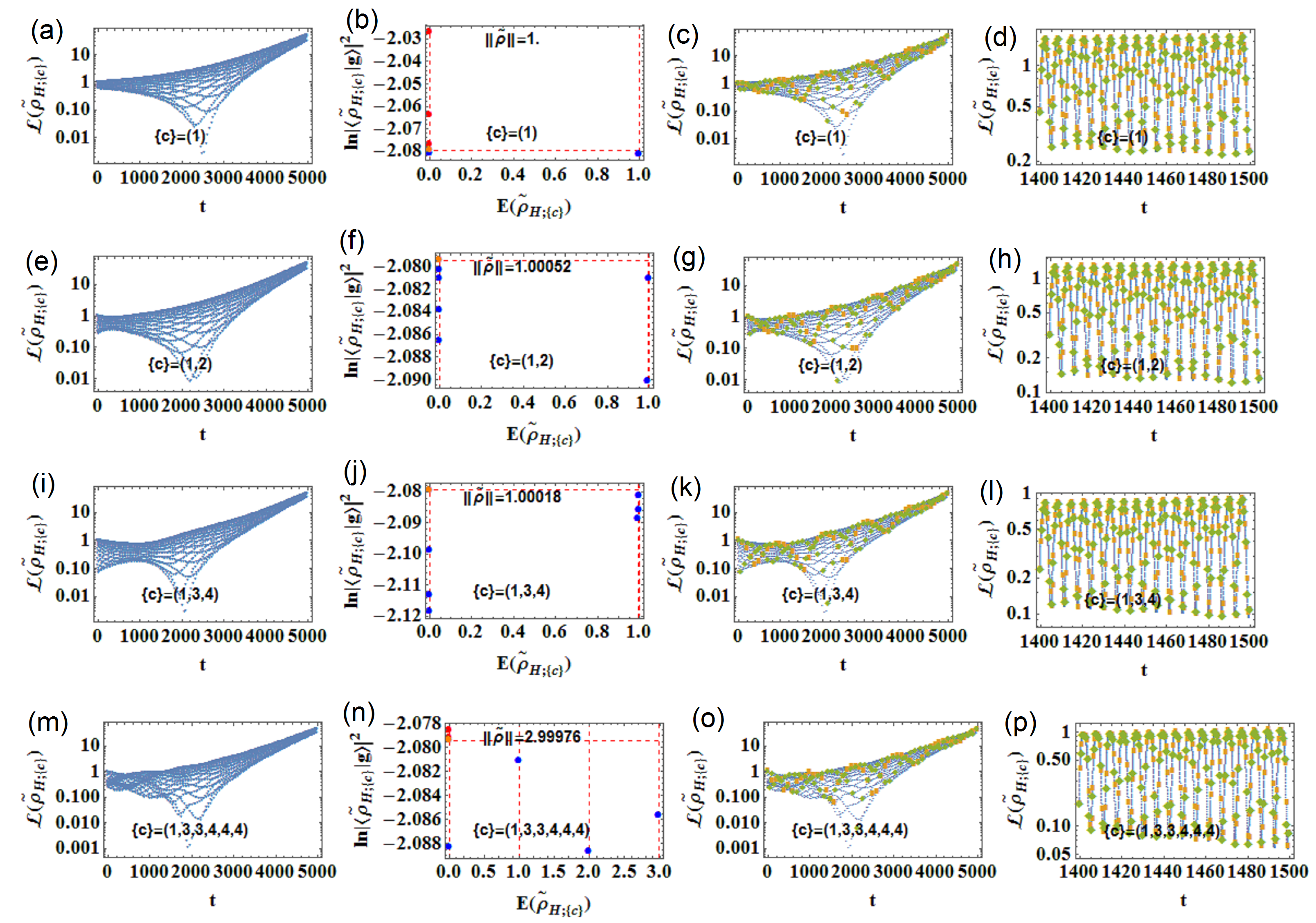}
	\captionsetup{font=small}
\caption{Echo for single subspace and their mixture in Hermitian basis
which can be regarded as the local macroobservables as the result of partition in Hilbert space,
where each mesh-like spectrum consist of only the entangled line-shape quantum states.
There are highly entangled quantum states within each subspace,
and a complete set of subspaces in Hermitian basis leads to ETH,
which is consistent with results of Ref.\cite{Anza}.
Unlike the non-Hermitian basis,
each macroobservable here can be diagonalized by the bases
which are exactly (very close to) the mutually-unbiased basis in (outside) the NLS region,
with high degeneracy as well as strong entanglement.
	}
\label{EO4}
\end{figure}

\begin{figure}
	\centering
	\includegraphics[width=0.8\linewidth]{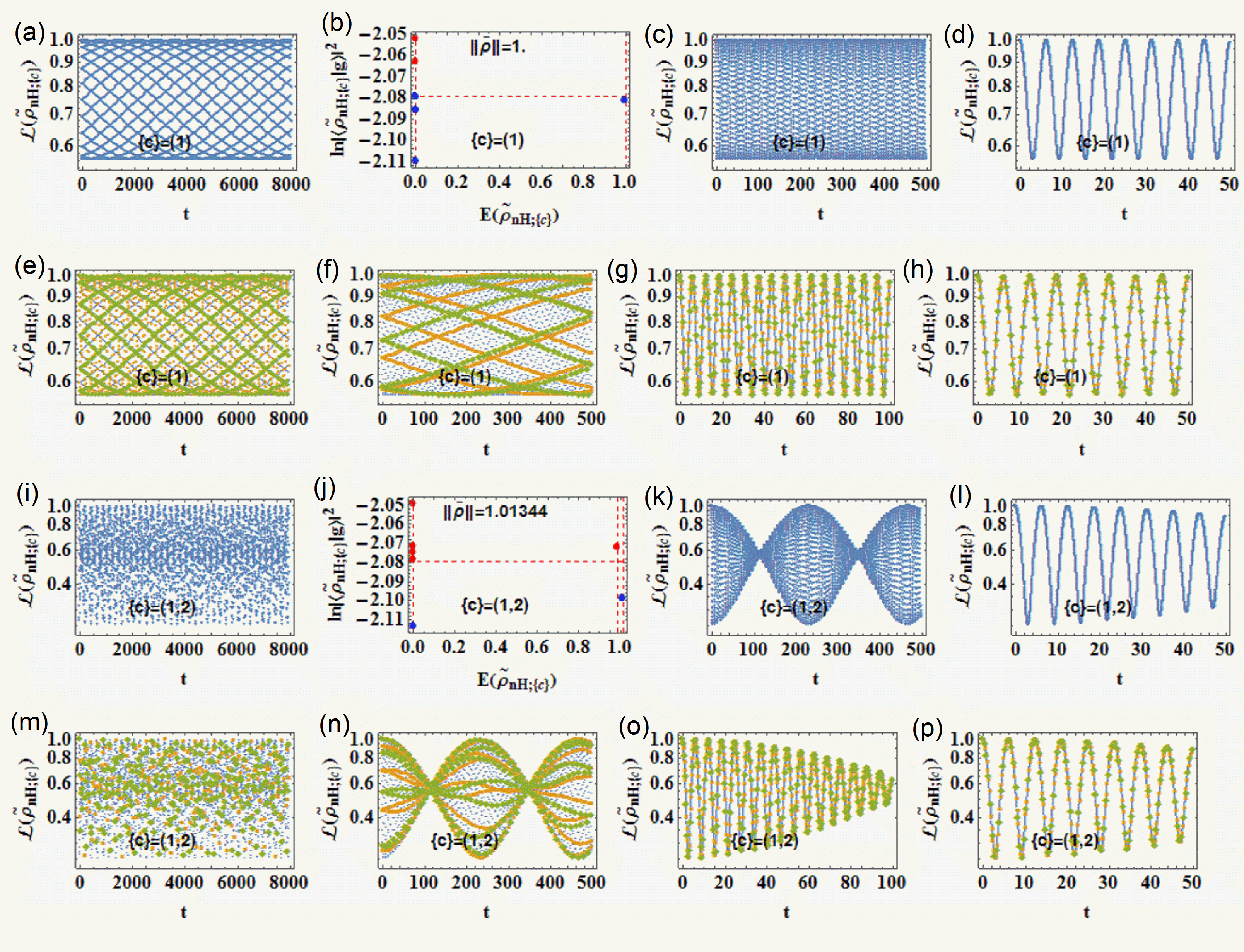}
\captionsetup{font=small}
\caption{
Echo for each subspace and their mixture in non-Hermitian basis,
which is the coherented mesh-like patterns that each one be formed by the superposition of 
line-shape quantum states as can be observed
 down to the UV limit (short-time cutoff).
Coherented mesh-like patterns cause warpped boundaries in the spectrum
which appear periodically such that it will not obstruct the formation of new conservation
through the further coherences.
More complex the coherence set is,
longer time is needed to exhibit periodicity in the echo spectrum,
which is to compatible with the conservations in shorter periodicity
(e.g., see Fig.\ref{echooverlap}(a)-(b)).
Unlike the Hermitian basis,
a complete set for coherence is inavailable to realize complete conservation
(many-body localization)
in which case the spectrum merge to a single line.
As shown in Fig.\ref{EO6}(f),
such localization requires a full (diagonal and off-diagonal)
coherence set (reduced density).
	}
\label{EO5}
\end{figure}

\begin{figure}
	\centering
	\includegraphics[width=0.8\linewidth]{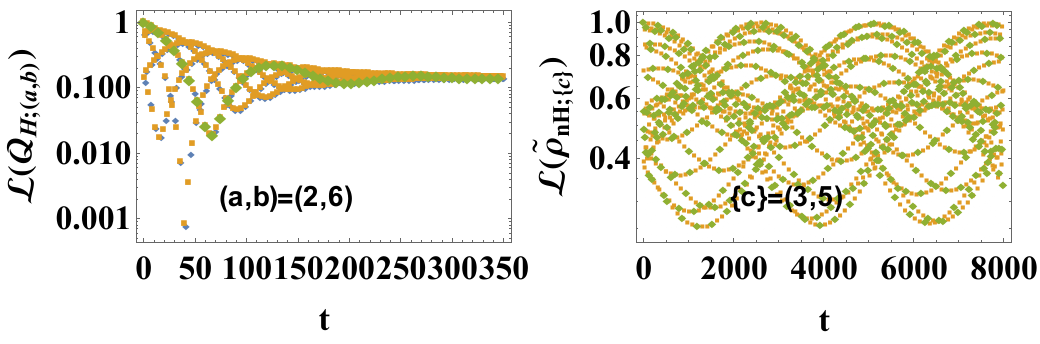}
	\captionsetup{font=small}
\caption{Echo for teh densities $\mathcal{Q}_{H;(a,b)=(2,6)}$ and 
$\tilde{\rho}_{nH;\{c\}=(3,5)}$.
	}
\label{EO7}
\end{figure}

\begin{figure}
	\centering
	\includegraphics[width=1\linewidth]{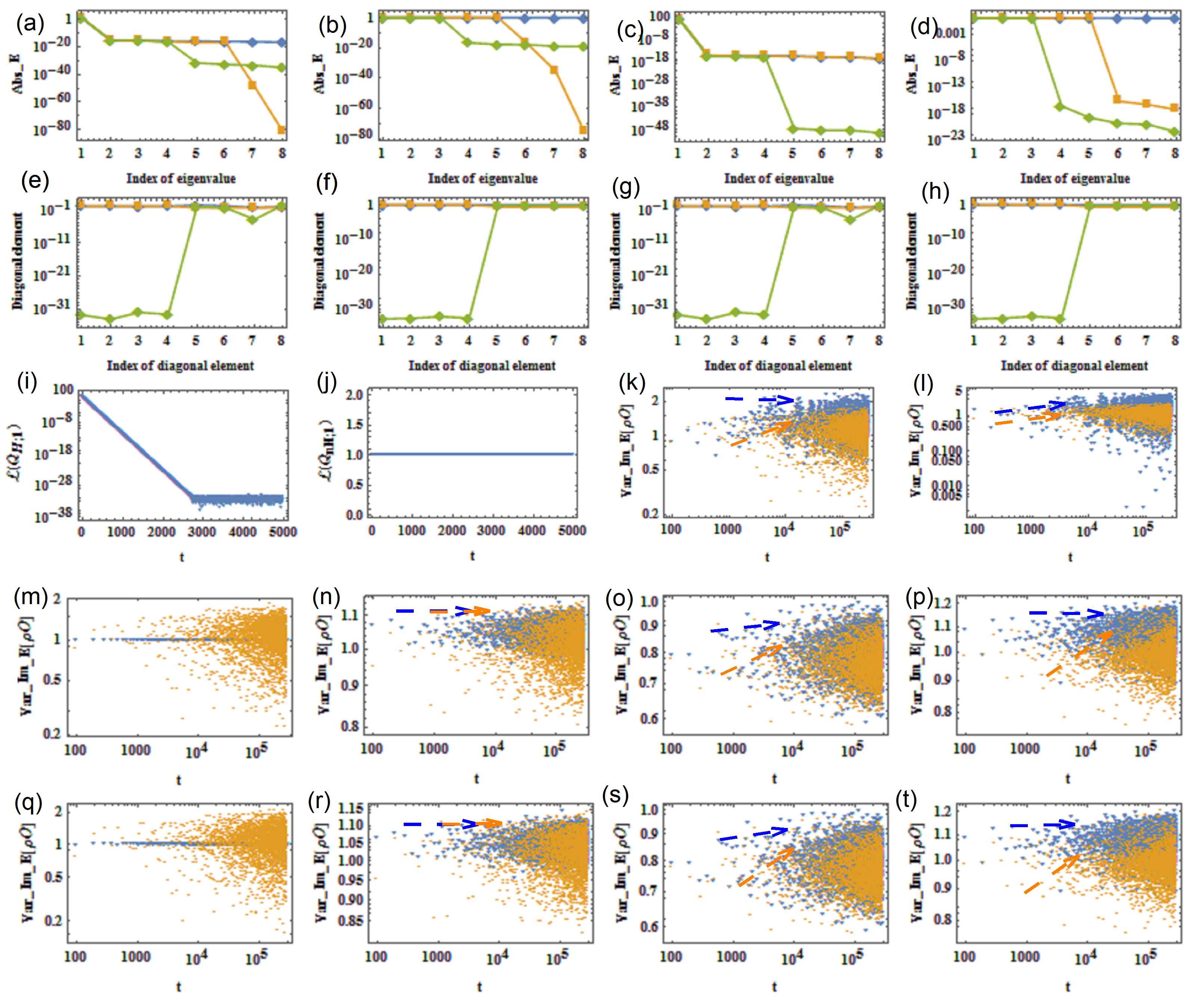}
\captionsetup{font=small}
\caption{
(a) Absolute eigenvalues for $\mathcal{Q}_{H;1}$ (blue), $\mathcal{Q}_{H;2}$ (orange)
and $\mathcal{Q}_{H;3}$ (green).
(b) Absolute eigenvalues for $\tilde{\rho}_{H;1}$ (blue), $\tilde{\rho}_{H;2}$ (orange)
and $\tilde{\rho}_{H;3}$ (green).
(c) The same with (a) but for non-Hermitian basis.
(d) The same with (b) but for non-Hermitian basis.
(e)-(h) Diagonal elements of the density operators following the legends of (a)-(d).
(i),(j) show the Echos for the densities $\mathcal{Q}_{H;1}$ and $\mathcal{Q}_{nH;1}$, respectively.
(k),(l) show the VIE for the densities $\mathcal{Q}_{H;1}$ and $\mathcal{Q}_{nH;1}$, respectively.
The orange spectrum corresponds to the local observable ${\bf J}^{T}_{\alpha_{1}}{\bf J}_{\alpha_{1}}$,
while the blue spectrum corresponds to its counterpart
(i.e., replacing the subscript $\alpha_{1}$ by $\alpha_{2}$).
(m)-(p) present a comparasion between the VIE for a part of mixed (blue) and 
reduced (orange) densities in Hermitian basis
in subscript of $\alpha_{1}$ with respect to the observable ${\bf J}^{T}_{\alpha_{1}}{\bf J}_{\alpha_{1}}$, for sum configurations
$\{c\}=1,2,(6,7),3$, respectively.
For example,
(m) shows a comparasion between the VIE's between the densities 
$\tilde{\rho}_{H;1}$ (blue) and $\mathcal{Q}_{H;1}$ (orange).
(q)-(t) present the results the same with (m)-(p) but for non-Hermitian basis.
The VIE in panel (k) for two subscripts $\alpha_{1}$ and $\alpha_{2}$ 
does not show well-performed level repulsion (not the thermal states),
which is in agree with the fact that $\mathcal{Q}_{H;1}$
does not corresponds to a restricted Hilbert space, 
despite the off-diagonal part cancels the disorder which cause the thermalization 
of its diagonal counterpart $\tilde{\rho}_{H;1}$ as evidenced by the quasi-steady-state in long-time limit (i).
While for $\mathcal{Q}_{nH;1}$, the VIE (l) exhibits well-performed level repulsion
as indicated by the two parallel and horizontal dashed arrows
(like what happen in Fig.\ref{34_41}(b)),
in agree with the static and initial-state-preserved echo shown in (j).
Regarding the comparasion between mixed/reduced densities,
it shows that,
only within NLS, the corresponding mixed state and reduced state exhibit complementary VIE
with coherence, which 
indicates the indistinguishable quantum evolutionaries between mixed and reduced densities.
This phenomenon is only possible in NLS,
as we discussed in Sec.3.
Such complementary VIE exhibits: one spectrum evolves toward the plateau and the other one reaches the same height after a dip-ramp-plateau process.
	}
\label{EO6}
\end{figure}

\section{Linear map in NLS block}

As shown in Fig.\ref{log}(a), 
the non-defective degeneracies in the NLS region as well as the non-Hermiticity relies on the 
coalesced but still robustly continued eigenvalues of the NLS block embeded into ${\bf \Psi}_{\alpha_{1}}$.
This is indeed necessary to keep the geometric multiplicity
of the NLS block (nullspace) equals to $D_{NLS}$.
Inevitably,
$\frac{1}{\lambda_{\alpha_{1};i}}
{\bf \Psi}_{\alpha_{1}}|{\bf \Psi}_{\alpha_{1};i}\rangle
\neq |{\bf \Psi}_{\alpha_{1};i}\rangle$ for $i\in NLS$.
In fact,
within the NLS region,
the above expression should becomes a functional derivative
with the function $\lambda_{\alpha_{1};i}$
\begin{equation} 
	\begin{aligned}		
\frac{\delta {\bf \Psi}_{\alpha_{1}}|{\bf \Psi}_{\alpha_{1};i}\rangle}
{\delta \lambda_{\alpha_{1};i}}
= |{\bf \Psi}_{\alpha_{1};i}\rangle_{f},
	\end{aligned}
\end{equation}
which
can be realized by an eigenvalue-reparameterization
transformation $\lambda_{\alpha_{1};i} \rightarrow f(\lambda_{\alpha_{1};i})$
with $f(\lambda_{\alpha_{1};i})$ a diffeomorphism
$\lambda_{\alpha_{1};i}\rightarrow \lambda_{\alpha_{1};j}$.
By considering the eigenvalues as the Lebesgue-Borel measure
on Borel sets (the two measurable sets corresponding to the linear map of $f(\lambda_{\alpha_{1};i})$),
and the derivative of $f(\lambda_{\alpha_{1};i})$ on $\lambda_{\alpha_{1};i}$ as a GL($D_{NLS}$,$\mathbb{R}$) mapping $\mathbb{R}^{D_{NLS}}\rightarrow \mathbb{R}^{D_{NLS}}$,
the requirement of (monotonic) differentiability of the function $f(\lambda_{\alpha_{1};i})$ can be replaced by the invertibility of $f(\lambda_{\alpha_{1};i})$:
\begin{equation} 
	\begin{aligned}		
		\frac{\delta {\bf \Psi}_{\alpha_{1}}|{\bf \Psi}_{\alpha_{1};i}\rangle_{f}}
		{\delta \lambda_{\alpha_{1};i}}
=		\frac{\partial {\bf \Psi}_{\alpha_{1}}|{\bf \Psi}_{\alpha_{1};i}\rangle_{f}}
{\partial f(\lambda_{\alpha_{1};i})}
	\frac{\partial f(\lambda_{\alpha_{1};i})}
	{\partial \lambda_{\alpha_{1};i}}
	=	\frac{\partial {\bf \Psi}_{\alpha_{1}}|{\bf \Psi}_{\alpha_{1};i}\rangle_{f}}
	{\partial f(\lambda_{\alpha_{1};i})}
	\frac{\lambda_{\alpha_{1};i}}{|{\rm Det}
	[	\partial_{\lambda_{\alpha_{1};i}}f(\lambda_{\alpha_{1};i})
		]|},
	\end{aligned}
\end{equation}
where
\begin{equation} 
	\begin{aligned}		
		&
 f(\lambda_{\alpha_{1};i})=
 	\frac{\lambda_{\alpha_{1};j}}{|{\rm Det}
 [	\partial_{\lambda_{\alpha_{1};i}}f(\lambda_{\alpha_{1};i})
 	]|\circ f(\lambda_{\alpha_{1};i})^{-1}},\\
 &
 |{\rm Det}
 [	\partial_{\lambda_{\alpha_{1};i}}f(\lambda_{\alpha_{1};i})
 ]|= f^{-1}(\lambda_{\alpha_{1};j})
 \lambda_{\alpha_{1};i}^{-1}.
	\end{aligned}
\end{equation}
Lebesgue-Borel measure on the Borel sets has the following property base on the invertible map $f$:
\begin{equation} 
	\begin{aligned}		
		f(\lambda_{\alpha_{1};i}(\upalpha))=\lambda_{\alpha_{1};i}
		(f^{-1}(\upalpha))
		=\sum^{\upalpha}
		 	\frac{\lambda_{\alpha_{1};j}}{|{\rm Det}
			[	\partial_{\lambda_{\alpha_{1};i}}f(\lambda_{\alpha_{1};i})
			]|\circ f^{-1}(\lambda_{\alpha_{1};i})},
			\end{aligned}
	\end{equation} 
where $\upalpha$'s denote the measurable subsets for the degenerated eigenvector labelled by $i$.
Indeed,
the diffeomorphism $f(\lambda_{\alpha_{1};i})$ as a bijection allow us to divide the eigenvalue $\lambda_{\alpha_{1};i}$
into subsets according to the emergent flavors within each degenerated eigenvector.

The reparameterization to the continuously differentiable map
allow us to explain the non-Hermiticity as well as the non-defective degeneracies in NLS region by considering the degenerated eigenvectors (virtual) into several components
that each represents a state with finite expectation (not the ground state) under the local symmetry with respect to the the eigenspace of this degenerated eigenvector.
An extensive number of such mutually commutative virtual components is allowed like the possible generalized canonical or Gibbs ensemble in a nonintegrable system with an extensive number of local symmetries\cite{Hamazaki}.
Each one of the components of a degenerated eigenvector
can be classified into a flavor uniquely by the diffeomorphism-asisted differention on the nullspace eigenvalues
whose distinct magnitudes can be seem in log scale (see Fig.\ref{log}(a)).
This is enforced by the smoothness requirement of the ETH (see, e.g., Ref.\cite{Anza}).

In perspective of the robustly continuous spectrum even down to the degenerated region,
which exhibit the persistive level repulsion,
there is indeed nonzero correlations between the subspace
containing all the integrable eigenstates and that containing all the degenerated eigenstates (the nullspace).
This is guaranteed by the extensive nature of these subspaces,
where both of them are made up by local terms,
and here the local terms refer to the mutually orthogonal eigenvectors (where the head elements plays an important role in keeping the level repulsion) in NLS block and the
discrete eigenvalues outside the NLS block.
Such continous feature exhibited in both the (rescaled) spectrum and the eigenvectors,
related to the topological symmetries,
is important in the study of topological materials,
like the concepts of exceptional point (where both the eigenvalues and eigenvectors coalesce) or non-defective
degeneracy point\cite{Yang Z}.

Despite the local components,
the head element ($\upalpha_{1}$) results in special relation between eigenvectors
$	| {\bf \Psi}_{\alpha_{1};i} \rangle$ ($i=\frac{D}{2}+1, \cdots ,D$),
i.e., the orthogonality as well as the level repulsion (in log scale) for the corrresponding eigenvalues.
However, the correlation for head elements does not cause
correlation between arbitary two degenerated eigenvectors
(which is for sure due to the extensive character of the subspace played by the whole nullspace).
That is the reason why we using the smooth map,
$f:\ 
\lambda_{\alpha_{1};i}\rightarrow \lambda_{\alpha_{1};j}$
($i,j\in NLS$),
to describe such relations between the degenerated eigenvectors. Correspondently, there is the differential
$df:\ 
T(\lambda_{\alpha_{1};i})\mathbb{R}_{i}
\rightarrow 
T(f(\lambda_{\alpha_{1};i}))\mathbb{R}_{j}$
($i,j\in NLS$),
which is the linear map as a result of the binary operation and at the best linear approximation.
Here $T$ denotes the tangent space as a result of the derivative for original subset (an arbitarty eigenvector)
to the $\mathbb{R}$,
and thus the above differential is isomorphic to 
$\mathbb{R}_{i}\rightarrow \mathbb{R}_{j}$.
Theoretically infinite-order derivative can be performed as the linear map in both subsets,
and results in 
$df:\ \mathbb{R}\rightarrow \mathbb{R}$.
In this process, the
differential as a linear map replace the smooth map by 
the pushforward between the elements of an abelian group.
With the order of differention move deeper
(along the direction of a single subset),
the isomorphism guarantees the resulting groups always be abelian.
This equivalents to restricting the bijections to exist only between two adjacent subsets,
which is to guarantees the commutative relation between bijections.
Also,
since the differential from the a subset to the next subset always equivalents to the derivation within the next subset, 
the invertibility of $f$ or $df$ (whose inversed forms are also continuously differentiable) 
will not cause the many-body localization of a single subset.
This is analogous to the reciprocity breaking in a non-Hermitian circuit.
In other word,
for an arbitary subset within the NLS region,
despite the invertible bijections,
the complete localization for each subset is suppressed by the next subset.

This requires $f$ to be $C^{\infty}$ (infinite-times continuously) differentiable,
and thus the resulting head elements of the degenerated eigenvectors will nearly exhibit not different
which corresponds to the limit of asympototic degeneracy 
for the continuous spectrum where the level repusion is persistive.
While in our case,
we consider $f$ to be $C^{1}$-differentiable,
and as a result there is level repulsion in log scale for the
spectrum (see Fig.\ref{log})
and the head element of the degenerated eigenvectors
(as well as that of the $(\frac{D}{2}+1)$-th eigenvector
which is adjacent to the NLS block)
are correlated through the nearby linear map between arbitary two adjacent subsets (eigenvectors)
and allow us to collect the degeneracies
 into an extensive quantity.
 Thus the linear map across single (or several) subsets equivalents to the derivative at the last subset where the best linear approximation sets in.
Thus,
 the level repulsion (in log scale) within the degenerated region
 is due to the accumulation of linear map to different ends.

The summation over all the $\upalpha$-labelled subset of a certain degenerated eigenvector is essential in guaranteeing the above invertibility-related relation
as well as the non-Hermiticity of whole system by
enforcing the non-defective degeneracies in NLS region
(this can be seem from Fig.\ref{log}(e)-(f) according to the comparasion of eigenvector elements of ${\bf \Psi}_{\alpha_{1}}$
and ${\bf \Psi}_{\alpha_{2}}$).

\begin{figure}
	\centering
	\includegraphics[width=0.5\linewidth]{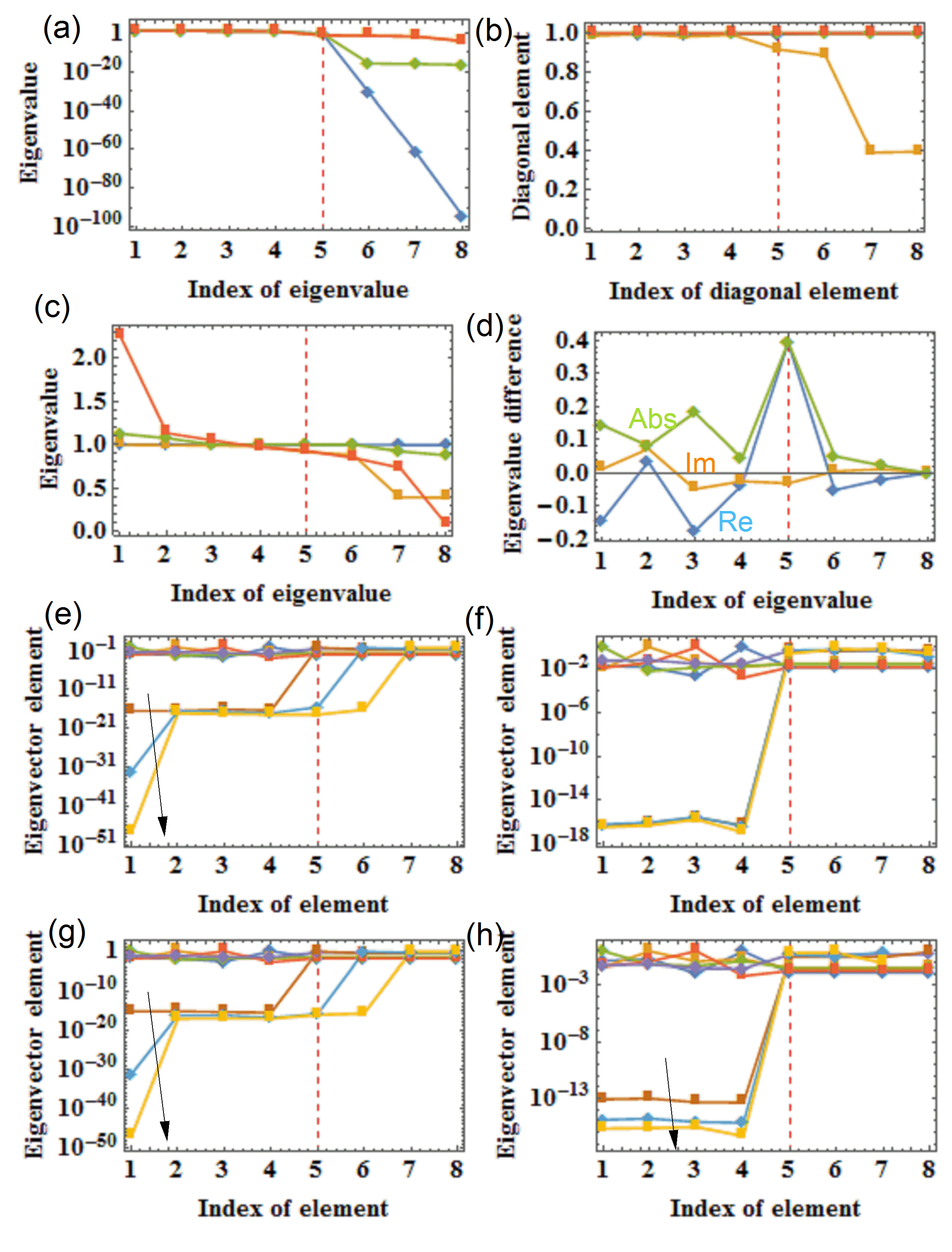}
	\captionsetup{font=small}
\caption{
		(a)		Absolute value of complex eigenvalues of  of ${\bf \Psi}_{\alpha_{1}}$,
		${\bf \Psi}_{\alpha_{2}}$, 
		$\pmb{\uppsi}^{-1}_{\alpha_{1}}\pmb{\lambda}_{\alpha_{1}}
		\pmb{\uppsi}_{\alpha_{1}}$,
		$\pmb{\uppsi}^{-1}_{\alpha_{2}}\pmb{\lambda}_{\alpha_{2}}
		\pmb{\uppsi}_{\alpha_{2}}$
	for blue, orange, green, wine red, respectively.
(b)
		Absolute value of diagonal elements of $\pmb{\uppsi}_{\alpha_{1}}\pmb{\uppsi}_{\alpha_{1}}^{T}$
			$\pmb{\uppsi}_{\alpha_{2}}\pmb{\uppsi}_{\alpha_{2}}^{T}$
			$\pmb{\uppsi}_{\alpha_{1}}\pmb{\uppsi}_{\alpha_{1}}^{H}$
			$\pmb{\uppsi}_{\alpha_{2}}\pmb{\uppsi}_{\alpha_{2}}^{H}$, 
				for blue, orange, green, wine red, respectively.
(c)
		Eigenvalues of $\pmb{\uppsi}_{\alpha_{1}}\pmb{\uppsi}_{\alpha_{1}}^{T}$
			$\pmb{\uppsi}_{\alpha_{2}}\pmb{\uppsi}_{\alpha_{2}}^{T}$
			$\pmb{\uppsi}_{\alpha_{1}}\pmb{\uppsi}_{\alpha_{1}}^{H}$
			$\pmb{\uppsi}_{\alpha_{2}}\pmb{\uppsi}_{\alpha_{2}}^{H}$, 	for blue, orange, green, wine red, respectively.
			We see that the conjugate operation results in the normalized diagonal element and larger variance of eigenvalue.
			For Hermitian products $\pmb{\uppsi}_{\alpha_{1}}\pmb{\uppsi}_{\alpha_{1}}^{H}$ and
			$\pmb{\uppsi}_{\alpha_{2}}\pmb{\uppsi}_{\alpha_{2}}^{H}$ the mean eigenvalue is 1,
	while for non-Hermitian products $\pmb{\uppsi}_{\alpha_{1}}\pmb{\uppsi}_{\alpha_{1}}^{T}$ and
			$\pmb{\uppsi}_{\alpha_{2}}\pmb{\uppsi}_{\alpha_{2}}^{T}$ the mean eigenvalues (real part) are 0.998387
			and 0.992396, respectively.
	Eigenvalues of
		$\pmb{\uppsi}_{\alpha_{1}}\pmb{\uppsi}_{\alpha_{1}}^{T}$,
		$\pmb{\uppsi}_{\alpha_{2}}\pmb{\uppsi}_{\alpha_{2}}^{T}$.
		Since they are diagonal matrix, their eigenvalues are also their diagonal elements.
		Correlation between real and imaginary parts are indicated in each plot.
		Elements of the eigenvectors of 
$ {\bf \Psi}_{\alpha_{1}}$,
	$\pmb{\uppsi}^{-1}_{\alpha_{1}}\pmb{\lambda}_{\alpha_{1}}
\pmb{\uppsi}_{\alpha_{1}}$.
		Eigenvalues of ${\bf \Psi}_{\alpha_{1}}$,
		${\bf \Psi}_{\alpha_{2}}$, and their difference.
		Here we can see that,
		the effect of edge burst reflected in eigenvalues does not happen in the boundary of degenerated
		region,but the edge of nondegenerated region that closest to the degenerated region,
		i.e., the node $5$, where the most obvious level repulsion happen.
	}
\label{log}
\end{figure}

Similarly, for the full-rank subsystem ${\bf \Psi}_{\alpha_{2}}$,
there will also be a $\frac{D}{2}\times \frac{D}{2}$ block with real entries 
$\frac{1}{D}$ (diagonal) and $\frac{D-4}{D^2}$ (off-diagonal).
Exactly diagonalization can be performed 
as $\pmb{\uppsi}^{T}_{\alpha_{2}} \pmb{\lambda}_{\alpha_{2}}
[\pmb{\uppsi}_{\alpha_{2}}^{T}]^{-1}$ and 
$\pmb{\uppsi}^{-1}_{\alpha_{2}}\pmb{\lambda}_{\alpha_{2}}
\pmb{\uppsi}_{\alpha_{2}}$,
although have the RIB which is distinct from ${\bf \Psi}_{\alpha_{2}}$,
have exactly the same eigenvectors with 
${\bf \Psi}_{\alpha_{2}}$,
i.e., $D$ mutually orthogonal eigenvectors,
and thus
$\pmb{\uppsi}_{\alpha_{2}}\pmb{\uppsi}_{\alpha_{2}}^{T}$
is a diagonal matrix with all diagonal entries smaller than one
(due to the absence of orthogonal diagonalizability).
It is easy to see that the absence of orthogonal
diagonalizability (available in Hermitian case) here is due to the existence of imaginary parts of both the ${\bf \Psi}_{\alpha_{1}}$ and ${\bf \Psi}_{\alpha_{2}}$
(or in other word, the mixture of real and imaginary parts)
as we are able to obtain a exact Hermitian basis
by deviding ${\bf \Psi}_{\alpha}$ into 
${\rm Re}[{\bf \Psi}_{\alpha}]$ and ${\rm Im}[{\bf \Psi}_{\alpha}]$,
in which case 
$\widetilde{\pmb{\uppsi}}_{\alpha}
\widetilde{\pmb{\uppsi}}_{\alpha}^{T}\equiv {\bf I}$
(here $\widetilde{\pmb{\uppsi}}_{\alpha}$ is the matrix whose rows are the eigenvectors of ${\rm Re}[{\bf \Psi}_{\alpha}]$ or ${\rm Im}[{\bf \Psi}_{\alpha}]$).

Thus it is the collective effect of the real and imaginary parts
guarantees the exact equivalence between 
${\bf \Psi}_{\alpha_{2}}$ and its diagonalized forms
(e.g., $\pmb{\uppsi}^{T}_{\alpha_{2}} \pmb{\lambda}_{\alpha_{2}}
[\pmb{\uppsi}_{\alpha_{2}}^{T}]^{-1}$ and 
$\pmb{\uppsi}^{-1}_{\alpha_{2}}\pmb{\lambda}_{\alpha_{2}}
\pmb{\uppsi}_{\alpha_{2}}$),
and prevent the generation of NLS by the random imaginary parts in RIB.
While such collective effect of the real and imaginary part is absent in rank deficient subsystem ${\bf \Psi}_{\alpha_{1}}$
due to the lacking of enough nonzero eigenvalues (expectations from observable ${\bf M}$) to avoiding the formation of orthogonal basis.
In other word,
the nondegenerated eigenvector corresponding to a nonzero eigenvalue, that doesnot form othrogonal basis by itself, 
identifying a certain mixture pattern between real and imaginary parts of ${\bf \Psi}_{\alpha_{2}}$.
Due to the same reason,
the fluctuation of minor imaginary part in the RIB of ${\bf \Psi}_{\alpha_{1}}$ results in non-negligible asymmetry effect,
i.e., ${\bf \Psi}_{\alpha_{1}}$ has a different eigenvector basis
from its diagonalization representations (its eigendecomposed form).

A comparasion between 
Hermitian basis
$|{\bf \Psi}_{\alpha;i}\rangle \langle {\bf \Psi}_{\alpha;i}|$
and the non-Hermitian basis
$|{\bf \Psi}_{\alpha;i}\rangle \langle {\bf \Psi}^{*}_{\alpha;i}|$,
and more information about the biorthogonality for the current bipartite system,
is presented in Appendix.

In terms of the basis of original Hermition operator $H_{D}$,
the fidelity can also be revealed through the spectrum of
phase factor $e^{iH_{D}t}$,
where we found that
\begin{equation} 
	\begin{aligned}
E(H_{D}):=
{\rm Tr}[\sum_{\{i\}}|E_{i}\rangle\sum_{j}\langle E_{j}|e^{iH_{D}t}]
=\sum_{\{i\}}\langle E_{i}|e^{iH_{D}t}\sum_{\{j\}}|E_{j}\rangle
=\frac{1}{{\rm dim}\{k\}}\sum_{\{k\}}e^{iE_{k}t}
\sum_{\{i\}}\langle E_{i}|\sum_{\{j\}}|E_{j}\rangle,
			\end{aligned}
	\end{equation}
where $\{i\},\{j\}=1,\cdots,D$, and
$\{k\}=\{i\}\lor\{j\}$ denote the overlap between groups $\{i\}$
and $\{j\}$.
Specifically,
for ${\rm dim}\{k\}=1$, both
the real and imaginary part of $E(H_{D})$ exhibit a fixed pattern in the evolutionary spectrum,
as a result of the coherence between themself
(which induce the restriction on time-independent absolute value
of the complex expectations).
In this case, the spectrum reads
$E(H_{D};i)=e^{iE_{i}t}=\langle E_{i}|e^{iH_{D}t}|E_{i}\rangle$.
Similarly, for ${\rm dim}\{k\}=2$,
the spectrum of the absolute expectations
is nomore static but evolve with fixed pattern until ${\rm dim}\{k\}\ge 3$.
Same conclusion can be obtained from the Loschmidt echo, i.e,
the overlap between initial-state-based exponential operator and the non-Hermitian basis.
The better the echo follows a fixed pattern (or periodicity), the higher the
initial state recurrence is.

\section{Conclusion}

In terms of the non-reciprocal linear map in the degenerated matrix block (towards the nondegenerated one)
and the derivative of Lebesgue measurement,
we verify the unidirectional accumulation effect of the 
boundary-dependence for the degenerated eigenvectors which form the nonlocal symmetry block
in the rank-deficient subsystem.
Compares to the defective degeneracies
where the degenerated eigenvectors loss the self-normalization and mutually orthogonality,
there is robust accumulation effect of non-Hermicity 
toward the nondegenerated block where each eigenvcetors (mutually orthogonal but not self-normalized)
corresponds to a distinct nonzero eigenvalue.
In log scale, the continued spectrum can be observed in the NLS region,
and as a result,
the correlations between an eigenvector $ |{\bf \Psi}_{\alpha_{1};i}\rangle$
and the collection of the rest part
$\sum_{j=i+1}^{D} |{\bf \Psi}_{\alpha_{1};j}\rangle$
is always proportional to the expectation of subsystem
${\bf \Psi}_{\alpha_{1}}$ on it.
As another evidence of the non-reciprocity,
the correlations between $ |{\bf \Psi}_{\alpha_{1};i}\rangle$
and its rest $\sum_{j=i+1}^{D} |{\bf \Psi}_{\alpha_{1};j}\rangle$
is always dominating over the 
$ |{\bf \Psi}_{\alpha_{1};i+1}\rangle$
and its rest $\sum_{j=i+2}^{D} |{\bf \Psi}_{\alpha_{1};j}\rangle$,
which is an unique phenomenon in the non-defective degenerated block.
In perspective of the Lebesgue measurement,
this can be explained by the longer time accumulation (toward the equilibrium) for the former compares to the latter.
The robust non-defective degeneration originates from the robust level repulsion between the local states located in the middle of spectrum.

For a non-Hermitian system that localized (thermalized) by itself,
where the localization as well as the initial state fidelity originates from the 
entangled thermal states as well as the disentangling effect caused by response to the edge
(non-Hermitian skin effect),
it's imaginary spectrum of unitary evolution always exhibit a single edge burst in the end of arbitary long-time limit.
Such significant reflection boundary in the long-time side
is due to the invariant distribution of the states drawned randomly
from the restricted Hilbert space, or even just from serveral "typical" random states. 
Importantly,
such character is inavailable from the real spectrum, i.e,, the real part of eigenvalues
either be steady under unitary evolution or lacking a invariant distribution under unitary transformation
(Haar measurement).
Such class of non-Hermitian systems are classified by the initial density
$\rho_{res}$ with a restricted Hilbert space,
including both the mixed or reduced states, like $\mathcal{P}_{D}'$, $\tilde{\rho}_{nH;\{c\}}\ (\forall \{c\})$, $\tilde{\rho}_{H;3}$, $\mathcal{Q}_{nH;\{c\}}\ (\forall \{c\})$, $\mathcal{Q}_{H;3}$, etc.
Specifically, for those densities have an constant Loschmidt echo equals to 1,
like $\mathcal{P}_{D}'$ and $\mathcal{Q}_{nH;3}$,
the expectations for local observables will exhibit level repulsion
which can be visualized by the VIE spectrum (Fig.\ref{EO6}(l), Fig.\ref{34_41}(b)).
Such level repulsion indicates the strong entanglement between the thermal states,
which behave like the mutually commutating macroobservables without any coherence between them.
and the edge burst (disentangling effect) nomore exist within each state.
This intrinsic property of macroobservable is more significant from the
echo of $\tilde{\rho}_{H;(c)}\ (\forall \{c\})$ as shown in Fig.\ref{echooverlap}(a),
where each the coherence as well as the disentangling effect is overwhelmed by the 
character of a thermal state.
This guarantees the entanglement between arbitary thermal states
(subspaces) will not modifies their individual well-defined eigenvalues,
i.e., only modifies the echo spectrum in the direction orthogonal to the 
effect of nonextensive boundaries (e.g., Fig.\ref{echooverlap3}(i)).
Such coherence along the direction orthogonal to where the entanglement happen,
enforce the thermal equilibrium,
where we can see the steady character from the echo
(Fig.\ref{EO5}(j) for non-Hermitian basis 
and Fig.\ref{echooverlap}(d) for Hermitian basis). 
It is the coherences happen in every static moment
that make the equilibrium possible without needing the thermodynamic limit.

For Hermitian system that eventually be thermalized 
a localized equilibrium state,
there is mixture-induced quantized spectrum,
whose ergodic thermalization timescale up to $T\rightarrow \infty$.
and there are edge bursts in the left side of each segment.
Indeed, such edge burst always appears in the side opposite to where localization
(or the effect of thermal inclusion) can be seen.
For initial densities with restricted Hilbert space like $\mathcal{P}_{D}'$,
the edge busrt appears in the long-time side,
which is in consistent with the fact that the initial state that able to be thermalized
contain a thermalized system initially and the only different to the version after long-time evolution
(dephasing) is the suppressed Hermicity (coherence)\cite{Rigol7}.
It is the initial state $\mathcal{P}_{D}'$ carries the information about the ergodic thermalization timescale, and decoupled with the disorders.

\begin{figure}
\centering
\includegraphics[width=0.8\linewidth]{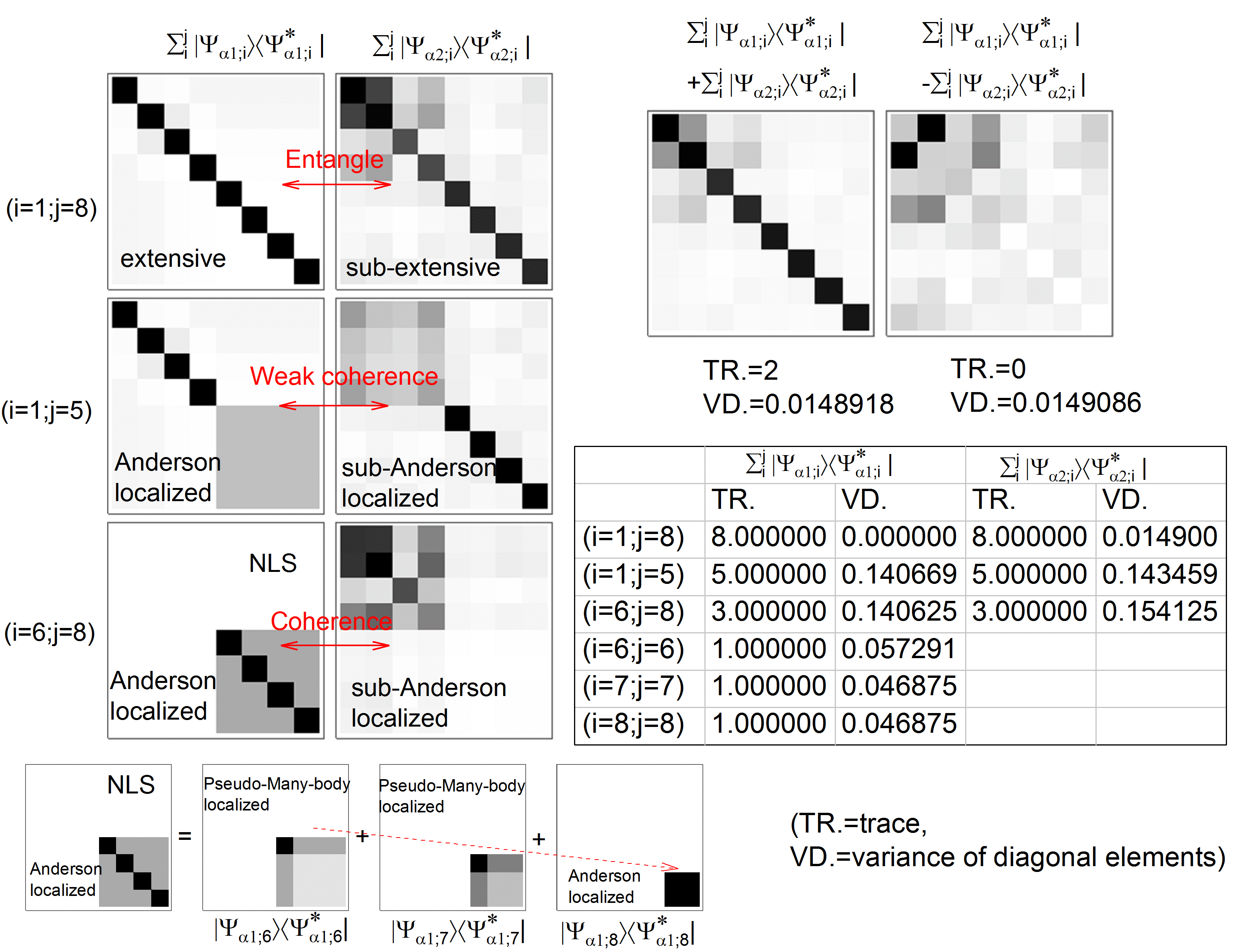}
\captionsetup{font=small}
\caption{Entries distribution of 
$\sum_{i=1}^{D}|{\bf \Psi}_{\alpha;i}\rangle\langle {\bf \Psi}^{*}_{\alpha;i}|$,
$\sum_{i\notin NLS}|{\bf \Psi}_{\alpha;i}\rangle\langle {\bf \Psi}^{*}_{\alpha;i}|$,
$\sum_{i\in NLS}|{\bf \Psi}_{\alpha;i}\rangle\langle {\bf \Psi}^{*}_{\alpha;i}|$
($\alpha=\alpha_{1},\alpha_{2}$).
This can also be verified by initial state localization
through the ratio
$\mathcal{C}(\rho)$
where we found that $\mathcal{C}(\rho)=1$ only for $\rho=
\sum_{i=1}^{D}|{\bf \Psi}_{\alpha;i}\rangle\langle {\bf \Psi}^{*}_{\alpha;i}|$.
The NLS block is shown in the bottom,
where the finite coherence exist between 
the maximally chaotic clusters $\sum_{i=6}^{8}|{\bf \Psi}_{\alpha_{1};i}\rangle
\langle {\bf \Psi}_{\alpha_{1};i}^{*}|$
and $\sum_{i=6}^{8}|{\bf \Psi}_{\alpha_{2};i}\rangle
\langle {\bf \Psi}_{\alpha_{2};i}^{*}|$.
The equation in bottom shows the superposition of pure states of the NLS block
into an extensive or Anderson localized block (extensive due to the homogeneous diagonal elements
and the absence of any reflecting boundary condition;
Anderson localized due to its restricted subspace dimension).
Thus its counterintuitive that a superposition results in the result that of the type of
statistic mixture.
However, this also consistent with the fact that
the mutually orthogonal states can still be physically similar
(in a semi-classical chaotic system\cite{Halder}):
The non-defectively degenerated states share the similar eigenvector structure
(compares to those out of NLS block)
as can be seem from Eq.(\ref{1225}),
but just distinguished by the subspace dimensions with nonzero entries.
Also, the nearest linear map (uni-directional forward scattering) make sure
there without any reflecting boundary within the NLS block.
In other word,
the subspaces generated by invariant toris in the energy surface completely 
neutralize the roles of edge and bulk.
}
\label{nHbasis}
\end{figure}

\section{Appendix}

\subsection{diagnosis in terms of diagonal ETH and diagonal ensemble}

Starting from the normalization,
$\sum_{j\le j'}\psi_{j}^{\alpha}(\psi_{j'}^{\alpha})^{*}=
\sum_{j\le j'}\psi_{j}^{\alpha}(\psi_{j'}^{\alpha})^{*}=1$.
we next denote 
$\pmb{\mathscr{S}}_{\alpha}$ as the square matrix whose rows are the eigenvectors $\psi_{j}^{\alpha}$ arranged in order of $j$
($1,\cdots,D$).
Then with respect to the symmetry matrix 
${\bf J}_{\alpha}^{T}{\bf J}_{\alpha}$,
\begin{equation} 
	\begin{aligned}
		\label{8111}
		&
		{\rm Rank}[\sum_{\alpha}^{\alpha_{m}}
		{\bf J}_{\alpha}{\bf J}_{\alpha}^{T}]=D,\\
		&
		{\rm Rank}[{\bf J}_{\alpha}]={\rm Rank}[{\bf J}^{T}_{\alpha}
		{\bf J}_{\alpha}]={\rm Rank}[{\bf J}_{\alpha}
		{\bf J}_{\alpha}^{T}]
		={\rm Rank}[\pmb{\mathscr{S}}_{\alpha}]
		={\rm Rank}[
		\pmb{\mathscr{S}}_{\alpha}\pmb{\mathscr{S}}^{T}_{\alpha}]
		={\rm Rank}[
		\pmb{\mathscr{S}}^{T}_{\alpha}
		\pmb{\mathscr{S}}_{\alpha}]
		={\rm Rank}[
		\pmb{\mathscr{S}}_{\alpha}\pmb{\mathscr{S}}^{H}_{\alpha}]
		={\rm Rank}[
		\pmb{\mathscr{S}}^{H}_{\alpha}
		\pmb{\mathscr{S}}_{\alpha}]
		\le D,\\
		&
		{\rm Tr}[{\bf J}_{\alpha}{\bf J}_{\alpha}^{T}]
		= {\rm Tr}[{\bf J}_{\alpha}^{T}{\bf J}_{\alpha}]
		={\rm Tr}[
		\pmb{\mathscr{S}}_{\alpha}
		\pmb{\mathscr{S}}^{T}_{\alpha}]
		={\rm Tr}[
		\pmb{\mathscr{S}}^{T}_{\alpha}
		\pmb{\mathscr{S}}_{\alpha}]
		=1,\ 
		{\rm Tr}[{\bf J}_{\alpha}{\bf J}_{\alpha}^{H}]
		= {\rm Tr}[{\bf J}_{\alpha}^{H}{\bf J}_{\alpha}]
		={\rm Tr}[
		\pmb{\mathscr{S}}_{\alpha}
		\pmb{\mathscr{S}}^{H}_{\alpha}]
		={\rm Tr}[
		\pmb{\mathscr{S}}^{H}_{\alpha}
		\pmb{\mathscr{S}}_{\alpha}]
		\neq 1,\ 
		\ (\forall \alpha)\\
		&
		{\rm Rank}[|E_{j}\rangle\langle E_{j'}|]=1,
		{\rm Rank}[\sum_{j}|E_{j}\rangle\langle E_{j}|]
		={\rm Tr}[\sum_{j}|E_{j}\rangle\langle E_{j}|]
		={\rm Tr}[\sum_{jj'}|E_{j}\rangle\langle E_{j'}|]
		={\rm Tr}[\sum_{j\le j'}|E_{j}\rangle\langle E_{j'}|]
		={\rm Rank}[\sum_{j\le j'}|E_{j}\rangle\langle E_{j'}|]
		=D,\\
		&
		{\rm Rank}[\sum_{jj'}|E_{j}\rangle\langle E_{j'}|]=1,\ 
		{\rm Tr}[\sum_{j}|E_{j}\rangle\langle E_{j}|]=D,\ 
		{\rm Rank}[\sum_{jj'}|E_{j}\rangle\langle E_{j'}|]=1,\ 
		{\rm Tr}[|E_{j}\rangle\langle E_{j'}|]=\delta_{jj'},
	\end{aligned}
\end{equation}
where for $j\neq j'$ the resulting matrix $|E_{j}\rangle\langle E_{j'}|$
is an nilpotent matrix,
as an outer product of two orthogonal eigenstates.
For convinience, next
we denote the matrix ${\bf \Psi}_{\alpha}:=
\pmb{\mathscr{S}}_{\alpha}\pmb{\mathscr{S}}_{\alpha}^{T}
(\neq 
\pmb{\mathscr{S}}_{\alpha}^{T}\pmb{\mathscr{S}}_{\alpha})$,
whose elements are $\psi_{j}^{\alpha}(\psi_{j'}^{\alpha})^{*}$.

Further, since $\sum_{j}(\psi_{j}^{\alpha})^{*}\psi_{j}^{\alpha}
=\pmb{\mathscr{S}}_{\alpha}^{T}\pmb{\mathscr{S}}_{\alpha}$,
\begin{equation} 
	\begin{aligned}
		\label{8112}
		&
		{\rm Rank}[\sum_{j}(\psi_{j}^{\alpha})^{*}\psi_{j}^{\alpha}]
		=
		{\rm Rank}[\sum_{j\le j'}(\psi_{j}^{\alpha})^{*}\psi_{j'}^{\alpha}]
		\le D,\ 
		{\rm Rank}[\sum_{jj'}(\psi_{j}^{\alpha})^{*}\psi_{j'}^{\alpha}]=1,\\ 
		&
		{\rm Tr}[\sum_{j}(\psi_{j}^{\alpha})^{*}\psi_{j}^{\alpha}]
		={\rm Tr}[\sum_{j\le j'}(\psi_{j}^{\alpha})^{*}\psi_{j'}^{\alpha}]=
		{\rm Tr}[\sum_{jj'}(\psi_{j}^{\alpha})^{*}\psi_{j'}^{\alpha}]=1,\ 
		(\forall \alpha)
	\end{aligned}
\end{equation}
For the rank-one matrices
$\sum_{jj'}(\psi_{j}^{\alpha})^{*}\psi_{j'}^{\alpha}$,
the only one nonzero eigenvalue is one,
and corresponding to a one-dimensional eigenspace.
As symmetry matrix which is diagonalizable,
each eigenvalue of
$\sum_{jj'}(\psi_{j}^{\alpha})^{*}\psi_{j'}^{\alpha}$
has the geometric multiplicity equals the algebraic multiplicity,
thus the eigenspace of zero eigenvalues is $(D-1)$.
\begin{equation} 
	\begin{aligned}
		&
		\sum_{jj'}(\psi_{j}^{\alpha})^{*}\psi_{j'}^{\alpha}
		=\ell_{i}^{T} (\frac{|\ell_{1}|}{|\ell_{i}|},\cdots, \frac{|\ell_{D}|}{|\ell_{i}|})
		=(\frac{|\ell_{1}|}{|\ell_{i}|},\cdots, \frac{|\ell_{D}|}{|\ell_{i}|})^{T}
		\ell_{i},\\
		&
		\ell_{i} (\frac{|\ell_{1}|}{|\ell_{i}|},\cdots, \frac{|\ell_{D}|}{|\ell_{i}|})^{T}
		=(\frac{|\ell_{1}|}{|\ell_{i}|},\cdots, \frac{|\ell_{D}|}{|\ell_{i}|})
		\ell_{i}^{T}=1,
	\end{aligned}
\end{equation}
where $\ell_{i}\ (i\le D)$ is the $i$-th row of matrix
$\sum_{jj'}(\psi_{j}^{\alpha})^{*}\psi_{j'}^{\alpha}$.
Then in terms of 
the eigenvector $\ell_{1}$ corresponding to the only one nonzero eigenvalue,
which can be expressed as linear scalings of arbitary $\ell_{i}$
or $(\frac{|\ell_{1}|}{|\ell_{i}|},\cdots, \frac{|\ell_{D}|}{|\ell_{i}|})$,
\begin{equation} 
	\begin{aligned}
		&
		\sum_{jj'}(\psi_{j}^{\alpha})^{*}\psi_{j'}^{\alpha}
		\ell_{1}^{T}=\ell_{1}^{T}
		=		\ell_{i} \ell_{1}^{T}
		(\frac{|\ell_{1}|}{|\ell_{i}|},\cdots, \frac{|\ell_{D}|}{|\ell_{i}|})^{T}
		=(\frac{|\ell_{1}|}{|\ell_{i}|},\cdots, \frac{|\ell_{D}|}{|\ell_{i}|})
		\ell_{1}^{T}\ell_{i}^{T}.
	\end{aligned}
\end{equation}

The results about the rank which is possible to smaller than $D$
in Eqs.(\ref{8111},\ref{8112})
indicates that the projector operators into the corresponding sector
(for each integrable eigenstates)
may not completely mutually orthogonal.
We also note that,
all the matrices ${\bf \Psi}_{\alpha}$ (for different $\alpha$) are symmetry,
and thus be orthogonally diagonalizable (non-defective, but no the normal matrix),
i.e., 
\begin{equation} 
	\begin{aligned}
		\label{9141}
		{\bf \Psi}_{\alpha}=
		\pmb{\uppsi}^{T}_{\alpha} \pmb{\lambda}_{\alpha}
		[\pmb{\uppsi}_{\alpha}^{T}]^{-1}
		=
		\pmb{\uppsi}^{-1}_{\alpha}\pmb{\lambda}_{\alpha}
		\pmb{\uppsi}_{\alpha}
	\end{aligned}
\end{equation}
where $\pmb{\uppsi}_{\alpha}$ is the nonsingular 
matrix whose rows are the eigenvectors of ${\bf \Psi}_{\alpha}$,
and $\pmb{\lambda}_{\alpha}$ is the diagonal matrix whose diagonal elements are the eigenvalues of ${\bf \Psi}_{\alpha}$.
Thus
the eigenvectors for each ${\bf \Psi}_{\alpha}$ form
an orthonormal basis $\mathds{R}^{D}$ even for those which containing zero eigenvalues (low rank ones).
In our simulations,
the eigenvectors $\psi_{j}^{\alpha}$ are parametrized into complex form,
and the resulting matrices ${\bf \Psi}_{\alpha}$ contains the elements of the complex inner product space,
thus the orthogonally diagonalizability of ${\bf \Psi}_{\alpha}$'s
can be verified according to
\begin{equation} 
	\begin{aligned}
		&
		{\bf \Psi}_{\alpha}^{H}{\bf \Psi}_{\alpha} 
		={\bf \Psi}_{\alpha}{\bf \Psi}_{\alpha}^{H}\neq {\bf I},\ 
		{\bf \Psi}_{\alpha}^{H}\neq 
		{\bf \Psi}_{\alpha}^{T}={\bf \Psi}_{\alpha}\neq 
		{\bf \Psi}_{\underline{\alpha}}^{-1},\ 
		{\bf \Psi}_{\underline{\alpha}}^{H}\neq 
		{\bf \Psi}_{\underline{\alpha}}^{-1},\ 
		{\bf \Psi}_{\alpha}^{T}{\bf \Psi}_{\alpha} 
		={\bf \Psi}_{\alpha}{\bf \Psi}_{\alpha}^{T}\neq {\bf I},\ 
		{\bf \Psi}_{\underline{\alpha}}^{-1}{\bf \Psi}_{\underline{\alpha}} 
		={\bf \Psi}_{\underline{\alpha}}{\bf \Psi}_{\underline{\alpha}}^{-1}={\bf I},\\
		&
		{\rm Rank}[{\bf \Psi}_{\underline{\alpha}}^{-1}{\bf \Psi}_{\underline{\alpha}}]
		={\rm Rank}[{\bf \Psi}_{\underline{\alpha}}{\bf \Psi}_{\underline{\alpha}}^{-1}]
		={\rm Rank} [{\bf \Psi}_{\underline{\alpha}}^{-1}]
		={\rm Rank} [{\bf \Psi}_{\underline{\alpha}}]=D,\\
		&
		{\rm Rank} {\bf \Psi}_{\alpha}=
		{\rm Rank} {\bf \Psi}_{\alpha}^{H}=
		{\rm Rank} {\bf \Psi}_{\alpha}^{T}=
		{\rm Rank} [{\bf \Psi}_{\alpha}^{T}{\bf \Psi}_{\alpha}]=
		{\rm Rank}[ {\bf \Psi}_{\alpha}{\bf \Psi}_{\alpha}^{T}]=
		{\rm Rank}[{\bf \Psi}_{\alpha}^{H}{\bf \Psi}_{\alpha}]=
		{\rm Rank}[{\bf \Psi}_{\alpha}{\bf \Psi}_{\alpha}^{H}]
		\le D.\\
		&
		{\rm Tr}{\bf \Psi}_{\alpha}={\rm Tr}{\bf \Psi}^{T}_{\alpha}= {\rm Tr}{\bf \Psi}^{H}_{\alpha}
		=1,\ 
		{\rm Tr}{\bf \Psi}_{\underline{\alpha}}^{-1}{\bf \Psi}_{\underline{\alpha}}
		={\rm Tr}{\bf \Psi}_{\underline{\alpha}}{\bf \Psi}_{\underline{\alpha}}^{-1}=D.
	\end{aligned}
\end{equation}
where superscript $H$ denote the Hermitian transpose.
Thus both the ${\bf \Psi}_{\alpha}$'s together with $\pmb{\uppsi}_{\alpha}$'s are neither the
normal matrix nor the unitary matrix ($	\pmb{\uppsi}_{\alpha}^{-1}\neq \pmb{\uppsi}_{\alpha}^{H}$).
Different to ${\bf \Psi}_{\alpha}$'s,
$\pmb{\uppsi}_{\alpha}$'s are not symmetry and thus donot share the orthogonally diagonalizability,
$\pmb{\uppsi}_{\alpha}^{H}\pmb{\uppsi}_{\alpha}\neq
\pmb{\uppsi}_{\alpha}\pmb{\uppsi}_{\alpha}^{H}$.
Unlike
$\pmb{\uppsi}_{\alpha}\pmb{\uppsi}_{\alpha}^{T}$,
both 
$\pmb{\uppsi}_{\alpha}\pmb{\uppsi}_{\alpha}^{H}$
and
$\pmb{\uppsi}_{\alpha}^{H}\pmb{\uppsi}_{\alpha}$
are not diagonal matrices.
Furthermore,
the diagonal elements for both 
$\pmb{\uppsi}_{\alpha}\pmb{\uppsi}_{\alpha}^{H}$
and $\pmb{\uppsi}_{\underline{\alpha}}\pmb{\uppsi}_{\underline{\alpha}}^{-1}$
(identity matrix)
consist of only the real number 1.
While for $\pmb{\uppsi}_{\alpha}^{H}\pmb{\uppsi}_{\alpha}$,
its diagonal elements are real and slightly fluctuate around 1.
\begin{equation} 
	\begin{aligned}
		\label{8171}
		&
		\pmb{\uppsi}_{\alpha}^{H}\pmb{\uppsi}_{\alpha}\neq 
		\pmb{\uppsi}_{\alpha}\pmb{\uppsi}_{\alpha}^{H},\ 
		\pmb{\uppsi}_{\alpha}^{T}\pmb{\uppsi}_{\alpha}\neq 
		\pmb{\uppsi}_{\alpha}\pmb{\uppsi}_{\alpha}^{T},\ 
		\pmb{\uppsi}_{\alpha}^{-1}\pmb{\uppsi}_{\alpha}= 
		\pmb{\uppsi}_{\alpha}\pmb{\uppsi}_{\alpha}^{-1}
		=(\pmb{\uppsi}_{\alpha}^{T})^{-1}\pmb{\uppsi}_{\alpha}^{T}=
		\pmb{\uppsi}_{\alpha}^{T}(\pmb{\uppsi}_{\alpha}^{T})^{-1}
		={\bf I},\\
		&
		{\rm Rank}[\pmb{\uppsi}_{\alpha}^{H}\pmb{\uppsi}_{\alpha}]
		=
		{\rm Rank}[\pmb{\uppsi}_{\alpha}\pmb{\uppsi}_{\alpha}^{H}]
		=
		{\rm Rank}[\pmb{\uppsi}_{\alpha}^{T}\pmb{\uppsi}_{\alpha}]
		=
		{\rm Rank}[\pmb{\uppsi}_{\alpha}\pmb{\uppsi}_{\alpha}^{T}]
		=D,\\
		&
		{\rm Tr}[\pmb{\uppsi}_{\alpha}^{H}\pmb{\uppsi}_{\alpha}]
		=
		{\rm Tr}[\pmb{\uppsi}_{\alpha}\pmb{\uppsi}_{\alpha}^{H}]
		=D,\ 
		{\rm Tr}[\pmb{\uppsi}_{\alpha}^{T}\pmb{\uppsi}_{\alpha}]
		=
		{\rm Tr}[\pmb{\uppsi}_{\alpha}\pmb{\uppsi}_{\alpha}^{T}]
		\le D,\\
		&
		\pmb{\uppsi}_{\alpha}^{T}\pmb{\uppsi}_{\alpha}\neq 
		\pmb{\uppsi}_{\alpha}\pmb{\uppsi}_{\alpha}^{T},\ 
		\pmb{\uppsi}_{\alpha}^{-1}\pmb{\uppsi}_{\alpha}= 
		\pmb{\uppsi}_{\alpha}\pmb{\uppsi}_{\alpha}^{-1}
		={\bf I},\\
	\end{aligned}
\end{equation}
Specifically, $\pmb{\uppsi}_{\underline{\alpha}}\pmb{\uppsi}_{\underline{\alpha}}^{T}$'s are always the diagonal matrix,
where diagonal elements are real and close to (but slightly lower that) 1.
For nonsingular matrices ${\bf \Psi}_{\underline{\alpha}}$,
each one of them are rank-$D$, where each one of the $D$ distinct nonzero eigenvalues corresponds to an one-dimensional eigenspace spanned by a single eigenvector.
While for singular matrix ${\bf \Psi}_{\alpha}$
whose rank is $(D-n)$, its corresponding product
$\pmb{\uppsi}_{\alpha}\pmb{\uppsi}_{\alpha}^{T}$ is still a diagonal matrix,
but with $(D-n)$ of the diagonal elements close but below one,
and $n$ of the diagonal elements be one.
Then there will be eigenvalues equal to one with an algebraic multiplicity $n$, corresponding to the eigenspace spanned by $n$ linearly independent eigenvectors,
for the (full rank) matrices $\pmb{\uppsi}_{\alpha}^{T}\pmb{\uppsi}_{\alpha},
\pmb{\uppsi}_{\alpha}^{H}\pmb{\uppsi}_{\alpha},
\pmb{\uppsi}_{\alpha}\pmb{\uppsi}_{\alpha}^{T},
\pmb{\uppsi}_{\alpha}\pmb{\uppsi}_{\alpha}^{H}$.
While for
$\pmb{\uppsi}_{\underline{\alpha}}^{T}\pmb{\uppsi}_{\underline{\alpha}},
\pmb{\uppsi}_{\underline{\alpha}}^{H}\pmb{\uppsi}_{\underline{\alpha}},
\pmb{\uppsi}_{\underline{\alpha}}\pmb{\uppsi}_{\underline{\alpha}}^{T},
\pmb{\uppsi}_{\underline{\alpha}}\pmb{\uppsi}_{\underline{\alpha}}^{H}$,
each one of them has $D$ distinct eigenvalues.
Note that the distinct traces for the products with transpose 
and Hermitian operations in
Eq.(\ref{8111}) and Eq.(\ref{8171}) reflect the special role of local observables.
Is worth to be noticed that,
the projector of $\mathcal{H}_{i}^{\alpha}
=\sum_{j}(\psi_{j}^{\alpha})^* \psi_{j}^{\alpha}$ has a same rank with ${\bf \Psi}_{\alpha}$.
In our numerical simulation,
${\rm Tr}[\pmb{\uppsi}_{\alpha_{1}}\pmb{\uppsi}_{\alpha_{1}}^{T}]=\sum_{i}\langle{\bf \Psi}_{\alpha_{1};i}|
|{\bf \Psi}_{\alpha_{1};i}\rangle
=(\sum_{i}\langle{\bf \Psi}_{\alpha_{1};i}|)
(\sum_{i}|{\bf \Psi}_{\alpha_{1};i}\rangle)
=7.97955 + 0.013448 i$,
${\rm Tr}[\pmb{\uppsi}_{\alpha_{2}}\pmb{\uppsi}_{\alpha_{2}}^{T}]
=\sum_{i}\langle{\bf \Psi}_{\alpha_{2};i}|
|{\bf \Psi}_{\alpha_{2};i}\rangle
=(\sum_{i}\langle{\bf \Psi}_{\alpha_{2};i}|)
(\sum_{i}|{\bf \Psi}_{\alpha_{2};i}\rangle)
=6.33035 - 0.690615 i$.

\subsection{Biorthogonality with Hermitian basis
$|{\bf \Psi}_{\alpha;i}\rangle \langle {\bf \Psi}_{\alpha;i}|$
and the non-Hermitian basis
$|{\bf \Psi}_{\alpha;i}\rangle \langle {\bf \Psi}^{*}_{\alpha;i}|$}

Due to the symmetry character of ${\bf \Psi}_{\alpha}$,
a nonorthogonal inner product in off-diagonal basis
is available from the nullspace eigenvectors of the ${\bf \Psi}_{\alpha_{1}}$ in decomposed representation
(denoted as $\widetilde{{\bf \Psi}}_{\alpha_{1}}$ hereafter).
\begin{equation} 
	\begin{aligned}
		&
		\langle\widetilde{{\bf \Psi}}_{\alpha_{1};i} | \widetilde{{\bf \Psi}}_{\alpha_{1}}^{\dag}=x_{1}\langle\widetilde{{\bf \Psi}}_{\alpha_{1};i} | ,\\
		&
		\langle\widetilde{{\bf \Psi}}^{\dag}_{\alpha_{1};i} | \widetilde{{\bf \Psi}}_{\alpha_{1}}=x_{2}\langle\widetilde{{\bf \Psi}}_{\alpha_{1};i}^{\dag} | ,
	\end{aligned}
\end{equation}
where $\langle\widetilde{{\bf \Psi}}_{\alpha_{1};i} | $
and $\langle\widetilde{{\bf \Psi}}_{\alpha_{1};i} ^{\dag}| $ are the $i$-th eigenvectors of $\widetilde{{\bf \Psi}}_{\alpha_{1}}$
and $\widetilde{{\bf \Psi}}^{\dag}_{\alpha_{1}}$, respectively.
The only solution here is $x_{1}=x_{2}=0$,
i.e., the eigenvectors labelled by $i$ are of the NLS (degenerated) part,
which corresponds to the case of 
$\langle\widetilde{{\bf \Psi}}_{\alpha_{1};i} | 
\widetilde{{\bf \Psi}}_{\alpha_{1};j}^{\dag}\rangle\neq 0\ (i\neq j)$\cite{Brody}.
The zero eigenvalues can be further verified in terms of the 
relation between the
nonorthogonal inner product 
and the real/imaginary part of $\widetilde{{\bf \Psi}}_{\alpha_{1}} $\cite{Brody} (for $i\neq j$),
\begin{equation} 
	\begin{aligned}
		&
		\langle\widetilde{{\bf \Psi}}_{\alpha_{1};i} | 
		\widetilde{{\bf \Psi}}_{\alpha_{1};j}\rangle=-2i
		\frac{ \langle\widetilde{{\bf \Psi}}_{\alpha_{1};i} |{\rm Im}\widetilde{{\bf \Psi}}_{\alpha_{1}}|
			\widetilde{{\bf \Psi}}_{\alpha_{1};j}\rangle}
		{x_{1}-x_{0}}
		=2
		\frac{ \langle\widetilde{{\bf \Psi}}_{\alpha_{1};i} |{\rm Re}\widetilde{{\bf \Psi}}_{\alpha_{1}}|
			\widetilde{{\bf \Psi}}_{\alpha_{1};j}\rangle}
		{x_{1}+x_{0}},\\
		&
		\langle\widetilde{{\bf \Psi}}_{\alpha_{1};i} ^{\dag}| 
		\widetilde{{\bf \Psi}}_{\alpha_{1};j}^{\dag}\rangle=-2i
		\frac{ \langle\widetilde{{\bf \Psi}}_{\alpha_{1};i}^{\dag} |{\rm Im}\widetilde{{\bf \Psi}}_{\alpha_{1}}|
			\widetilde{{\bf \Psi}}_{\alpha_{1};j}^{\dag}  \rangle}
		{x_{2}-x_{0}}
		=2
		\frac{ \langle\widetilde{{\bf \Psi}}_{\alpha_{1};i}^{\dag} |{\rm Re}\widetilde{{\bf \Psi}}_{\alpha_{1}}|
			\widetilde{{\bf \Psi}}_{\alpha_{1};j}^{\dag}  \rangle}
		{x_{2}+x_{0}}.
	\end{aligned}
\end{equation}
We can conclude that (for $i\neq j$)
\begin{equation} 
	\begin{aligned}
		&
		\langle\widetilde{{\bf \Psi}}_{\alpha_{1};i} | 
		\widetilde{{\bf \Psi}}_{\alpha_{1};j}\rangle =
		\langle\widetilde{{\bf \Psi}}_{\alpha_{1};i}^{\dag} | 
		\widetilde{{\bf \Psi}}_{\alpha_{1};j}^{\dag}\rangle 
		\left\{
		\begin{array}{ll}
			=0,\ i.j\notin NLS\\
			\neq 0,\ i,j\in NLS
		\end{array}
		\right.,\\
		&
		\langle\widetilde{{\bf \Psi}}_{\alpha_{1};i} | 
		\widetilde{{\bf \Psi}}_{\alpha_{1};j}^{\dag}\rangle 
		\left\{
		\begin{array}{ll}
			\neq 0,\ i.j\notin NLS \lor i,j\in NLS\\
			= 0,\ else
		\end{array}
		\right..
	\end{aligned}
\end{equation}
A significant feature for the subsystem containing the NLS block is
$\langle\widetilde{{\bf \Psi}}_{\alpha_{1};i} | 
\widetilde{{\bf \Psi}}_{\alpha_{1};i}^{*}\rangle =1$
while
$\langle\widetilde{{\bf \Psi}}_{\alpha_{1};i} | 
\widetilde{{\bf \Psi}}_{\alpha_{1};i}^{\dag}\rangle \neq 1$.
As an example,
we have
$\langle\widetilde{{\bf \Psi}}_{\alpha_{2};i} | 
\widetilde{{\bf \Psi}}_{\alpha_{2};i}^{*}\rangle =
\langle\widetilde{{\bf \Psi}}_{\alpha_{2};i} | 
\widetilde{{\bf \Psi}}_{\alpha_{2};i}^{\dag}\rangle= 1$ for the full rank subsystem.
That means the random imaginary part in RIB is able to (not able to) cause the asymmetry character of $\widetilde{{\bf \Psi}}_{\alpha_{1}}$ ($\widetilde{{\bf \Psi}}_{\alpha_{2}}$).
For $i,j\in NLS$,
$\langle\widetilde{{\bf \Psi}}_{\alpha_{1};i} | 
\widetilde{{\bf \Psi}}_{\alpha_{1};j}^{*}\rangle =0$ is only possible for $x_{1}\neq x_{2}$, which requires more than two subsystems.

Another significant character of the NLS in rank deficient subsystem can be seem from the combination effect between real and imaginary parts (which have nonzero correlations only in NLS block).
For the rank-deficient subsystem ${\bf \Psi}_{\alpha_{1}}$,
$\pmb{\uppsi}_{\alpha_{1}} 
\pmb{\uppsi}_{\alpha_{1}}^{T}$
is a diagonal matrix whose diagonal entries are bounded below 1:
\begin{equation} 
\begin{aligned}
&
	\langle{\bf \Psi}_{\alpha_{1};i} | 
{\bf \Psi}_{\alpha_{1};j}\rangle
	\left\{
\begin{array}{ll}
=1,\ i=j \land i,j\in NLS,\\
\lesssim 1,\ i=j \land i,j\notin NLS,\\
= 0,\ i\neq j
\end{array}
\right.\\
&
	\langle{\bf \Psi}_{\alpha_{1};i} | 
		{\bf \Psi}_{\alpha_{1};j}^{*}\rangle
		\left\{
		\begin{array}{ll}
			=1,\ i=j,\\
			\neq 0,\ i,j\notin NLS \\
			= 0,\ else
		\end{array}
		\right.\\
		&
		\langle\widetilde{{\bf \Psi}}_{\alpha_{1};i} | 
		\widetilde{{\bf \Psi}}_{\alpha_{1};j}^{*}\rangle 
		\left\{
		\begin{array}{ll}
			=1,\ i=j,\\
			\neq 0,\ else
		\end{array}
		\right.
	\end{aligned}
\end{equation}
This allow us to coarse-graine the eigenvectors to obtain the following relations
\begin{equation} 
\begin{aligned}
&\sum_{i\notin NLS}
			\langle{\bf \Psi}_{\alpha_{1};i} | \cdot
	\sum_{j\notin NLS}|	{\bf \Psi}_{\alpha_{1};j}^{*}\rangle 
	=		\sum_{i,j\notin NLS}
	\langle{\bf \Psi}_{\alpha_{1};i} | {\bf \Psi}_{\alpha_{1};j}^{*}\rangle
= \frac{D}{2}+1,\\
&\sum_{i\notin NLS}
			\langle{\bf \Psi}_{\alpha_{1};i} | \cdot
	\sum_{j\notin NLS}|	{\bf \Psi}_{\alpha_{1};j}\rangle 
	=		\sum_{i,j\notin NLS}
	\langle{\bf \Psi}_{\alpha_{1};i} | {\bf \Psi}_{\alpha_{1};j}\rangle
=\sum_{i\notin NLS}
			\langle{\bf \Psi}_{\alpha_{1};i} | \cdot
	\sum_{j\notin NLS}|	{\bf \Psi}_{\alpha_{1};j}\rangle 
\approx \frac{D}{2}+1,\\
	&
			\sum_{i\notin NLS}
	\langle{\bf \Psi}_{\alpha_{1};i} | \cdot
	\sum_{j\in NLS}|	{\bf \Psi}_{\alpha_{1};j}^{*}\rangle 
	=			\sum_{i\in NLS}
	\langle{\bf \Psi}_{\alpha_{1};i} | \cdot
	\sum_{j\notin NLS}|	{\bf \Psi}_{\alpha_{1};j}^{*}\rangle 
	=		\sum_{i\notin NLS,j\in NSL}
	\langle{\bf \Psi}_{\alpha_{1};i} | {\bf \Psi}_{\alpha_{1};j}^{*}\rangle
		=		\sum_{i\in NLS,j\notin NSL}
	\langle{\bf \Psi}_{\alpha_{1};i} | {\bf \Psi}_{\alpha_{1};j}^{*}\rangle
	=0,\\
&		\sum_{i\in NLS}
		\langle{\bf \Psi}_{\alpha_{1};i} | \cdot
		\sum_{j\in NLS}|	{\bf \Psi}_{\alpha_{1};j}^{*}\rangle 
		=		\sum_{i,j\in NLS}
		\langle{\bf \Psi}_{\alpha_{1};i} | {\bf \Psi}_{\alpha_{1};j}^{*}\rangle=
		\sum_{i\in NLS}
		\langle{\bf \Psi}_{\alpha_{1};i} | \cdot
		\sum_{j\in NLS}|	{\bf \Psi}_{\alpha_{1};j}\rangle 
		=		\sum_{i,j\in NLS}
		\langle{\bf \Psi}_{\alpha_{1};i} | {\bf \Psi}_{\alpha_{1};j}\rangle
= \frac{D}{2}-1,
	\end{aligned}
\end{equation}
While the non-defective degeneracy is absent for the basis of $|\widetilde{{\bf \Psi}}_{\alpha_{1};i}
\rangle$
\begin{equation} 
\begin{aligned}
		\sum_{i\in NLS}
		\langle\widetilde{{\bf \Psi}}_{\alpha_{1};i} | \cdot
		\sum_{j\in NLS}|	\widetilde{{\bf \Psi}}_{\alpha_{1};j}^{*}\rangle 
		=		\sum_{i,j\in NLS}
		\langle\widetilde{{\bf \Psi}}_{\alpha_{1};i} | 	\widetilde{{\bf \Psi}}_{\alpha_{1};j}^{*}\rangle 
		\neq \frac{D}{2}-1.
	\end{aligned}
\end{equation}

In $|{\bf \Psi}_{\alpha_{1};i}\rangle$ basis,
each eigenstate outside the NLS block is more localized and as a result the collection of them
(like $\tilde{\rho}_{H;1}$) is
more extensive.
As an evidence, the vanishing off-diagonal ($i\neq j$) product,
$\langle{\bf \Psi}_{\alpha_{1};i}|{\bf \Psi}_{\alpha_{1};j}\rangle
=
\langle{\bf \Psi}_{\alpha_{1};i}|
(|{\rm Re}\ {\bf \Psi}_{\alpha_{1};j}\rangle
+|i{\rm Im}\ {\bf \Psi}_{\alpha_{1};j}\rangle)=0$
for every $i,j\notin NLS$
results from 
$\langle{\bf \Psi}_{\alpha_{1};i}|{\rm Re}\ {\bf \Psi}_{\alpha_{1};j}\rangle
=-\langle{\bf \Psi}_{\alpha_{1};i}||
i{\rm Im}\ {\bf \Psi}_{\alpha_{1};j}\rangle)\neq 0$.
While for $i,j\in NLS$,
we have
$\langle{\bf \Psi}_{\alpha_{1};i}|{\rm Re}\ {\bf \Psi}_{\alpha_{1};j}\rangle
=\langle{\bf \Psi}_{\alpha_{1};i}||
i{\rm Im}\ {\bf \Psi}_{\alpha_{1};j}\rangle)= 0$.

The non-defective degeneracies within the RIB of ${\bf \Psi}_{\alpha_{1}}$
work to maintain the integrability of each orthogonal eigenvector
and non-Hermiticity of ${\bf \Psi}_{\alpha_{1}}$.
Thus in non-degenerated region,
the level repulsion is guaranteed by distinct eigenvalues
while in degenerated region the level repulsion is realized 
by "squeezing" (from a single geometry side) 
the replicas of a single non-Hermitian spectrum arranged in acsending order.
$\mathcal{Q}_{ij}:=|{\bf \Psi}_{\alpha_{1};i}\rangle\langle{\bf \Psi}^{*}_{\alpha_{1};j}|$,
($i,j\in$ NLS),
$\mathcal{Q}_{i}^{2}=\mathcal{P}_{i}$,
the nonreciprocal can be seem from $\langle {\bf \Psi}_{\alpha_{1};i}|
\mathcal{Q}_{ij}-\mathcal{Q}_{ji}|{\bf \Psi}_{\alpha_{1};j}\rangle=(1-\delta_{ij})$.
The non-feficient degeneracies as well as the thermal inclusion effect on NLS block
can be seem from the $(\frac{D}{2}-1)$-fold degeneracy
from the $\sum_{i\in NLS}\mathcal{Q}_{ii}$.

	The non-defective degeneracies is pretected by the trace classes 
		where we numerically verify that (for $i\neq j$ and $i,j\in NLS$)
		\begin{equation} 
			\begin{aligned}
				&
			{\rm Tr}[\mathcal{Q}_{ij}-\mathcal{Q}_{ji}]=0,
					{\rm Rank}[	\mathcal{Q}_{ij}-\mathcal{Q}_{ji}]=
									{\rm Rank}[	(\mathcal{Q}_{ij}-\mathcal{Q}_{ji})^2]=2,\\
				&
	(\mathcal{Q}_{ij}-\mathcal{Q}_{ji})|{\bf \Psi}_{\alpha_{1};i} \rangle
		=-	|{\bf \Psi}_{\alpha_{1};j} \rangle,
	(\mathcal{Q}_{ij}-\mathcal{Q}_{ji})|{\bf \Psi}_{\alpha_{1};j} \rangle
		=	|{\bf \Psi}_{\alpha_{1};i} \rangle,
		(\mathcal{Q}_{ij}-\mathcal{Q}_{ji})^2
	|	{\bf \Psi}_{\alpha_{1};j} \rangle
		=-	|{\bf \Psi}_{\alpha_{1};j} \rangle,\\
&
		(\mathcal{Q}_{ij}-\mathcal{Q}_{ji})^2
	|	{\bf \Psi}_{\alpha_{1};i} \rangle
		=-	|{\bf \Psi}_{\alpha_{1};i} \rangle,\\
						&
		{\rm Tr}[(\mathcal{Q}_{ij}-\mathcal{Q}_{ji})^2]=
			{\rm Tr}[(\mathcal{Q}_{ij}-\mathcal{Q}_{ji})
		(\mathcal{Q}_{ij}-\mathcal{Q}_{ji})^{*}
		]=-2
		=-		{\rm Tr}[(\mathcal{Q}_{ij}-\mathcal{Q}_{ji})
		(\mathcal{Q}_{ij}-\mathcal{Q}_{ji})^{\dag}
		].
			\end{aligned}
	\end{equation}
whereas in basis of defective degeneracy $
	{\rm Tr}[(	|\widetilde{{\bf \Psi}}_{\alpha_{1};i} \rangle\langle 
\widetilde{{\bf \Psi}}_{\alpha_{1};j}^{*} |-
	|\widetilde{{\bf \Psi}}_{\alpha_{1};j} \rangle\langle 
\widetilde{{\bf \Psi}}_{\alpha_{1};i}^{*} |)^2]
\approx  -2$.
Another feature related to the non-Hermitian physics,
is the degenerated eigenstates (non-defective degeneracy)
contributing to the
continuous spectrum of the scattering states
(rather than the completely coalesced defective degeneracy),
where in
 large N limit the non-fermi liquid (replica-diagona) will lead to coalesced IR (long-time) poles that forming a branch cut along imaginary axis.
 Such singularity as well as the absence of fermi liquid (well-defined quasi particle) is related to the existence of strong scatterers\cite{Yurovsky}.
 Also, the IR pole is cuted off before vanish by the 
 continuum spectrum which means the system owns a self oscillation frequency (at the saddle point).

\subsection{${\bf J}_{\alpha}^{T}{\bf J}_{\alpha}$}

Then in terms of diagonal ETH applied with respect to the observable ${\bf J}_{\alpha}^{T}{\bf J}_{\alpha}$,
\begin{equation} 
	\begin{aligned}
		\label{9121}
		{\rm Tr}[{\bf J}_{\alpha}^{T}{\bf J}_{\alpha}]
		&=
		{\rm Tr}[\sum_{j}|E_{j} \rangle \langle E_{j}|
		{\bf J}_{\alpha}^{T}{\bf J}_{\alpha}
		]
		=\sum_{j}\langle E_{j}| {\bf J}_{\alpha}^{T}{\bf J}_{\alpha} |E_{j}\rangle \\
		&=
		{\rm Tr}[e^{iH_{D}t}
		\mathcal{P}_{D} {\bf J}_{\alpha}^{T}{\bf J}_{\alpha}
		e^{-iH_{D}t}]
		={\rm Tr}[e^{iH_{D}t} \mathcal{P}_{D} e^{-iH_{D}t}
		{\bf J}_{\alpha}^{T}{\bf J}_{\alpha}]\\
		&
		={\rm Tr}\left[e^{iH_{D}t} \sum_{j}|E_{j}\rangle 
		\sum_{j}\langle E_{j}|
		{\bf J}_{\alpha}^{T}{\bf J}_{\alpha} e^{-iH_{D}t}\right]\\
		&
		={\rm Tr}\left[e^{iH_{D}t} {\bf J}_{\alpha}^{T}{\bf J}_{\alpha} 
		\sum_{j}|E_{j}\rangle  \sum_{j}\langle E_{j}|
		e^{-iH_{D}t}\right]\\
		&=
		\sum_{j}
		\langle E_{j}|e^{iH_{D}t}e^{-iH_{D}t}
		{\bf J}_{\alpha}^{T}{\bf J}_{\alpha} 
		\sum_{j}|E_{j}\rangle\\
		&
		=\lim_{T\rightarrow\infty}\frac{1}{T}\int^{T}_{0}dt
		{\rm Tr}\left[e^{iH_{D}t} \sum_{j}|E_{j}\rangle \sum_{j}\langle E_{j}|
		e^{-iH_{D}t}  {\bf J}_{\alpha}^{T}{\bf J}_{\alpha} \right]\\
		&
		=\lim_{T\rightarrow\infty}\frac{1}{T}\int^{T}_{0}dt
		{\rm Tr}\left[ {\bf J}_{\alpha}^{T}{\bf J}_{\alpha}
		e^{iH_{D}t} \sum_{j}|E_{j}\rangle \sum_{j}\langle E_{j}|
		e^{-iH_{D}t} \right]\\
		&
		=\lim_{T\rightarrow\infty}\frac{1}{T}
		\int^{T}_{0}dt
		\sum_{j}
		\langle E_{j}|e^{iE_{j}t} 
		{\bf J}_{\alpha}^{T}{\bf J}_{\alpha}
		\sum_{j} e^{-iE_{j}t} |E_{j}\rangle\\
		&
		=\lim_{T\rightarrow\infty}\frac{1}{T}
		\int^{T}_{0}dt
		\sum_{j}
		\langle E_{j}|e^{iH_{D}t} 
		{\bf J}_{\alpha}^{T}{\bf J}_{\alpha}e^{-iH_{D}t} 
		\sum_{j} |E_{j}\rangle
		=1.
	\end{aligned}
\end{equation}

\begin{figure}
	\centering
		\includegraphics[width=0.8\linewidth]{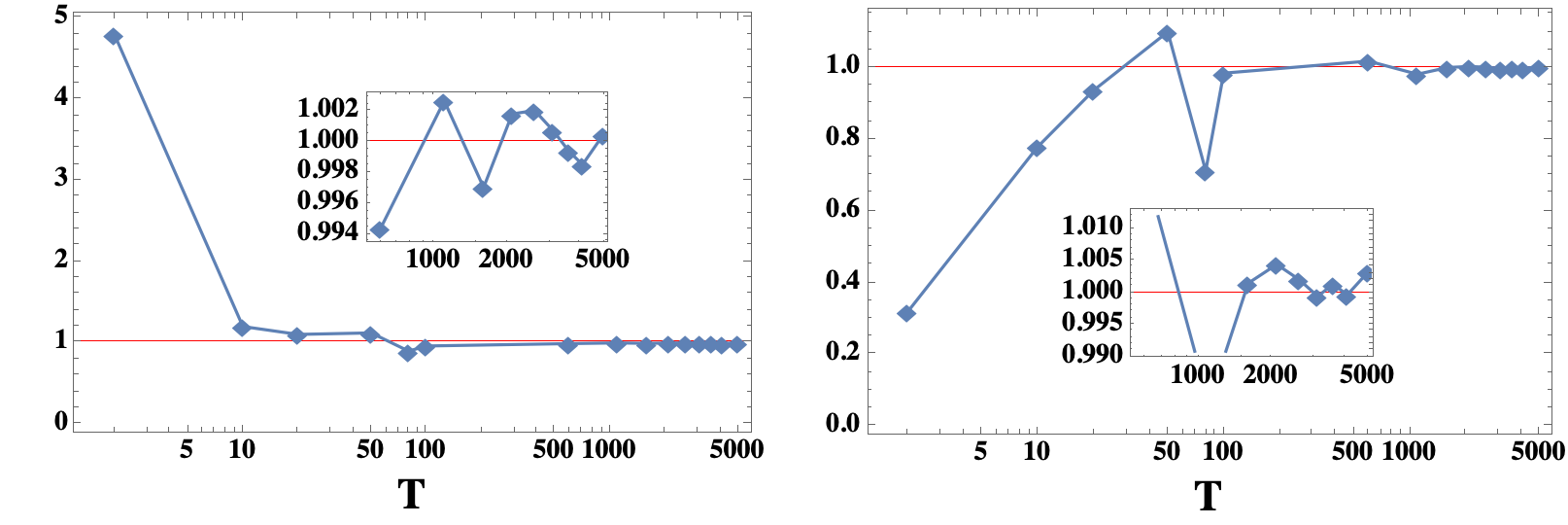}
	\caption{
		Time fluctuation of the integral in Eq.(\ref{9121}),
		$\frac{1}{T}\int^{T}_{0}dt
		{\rm Tr}\left[e^{iH_{D}t}\mathcal{P}_{D}'
		e^{-iH_{D}t}  {\bf J}_{\alpha}^{T}{\bf J}_{\alpha} \right]$
		($\alpha=\alpha_{1}$ for left panel and $\alpha=\alpha_{2}$ for right panel),
		as a function of $T$.
	}
\end{figure}
Since rank-1 density operators of initial state satisfy
$\sum_{j}|E_{j}\rangle\langle E_{j}|={\bf I}$,
it is of the quantum non-local case unlike the thermal case 
which take into account only local configuration\cite{Chung}.
Then for the expectation of observables 
${\bf J}^{T}_{\alpha}{\bf J}_{\alpha}$ and
${\bf J}_{\alpha}{\bf J}^{T}_{\alpha}$,
with respect to initial states,
we obtain
\begin{equation} 
	\begin{aligned}
		&
h_{1}(t):={\rm Tr}[ \mathcal{P}_{D}'(t){\bf J}^{T}_{\alpha}{\bf J}_{\alpha}]
={\rm Tr}[{\bf J}^{T}_{\alpha}{\bf J}_{\alpha}\mathcal{P}_{D}'(t)]
={\rm Tr}[{\bf J}_{\alpha}\mathcal{P}_{D}'(t){\bf J}^{T}_{\alpha}]
=
\sum_{j}e^{iE_{j}t}\langle E_{j}|
{\bf J}^{T}_{\alpha}{\bf J}_{\alpha}
\sum_{j}e^{-iE_{j}t}|E_{j}\rangle,\\
&
h_{2}(t):={\rm Tr}[ \mathcal{P}_{D}'(t){\bf J}_{\alpha}{\bf J}_{\alpha}^{T}]
={\rm Tr}[{\bf J}_{\alpha}{\bf J}_{\alpha}^{T}\mathcal{P}_{D}'(t)]
={\rm Tr}[{\bf J}^{T}_{\alpha}\mathcal{P}_{D}'(t){\bf J}_{\alpha}]
=\sum_{j}e^{iE_{j}t}\langle E_{j}|
{\bf J}_{\alpha}{\bf J}_{\alpha}^{T}
\sum_{j}e^{-iE_{j}t}|E_{j}\rangle,
	\end{aligned}
\end{equation}
where we define
\begin{equation} 
	\begin{aligned}
&
\mathcal{P}_{D}':=\sum_{j}|E_{j}\rangle \sum_{j}\langle E_{j}|,\\
&
\mathcal{P}_{D}'(t):=
\sum_{j}e^{iE_{j}t}|E_{j}\rangle
\sum_{j}e^{-iE_{j}t}\langle E_{j}|
=e^{iH_{D}t}\mathcal{P}_{D}'e^{-iH_{D}t}.
	\end{aligned}
\end{equation}
and here $h_{2}$ ($h_{1}$) corresponds to the unrestricted (restricted) Hilbert space.
As shown in Fig.\ref{34_41},
for unrestricted Hilbert space,
the evolutionary spectrum for the observables of two subsystems initially have different expectation
with respect to initial state $\mathcal{P}_{D}'$,
and
the two expectations are coherented to each other:
The subsystem with higher initial value is bounded by its initial value all through the evolution,
while the subsystem with lower initial value 
guaduatelly deviates from the initial value and approach the some upper boundary at long time.
Also,
we can see the expectation spectrums for unrestricted Hilbert space always exhibit dip-ramp-plateau character,
as the result of maximal chaos of the whole (final) system (GOE with full chaoticity\cite{Cao}).
Due to the same reason,
their will not be level repulsion in the variance (or mean value)
of imaginary part of the expectation.
While for $h_{1}$,
the dip-ramp-plateau structure is absent
and the resulting imaginary spectrums exhibit level repulsion
between $\alpha_{1}$ and $\alpha_{2}$.
This again prove that, in our system,
the imaginary part of eigenvalues plays an essential role in distinguishing the statistical mixture
and superposition of the wave functions.

\subsection{Comparasion between $\mathcal{P}_{D}'$ and $|\mathcal{J}_{\alpha}\rangle\langle\mathcal{J}_{\alpha}|$}

For $\rho_{unres}$
the final (integrable) Hamiltonian has definite eigenvalues 
and there is a clear correspondence (in terms of superposition)
between the initial states and final states.
For $\rho_{res}$, it is possible to find a equivalence between the infinite time average
with an ensemble-like average
$
\lim_{T\rightarrow\infty}
\frac{1}{T}\int^{T}_{0}dt {\rm Tr}[\rho(t)\mathcal{O}]
=
{\rm Tr}[\rho_{0}\mathcal{O}]$.
The restriction on Hilbert space here can be related to the many-body localization on the degenerated
part of the rank-1 initial density $\rho_{res}$.
Similar discussion is reported in Ref.\cite{Shiraishi},
where the zero expectation of local projector (with respect to the embeded integrable eigenstates)
indicate the possible rank-deficient structure of
global projector.
For example,
$\mathcal{P}_{D}'$ corresponds to an unrestricted (restricted) Hilbert space with respect to the observable ${\bf J}_{\alpha}^{T}{\bf J}_{\alpha}$ (${\bf J}_{\alpha}{\bf J}_{\alpha}^{T}$),
we have
$
\lim_{T\rightarrow\infty}
\frac{1}{T}\int^{T}_{0}dt 
h_{1}(t)
=
{\rm Tr}[ \mathcal{P}_{D}'{\bf J}_{\alpha}^{T}{\bf J}_{\alpha}]=1$,
which is invalid for $h_{2}(t)$.
While 
for initial state $|\mathcal{J}_{\alpha}\rangle\langle\mathcal{J}_{\alpha}|$ which can be fully thermalized, the long-time average results in the 
expectation that can be understood as a collection of connect subclusters
which can be intuitively seem from the seperated (overlaped)
initial value (long-time upper boundary) in the evolutionary spectrum,
where
the local character of spectrum boundary guarantees the localization for the collection of
all the subclusters included in this boundary.

When dealing with the long-time average of density matrix representing the initial state,
it is numerically hard to perform the exact integral on the unitarily transformed projectors,
but the averaged result always keep the same trace with the initial one,
and we have numerically proved that 
$\mathcal{P}_{D}'(t)\approx
\mathcal{P}'_{D}+t[H_{D},\mathcal{P}'_{D}]
\approx
\mathcal{P}'_{D}(1-it[H_{D},{\rm ln}\mathcal{P}_{D}^{'i}])
\approx 
\mathcal{P}'_{D}e^{-it[H_{D},{\rm ln}\mathcal{P}_{D}^{'i}]}$.
Then the long-time average can be simply
performed by
$\frac{1}{T}\int^{T}_{0}dt \mathcal{P}_{D}'e^{-it[H_{D},{\rm ln}\mathcal{P}_{D}^{'i}]}
=\mathcal{P}_{D}'$,
as long as $[H_{D},{\rm ln}\mathcal{P}_{D}^{'i}]\ll \frac{1}{T}$.
Physically, that means as long as the commutation 
$[H_{D},{\rm ln}\mathcal{P}_{D}^{'i}]$ is much smaller than the inverse temperature
(a typical case is the zero-temperature limit),
there is a (non-Bloch-type) topological invariance which can be represented by
the winding number $\frac{1}{T}\int^{T}_{0}
dt e^{-it[H_{D},{\rm ln}\mathcal{P}_{D}^{'i}]}
={\rm Re}
[e^{{\rm ln}(i
	[H_{D},{\rm ln}\mathcal{P}_{D}^{'i}]^{-1})
-i[H_{D},{\rm ln}\mathcal{P}_{D}^{'i}]T}]
/T\equiv 1$
(for $T\gg [H_{D},{\rm ln}\mathcal{P}_{D}^{'i}]^{-1}$;
for $T= [H_{D},{\rm ln}\mathcal{P}_{D}^{'i}]^{-1}$
the above winding number is 0.841471).
Thus the topological invariance here which is of a unrestricted Hilbert space,
can be understood by a general rule,
the Fourier transform of Gaussian is still a Gaussian,
and the intrinsic property of Kernel function
(the projector here) is related to the $H_{D}$
and temperature.

\begin{figure}
	\centering
	\includegraphics[width=0.3\linewidth]{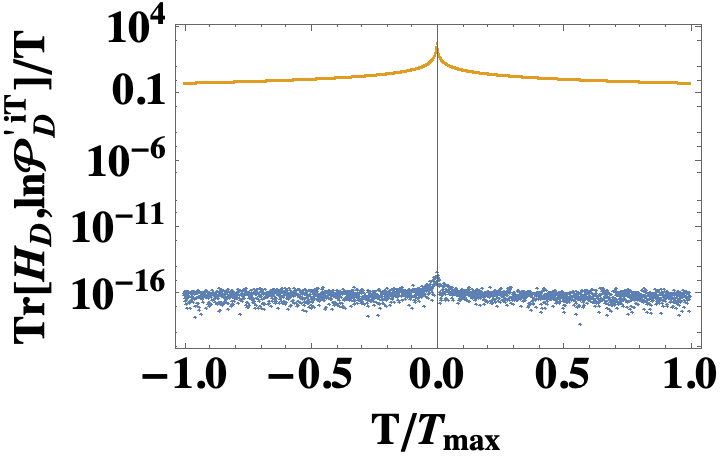}\captionsetup{font=small}
	\caption{$T_{max}$ is the maximal temperature that $\mathcal{P}_{D}^{'iT}$ could be full-rank. 
	}
	\label{O1}
\end{figure}

Here the condition $[H_{D},{\rm ln}\mathcal{P}_{D}^{'i}]\ll \frac{1}{T}$
with ${\rm Rank}\mathcal{P}_{D}'=1$
can be understood by
$[H_{D},{\rm ln}\mathcal{P}_{D}^{'i}]
=\frac{\delta {\rm ln}\mathcal{P}_{D}^{'i}}
{\delta H_{D}}
= -{\rm ln}\mathcal{P}_{D}^{'i}
\frac{\delta ({\rm ln}\mathcal{P}_{D}^{'i})^{-1}}
{\delta H_{D}}
 {\rm ln}\mathcal{P}_{D}^{'i}
<{\bf I} O(1)$
\footnote{
	With the rank-raised projector ${\rm ln}\mathcal{P}_{D}^{'i}$,
	we have another transformation
$\frac{\delta {\rm ln}{\rm ln}\mathcal{P}_{D}^{'i}}
{\delta H_{D}}
= -{\rm ln}\mathcal{P}_{D}^{'-i}
\frac{\delta ({\rm ln}\mathcal{P}_{D}^{'i})^{-1}}
{\delta H_{D}}
 {\rm ln}\mathcal{P}_{D}^{'i}
=\frac{\delta }{\delta H_{D}}
\int ({\rm ln}\mathcal{P}_{D}^{'i})^{-1} d({\rm ln}\mathcal{P}_{D}^{'i})
$.  Here the role of ${\rm ln}\mathcal{P}_{D}^{'i}$
can be seem from the same diagonal elements of the commutator such that the zero trace of the commutator without forms any conservation sector through the diagonal elements.
}
, as a result of the restrictive projection of ${\rm ln}\mathcal{P}_{D}^{'i}$.
As the quantity $O(1)=\frac{1}{T}$ vanishes in thermodynamic limit ($T\rightarrow\infty$),
the commutator here vanishes faster than the inverse temperature,
and thus numerically,
for $H_{D}\in O(D)$, we have ${\rm ln}\mathcal{P}_{D}^{'i}<O(D)$
(thus $\mathcal{P}_{D}^{'i}$ is superexponentially small in system size).
Related discussions about the orignal and thermal ETH was studied in Ref.\cite{Anza},
where for the original ETH the global chaotic system has
the off-diagonal elements (or the difference of neighboring diagonal elements)
in bulk part be exponentially small in system size.
The reason cause such difference is the lower rank of $\mathcal{P}'_{D}$.
Here we note that 
the modified projector $\mathcal{P}_{D'}$
can be regarded as a continuous variance of temperature away from zero, as $\mathcal{P}_{D}'=\lim_{T\rightarrow 0}e^{-\frac{i}{T}{\rm ln}\mathcal{P}_{D}^{'iT}}$.
Numerics show that there will be a range of $T$ allow the resulting projector to be full-rank
which is only determined by the matrix structure.
Further, due to the same reason,
the commutator $[H_{D},{\rm ln}\mathcal{P}_{D}^{'i}]$ coincident with the $O(1)$ quantity only in the low
temperature limit,
but turns to be a plateau with lower value for higher temperature,
as shown in Fig.\ref{O1}.

For $\rho_{unres}$ the diagonal ensemble requires finite energy density (or finite temperature)
during the average over the densities,
$\rho_{0}(\rho_{unres})=\frac{1}{T'}\int^{T'}_{0}dt \rho_{unres}(t)=\rho_{unres}(T'):=|a\rangle\langle a|
=\lim_{T\rightarrow\infty}\int^{T}_{0}dt [\sum_{s}|a\rangle\langle a|_{s}](t)=\sum_{s}|a\rangle\langle a|_{s}$,
where $|a\rangle\langle a|(\neq \rho_{unres})$ is a specifically selected state (for $t_{max}=T'$) away from initial time whose behavior coincides with the that of the restricted initial state,
and we collect all such initial states (to deplete the dependence on $T'$'s)
into a higher rank superposition which turns to be the same result:
${\rm Tr}[|a\rangle\langle a|\mathcal{O}]
={\rm Tr}[\sum_{s}|a\rangle\langle a|_{s}\mathcal{O}]
={\rm Tr}[\rho_{0}\mathcal{O}]$.
Unlike the case of $\rho_{res}$,
here the rank-deficient structure of $|a\rangle\langle a|$ does not indiactes the 
local conserved quantity,
as the equivalent effect between $|a\rangle\langle a|$ and $\sum_{s}|a\rangle\langle a|_{s}$
is not due to the many-body fidelity of $|a\rangle\langle a|$ but the specifically selected configuration $\{s\}$.
However, due to the case distinct to $\rho_{res}$,
here we cannot decompose $\sum_{s}|a\rangle\langle a|_{s}$ into a pure state,
and thus a state $|b\rangle$ is unavailable such that
$\langle a|\mathcal{O}|a\rangle=\langle b|\mathcal{O}|b\rangle$,
i.e., there is not degenerated eigenstates which can form local symmetry by spanning a Hilbert subspace\cite{Shiraishi}.
For $\rho_{unres}$,
the diagonal ensemble is also available through the Kernel Fourier transform,
$\lim_{T\rightarrow\infty}
\frac{1}{T}\int^{T}_{0}dt \rho_{unres}(t)
=\frac{1}{T}\int^{T}_{0}dt \int ds
\rho^{-1}_{unres}(s)e^{-it[H_{D},{\rm ln}\rho_{unres}^{i}(s)]}
=\int ds \delta_{s}\rho_{unres}^{-1}(s)
={\rm ln}\rho_{unres}(s)\bigg|_{s}
=\frac{\overline{\mathcal{O}} \mathcal{O}^{-1}}
{{\rm dim}H_{D}}$,
where the selective role of $\delta_{s}$ can be seem from the constriction
$\int \int ds ds' \delta_{s}\rho^{-1}_{unres}(s')
(1-\delta_{ss'})=0$, and
the rank-deficent density $\rho_{unres}(s)$
is being reshaped into a full-rank one through its physics correspondence,
the fermi-Dirac distribution in zero-temperature limit
(in which case it becomes a step function),
such that
$\rho_{unres}(s)=\lim_{T\rightarrow 0}\frac{1}{1+e^{s/T}}$,
and ${\rm ln}\rho_{unres}(s)={\rm ln}\ \Theta(-s){\bf I}$.
For $\mathcal{O}_{s}$ of the $s$-labelled cluster,
we have
$\partial_{s}\mathcal{O}^{-1}(s)=\frac{\delta_{s}}
{\rho_{unres}(s) \overline{\mathcal{O}}}{\rm dim}H_{D}$,
where 
$\mathcal{O}^{-1}(s)
={\rm ln}\ \rho_{unres}^{\overline{\mathcal{O}}^{-1}}(s)
{\rm dim}H_{D}$.
Physically,
the procedure of rank-rasing here
(through a power of the density or the multiple of its logarithm)
is endowing the rank-deficent densities a phase (or symmetry)
that is persistive under continuous transformation
(which is the temperature turns toward the zero limit here).
Methodologically,
this is similar to the change of exponential factor within the Fourier transform,
where the non-Bloch wavevector acquires an imaginary part differet to that of Bloch case\cite{Yao}.
While the additional imaginary part is the $s$-dependent part here.

During the Kernel Fourier transform,
the rank-1 density $|\mathcal{J}_{\alpha}\rangle\langle\mathcal{J}_{\alpha}|$
of the cluster $s$ can be rank-raised by relating it (or its corresponding symmetry of the cluster)
to the inversed fermi-Dirac distribution,
such that its rank-deficient structure can be reasonably "filled" through the
continuous variance of fermi-Dirac distribution with the lowering temperature toward zero,
\begin{equation} 
	\begin{aligned}
		\label{1251}
		&
		|\mathcal{J}_{\alpha}\rangle\langle\mathcal{J}_{\alpha}|_{s}
= {\rm diag}(1,0,\cdots,0)_{s}
\rightarrow {\rm diag}(\Theta(x_{1}),\Theta(x_{2}),\cdots,\Theta(x_{D}))_{s}
		=\lim_{T\rightarrow 0}
		\frac{{\bf I}}{1+e^{s/T}}
		=\Theta(-s){\bf I}.
	\end{aligned}
\end{equation}
where the configuration $\{s\}$ is defined as a collection of 
selections of $x_{1},\cdots,x_{D}$
such that the diagonal elements can be the functions of a common variable $s$.
As the labels $\{s\}$ are distinguished by the temperature
(or the largest time during the average).
Here we have 
$|a\rangle\langle a |\mathcal{O}|a\rangle\langle a| =\overline{\mathcal{O}}
|a\rangle\langle a |$, and the rank-rasing is completed by
$|a\rangle\langle a | \rightarrow 
		\sum_{s}|a\rangle\langle a|_{s}=
		e^{-iT{\rm ln}|a\rangle\langle a|^{i/T}}$, where $T$ is generally a small value that depends on the detail form of $\rho_{unres}$,
and it is precisely determined by ${\rm Rank}[e^{-iT{\rm ln}|a\rangle\langle a|^{i/T}}]=
{\rm dim}H_{D}$
and 
$e^{iT{\rm ln}|a\rangle\langle a|^{i/T}}\rho_{0}={\bf I}$.
Compares to Eq.(\ref{1251}),
the above rank-raised form can be understood as a collection over all clusters
labelled by $s$'s.
Thus in the mean time,
the observable should also be reshaped into a collection over all clusters,
		$\mathcal{O}\rightarrow \sum_{s}\mathcal{O}_{s}
		={\rm dim}H_{D}\mathcal{O}$.
Base on above transitions,
during which $\tilde{\rho}_{DE}$ becomes a restrictive projector,
while $\mathcal{O}$ is simply multipled by the Hilbert space dimension, we can further obtain
$		\overline{O}\sum_{s}|a\rangle\langle a|_{s}
		=\sum_{s}|a\rangle\langle a|_{s}\sum_{s}\mathcal{O}_{s}
		\sum_{s}|a\rangle\langle a|_{s}$,
where there is a global restrictive projection\cite{Choi}
$
		\sum_{\{s,s',s''\}}|a\rangle\langle a|_{s}\mathcal{O}_{s'}
		|a\rangle\langle a|_{s''}=0$,
where $\{s,s',s''\}$ including all the cases other than $s=s'=s''$.
\begin{equation} 
	\begin{aligned}
		\overline{O}(\sum_{s}|a\rangle\langle a|_{s})^{-1}
		=|a\rangle\langle a|{\rm dim}H_{D}\mathcal{O}|a\rangle\langle a|^{-1}
		={\rm dim}H_{D}\mathcal{O},
	\end{aligned}
\end{equation}
which can be verified by substituting it back to 
${\rm Tr}[\rho_{0}\mathcal{O}]=\overline{O}$.
The summation over all clusters labelled by $s$'s
built a "virtue" constriction on the Hilbert space,
and such restriction is reflected by $\sum_{s}\frac{\mathcal{O}_{s}}{\mathcal{O}}={\rm dim}H_{d}$,
while for each cluster,
its localization is guaranteed by the step function which is asymptotically 
approached by the fermi-Dirac distribution at zero temperature limit. 
Again, here the (virtuelly) restricted Hilbert space can be seem by comparing the
inifinite one (i.e., infinite number of basis
$|a\rangle\langle a|_{s}$ as each $s$ corresponds to an arbitary end-time)
and the one whose dimensionality cannot be inifinite but, instead,
governed by $\sum_{s}\frac{\mathcal{O}_{s}}{\mathcal{O}}={\rm dim}H_{d}$.
In other word,
the restricted configuration $\{s\}$,
making $\sum_{s}|a\rangle\langle a|_{s}$ be a constant
(preclude the case of randomly selected $s$),
and as a result, 
the Hilbert space is restricted
(see Ref.\cite{Shiraishi} for similar discussion on this).
For unrestricted Hilbert space,
the corresponding $\rho_{0}$ has a rank higher than 1,
but
due to the persistive correlation between the unrestrictive projection 
of $\rho_{unres}$ to the observable,
we can make a collective approximation:
$\rho_{0}(\rho_{unres})\rightarrow |a\rangle\langle a|$,
$\mathcal{O}\rightarrow \overline{\mathcal{O}(\rho_{unres})}|a\rangle\langle a|^{-1}$,
such that $\rho_{0}(\rho_{unres})$ can be reshape into a restrictive projection
$|a\rangle\langle a|$,
and thus
there is kinetic-type restriction\cite{Popescu,Choi}
$|a\rangle\langle a|\frac{ \mathcal{O}}{\overline{\mathcal{O}(\rho_{unres})}} |a\rangle\langle a|
=|a\rangle\langle a|$
where $\mathcal{O}$ is $s$-independent here.
Also,
due to the property of restrictive projection,
which is, e.g., $[[H_{D},{\rm ln}\mathcal{P}_{D}^{-i}],\mathcal{P}_{D}]=0$
which equivalents to $[H_{D},{\rm ln}\mathcal{P}_{D}^{-i}]=0$,
the $|a\rangle\langle a|$ allows to be diagonalized
such that 
$|a\rangle\langle a|\frac{ \mathcal{O}}{\overline{\mathcal{O}(\rho_{unres})}} |a\rangle\langle a|
=\frac{ \mathcal{O}}{\overline{\mathcal{O}(\rho_{unres})}} |a\rangle\langle a|^2$,
and the corresponding 
thermal-type restriction (which leads to entropy)
can be seem from the conservation $[{\rm ln}|a\rangle\langle a|^{-i/s},{\rm ln}\mathcal{O}]
=0$.

Thus
the states taken into account are always of the nonlocal configuration
(usually corresponds to the quantum phases at zero temperature limit).
While for those initial states of a restricted Hulbert space such 
that unable to be fully thermalized,
the states taken into account are always only of a local configuration
(where the most significant signal is the emergent level repulsion)
and the final nonintegrable Hamiltonian can be well described by the thermal phases
(where the temperature or effective temperature play important role).
This is also why for a strongly localized state only a limited subset
of coefficients (projections of initial states among the eigenstates) contribute,
where there is large inverse participation ratio
as well as the participation entropy independent of system size\cite{Sousa,Ganeshan,van}.

\begin{figure}
	\centering
		\includegraphics[width=0.9\linewidth]{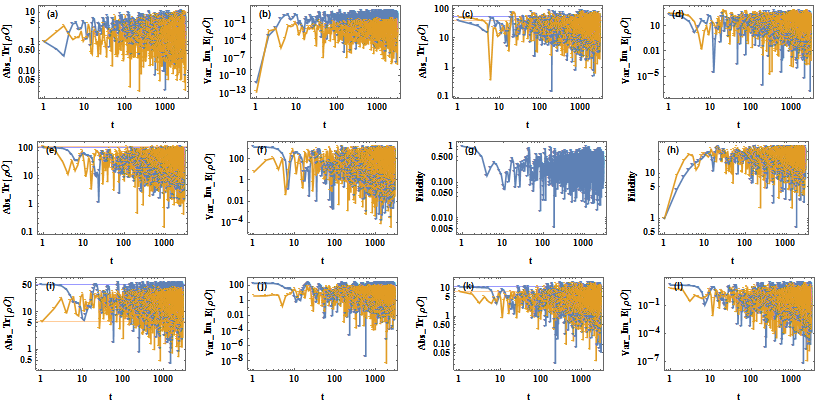}\captionsetup{font=small}
\caption{
(a) Absolute trace ${\rm Tr}[\mathcal{P}_{D}'{\bf J}_{\alpha_{1}}^{T}{\bf J}_{\alpha_{1}}]$ (orange)
and ${\rm Tr}[\mathcal{P}_{D}'{\bf J}_{\alpha_{2}}^{T}{\bf J}_{\alpha_{2}}]$ (blue).
(b) Variance of imaginary part of eigenvalues (VIE) for
$\mathcal{P}_{D}'{\bf J}_{\alpha_{1}}^{T}{\bf J}_{\alpha_{1}}$ (orange)
and $\mathcal{P}_{D}'{\bf J}_{\alpha_{2}}^{T}{\bf J}_{\alpha_{2}}$ (blue).
(c) Absolute trace ${\rm Tr}[\mathcal{P}_{D}'{\bf J}_{\alpha_{1}}{\bf J}^{T}_{\alpha_{1}}]$ (orange)
and ${\rm Tr}[\mathcal{P}_{D}'{\bf J}_{\alpha_{2}}{\bf J}^{T}_{\alpha_{2}}]$ (blue).
(d) Variance of imaginary part of eigenvalues (VIE) for
$\mathcal{P}_{D}'{\bf J}_{\alpha_{1}}{\bf J}^{T}_{\alpha_{1}}$ (orange)
and $\mathcal{P}_{D}'{\bf J}_{\alpha_{2}}{\bf J}^{T}_{\alpha_{2}}$ (blue).
(e) Absolute trace ${\rm Tr}[\mathcal{P}_{D}'\mathcal{J}_{\alpha_{1}}^{T}\mathcal{J}_{\alpha_{1}}]$ (orange)
and ${\rm Tr}[\mathcal{P}_{D}'\mathcal{J}_{\alpha_{2}}^{T}\mathcal{J}_{\alpha_{2}}]$ (blue).
(f) Variance of imaginary part of eigenvalues (VIE) for
$\mathcal{P}_{D}'\mathcal{J}_{\alpha_{1}}^{T}\mathcal{J}_{\alpha_{1}}$ (orange)
and $\mathcal{P}_{D}'\mathcal{J}_{\alpha_{2}}^{T}\mathcal{J}_{\alpha_{2}}$ (blue).
(g)
Many-body fidelity of initial state $\mathcal{P}_{D}'$.
(h) 
Many-body fidelity of initial state $|\mathcal{J}_{\alpha}(t)\rangle\langle\mathcal{J}_{\alpha}(t)|$
with $\alpha=\alpha_{1}$ (blue) and $\alpha=\alpha_{2}$ (orange).
(i) Absolute trace ${\rm Tr}[|\mathcal{J}_{\alpha_{1}}\rangle\langle\mathcal{J}_{\alpha_{1}}|{\bf J}_{\alpha_{1}}^{T}{\bf J}_{\alpha_{1}}]$ (orange)
and ${\rm Tr}[|\mathcal{J}_{\alpha_{2}}\rangle\langle\mathcal{J}_{\alpha_{2}}|{\bf J}_{\alpha_{2}}^{T}{\bf J}_{\alpha_{2}}]$ (blue).
(j) Variance of imaginary part of eigenvalues (VIE) for
$|\mathcal{J}_{\alpha_{1}}\rangle\langle\mathcal{J}_{\alpha_{1}}|{\bf J}_{\alpha_{1}}^{T}{\bf J}_{\alpha_{1}}$ (orange)
and $|\mathcal{J}_{\alpha_{2}}\rangle\langle\mathcal{J}_{\alpha_{2}}|{\bf J}_{\alpha_{2}}^{T}{\bf J}_{\alpha_{2}}$ (blue).
(k) Absolute trace 
${\rm Tr}[|\mathcal{J}_{\alpha_{2}}\rangle\langle\mathcal{J}_{\alpha_{2}}|{\bf J}_{\alpha_{1}}^{T}{\bf J}_{\alpha_{1}}]$ (blue)
and ${\rm Tr}[|\mathcal{J}_{\alpha_{1}}\rangle\langle\mathcal{J}_{\alpha_{1}}|{\bf J}_{\alpha_{2}}^{T}{\bf J}_{\alpha_{2}}]$ (orange).
(l) Variance of imaginary part of eigenvalues (VIE) for
$|\mathcal{J}_{\alpha_{2}}\rangle\langle\mathcal{J}_{\alpha_{2}}|{\bf J}_{\alpha_{1}}^{T}{\bf J}_{\alpha_{1}}$ (blue)
and $|\mathcal{J}_{\alpha_{1}}\rangle\langle\mathcal{J}_{\alpha_{1}}|{\bf J}_{\alpha_{2}}^{T}{\bf J}_{\alpha_{2}}$ (orange).
In the spectrums of the observable's expectation,
the dip-ramp-plateau structure appear only for the observables ${\bf J}_{\alpha}{\bf J}_{\alpha}^{T}$
and $|\mathcal{J}_{\alpha}(t)\rangle\langle\mathcal{J}_{\alpha}(t)|$
(which signals the unrestricted initial state Hilbert space and the integrable final state).
While for fidelity,
the dip-ramp-plateau structure appear only for the observables ${\bf J}_{\alpha}^{T}{\bf J}_{\alpha}$
(which signals the restricted initial state Hilbert space and the nonintegrable final state).
The horizon solid or dashed lines are used to distinguishing these two cases.
For $\rho_{unres}$,
the expectations for two subsystems
always started with different initial values and
overlap at long time (large energy density), during which process there is a formation of 
(topological-symmetry-protected) coherence.
Such time-dependence on the many body fidelity
(distingushable in short $t$ and nondistinguishable in long $t$)
is a unique character of the nonergodic system.
Also, this will introduce the bounded oscillatory dynamics in Loschmidt echo,
in a restricted Hilbert space with parent Hamiltonian. 
}
\label{34_41}
\end{figure}

For nonergodic system,
the initial local information is recoverable due to the persistent many-body revivals,
and the most stable revivals happen near $t=0$ where the spacings between 
quantized segments are nearly the same.
In this stage the many-body fidelity is highest as can be evidenced by the 
smallest distance to the initial trace (similar to the concept of statistical Bures distance)
8;
While in the next stage (the bulk part which still far away from the thermodynamic limit)
there is a linear increase of spacing which exhibits equal-spacing character in log scale,
In this stage the distinguishability between initial state and evoluted state is linearly increased,
which originate from the an ergodictype effect that the dominating edge burst in thermodynamic limit
enforce an extensive overlap between the high-entangled state and low-entangled state\cite{Choi}
(meanwhile the coherence effect is being faded).
Within each segment,
the edge burst strictly happen in the left side
which corresponds to the evolved initial state that has lagerest overlap
with initial state (over all the evolved states within this segment).
But the precondition for these stagements is that the thermodynamic limit does not taken into account,
which guarantees that the
effect from higher-energy-density states is attenuated
by the exponentially large Hilbert space.
In the last stage,
the spacing become exponentially increased,
and the eigenvalue variance for both the evolved $\mathcal{P}_{D}$
and $\mathcal{P}_{D}'$ behave in the same way.
That means, despite the persistent many-body revivals,
it becomes indistinguishable since the edge burst of each segment 
always be overwhelmed by the segment in latter time.
As a result,
the whole spectrum is being squeezed such that the bulk part is totally degenerated
due to the overwhelming edge at thermodynamic limit.
Also,
in the intermediate range between stage one and stage two,
there would be weakly thermalized eigenstates,
which means the observed entanglement is dominating over the coherence,
in which case the higher-energy-density eigenstates cannot be regarded as random vector due to the 
observation-enforced restriction on Hilbert space and such that
the competition effect from the coherence against entanglement effect can be 
effectively filtered.
The key point to realizing this is the increasing entanglement entropy
faster than linear grownth.
While the proximity to integrability from an ergodic system can be realized
by the subvolume entanglement law through the state merging (cause the saturation of mutual information)\cite{Brand}.
Such squeezing-induced proximity to thermalization is possible to be 
realized experimentally through, e.g., Hall measurement,
by the modification of carrier density
such that the conductivity can be effectively reduced
and lower the quantization effect.
where there is a transition from imcompressible to compressible for 
the effective density of state and 
in competition with the quantized conductance 
due to the absence of 
topological symmetry and the possible compressible (or entanglement) effect is
avoided by the insulating bulk.

Although the Hermitian matrices $\mathcal{P}_{D}$
and $\mathcal{P}_{D}'$
have the same trace $D$,
they have different ranks, where the former is always full rank and be a identity matrix,
while the latter is a rank-1 matrix.
In terms of the quantum evolution,
$e^{iH_{D}t}\mathcal{P}_{D}e^{-iH_{D}t}$ and
$e^{iH_{D}t}\mathcal{P}_{D}'e^{-iH_{D}t}$ are still the Hermitian matrices
and their eigenvalues are independent of time,
but we notice that they have quite different variances for the small imaginary part of
eigenvalues which can be observed only in log scale,
and also, their eigenvectors behave quite differently during the evolution of time.
$e^{iH_{D}t}\mathcal{P}_{D}'e^{-iH_{D}t}$ exhibits more significant character in non-Hermiticity,
that its eigenvectors keep mutually independent (do not forced to be mutually orthogonal and self-normalized with the evolution of time)
while its imaginary parts of eigenvalues exhibit persistive level repulsion
(in log scale) with the time evolution, 
which cannot be found in $e^{iH_{D}t}\mathcal{P}_{D}e^{-iH_{D}t}$.
Comparing Fig.\ref{44}(e) and Fig.\ref{45}(e),
for the mean value of the imaginary part of eigenvalues,
$e^{iH_{D}t}\mathcal{P}_{D}'e^{-iH_{D}t}$ exhibit larger fluctuation 
and a more mild approching to the zero axis,
compares to $e^{iH_{D}t}\mathcal{P}_{D}e^{-iH_{D}t}$.

Thus $e^{iH_{D}t}\mathcal{P}_{D}e^{-iH_{D}t}$ behaves as a
extensive sum of the subspaces
each one corresponds to a integrable eigenstate with well-defined eigenvalue.
While for the projector in reduced-density-form, 
the off-diagonal part contributes to the nonlocal symmetries within $e^{iH_{D}t}\mathcal{P}_{D}'e^{-iH_{D}t}$.
While 
$e^{iH_{D}t}|\mathcal{J}_{\alpha_{1}}\rangle\langle\mathcal{J}_{\alpha_{1}}| e^{-iH_{D}t}$ and
$e^{iH_{D}t}|\mathcal{J}_{\alpha_{2}}\rangle\langle\mathcal{J}_{\alpha_{2}}| e^{-iH_{D}t}$
exhibit stacked (or quantized) spectrum
where the contribution from off-diagonal as well as the single boundary correspondence
is absent.
This should be classified into the case of strong disorder limit where
the non-Hermicity is destoried and the off-diagonal part plays no role.
Also, the Hermicity can be seem from the linearly increased
variance as shown in Fig.,
which follows a fixed slope without any saturation
(i.e., an extensive type accumulation instead of scaling base on the corresponding expectations of each
subsystem). One of the direct result of absence of saturation is the absence of level statistics.

Also,
$e^{iH_{D}t}\mathcal{P}_{D}' {\bf J}^{T}_{\alpha}{\bf J}_{\alpha} e^{-iH_{D}t}$ or
$e^{iH_{D}t}\mathcal{P}_{D}
|\mathcal{J}_{\alpha}\rangle\langle\mathcal{J}_{\alpha}| e^{-iH_{D}t}$ behave in the same way with
$e^{iH_{D}t}|\mathcal{J}_{\alpha}\rangle\langle\mathcal{J}_{\alpha}| e^{-iH_{D}t}$.
Thus, except $\mathcal{P}_{D}'$ and $|\mathcal{J}_{\alpha}\rangle\langle\mathcal{J}_{\alpha}|$,
$\mathcal{P}_{D}'{\bf J}^{T}_{\alpha}{\bf J}_{\alpha}$
and $\mathcal{P}_{D}
\mathcal{J}_{\alpha}\rangle\langle\mathcal{J}_{\alpha}| $ can also be thermalized with respect to some certain observable.

Thus the investigation here contributes to the deeper study of nonintegrable initial state,
i.e., whether it can be thermalized to the integrable final state
(which corresponds to the case of successfully thermalized) after the proper quench
base on the Hamiltonian $H_{D}$,
or unable to do so (which corresponds to the incomplete thermalization or unrestricted Hilbert space
as mentioned above).
Also, there is a most trivial case,
the integrable initial state turns to another initegrable final state,
i.e.,
the case for the evolution of $e^{iH_{D}t}\mathcal{P}_{D}e^{-iH_{D}t}$.
While for the more interesting cases,
the nonintegrable initial state quenches to integrable final state and nonintegrable final state
(where the level repulsion is still available)
correspond to the evolutions of $e^{iH_{D}t}|\mathcal{J}_{\alpha}\rangle\langle\mathcal{J}_{\alpha}|
e^{-iH_{D}t}$
and $e^{iH_{D}t}\mathcal{P}'_{D}e^{-iH_{D}t}$, respectively.
For $\rho_{unres}$\cite{rig},
the initial state able to be fully thermalized should be
the superpositions of final eigenstates and the expectations of observable
are of the Haar measure,
while here we provide deeper insign to this interesting question.
Except the evolutionary spectrum,
the many-body fidelity as well as the imaginary part of expectation of some proper observable
is helpfull in distingushing them.
As shown in Fig.\ref{34_41}(i)-(j),
the many-body fidelity is bounded under 1 and exhibit the dip-ramp-plateau character for for the density $e^{iH_{D}t}\mathcal{P}'_{D}e^{-iH_{D}t}$,
which means it is coherented with its counterpart
and thus contains some symmetry which is robust to the quenches,
while for initial state $|\mathcal{J}_{\alpha}\rangle\langle\mathcal{J}_{\alpha}|$,
it is less suggresting the revival of initial states.

\begin{figure}
	\centering
	\includegraphics[width=0.6\linewidth]{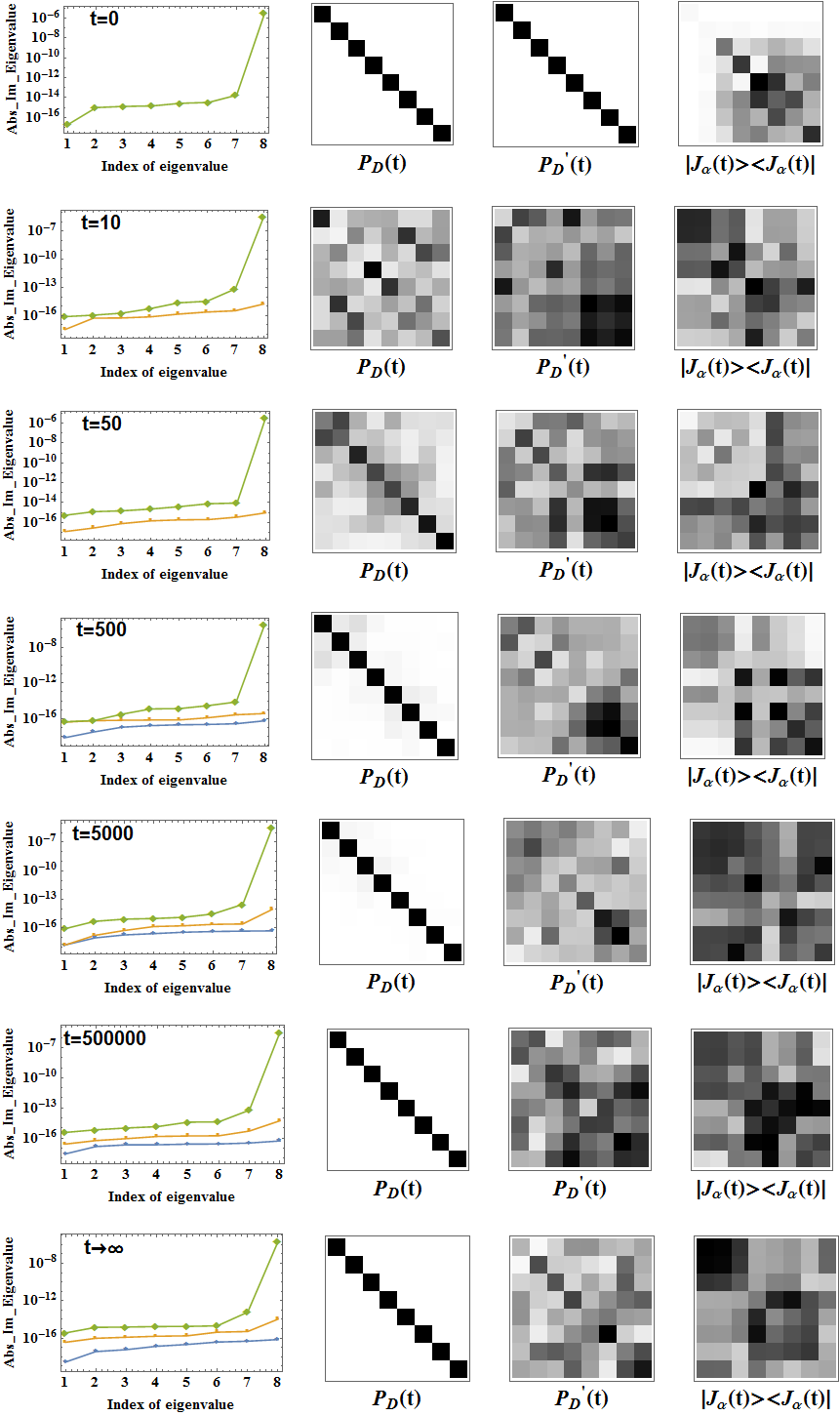}\captionsetup{font=small}
	\caption{
		Absolute values of the imaginary part of eigenvalues of $e^{iH_{D}t}\mathcal{P}_{D}e^{-iH_{D}t}$ (blue), $e^{iH_{D}t}\mathcal{P}'_{D}e^{-iH_{D}t}$ (orange),
		and $e^{iH_{D}t}|\mathcal{J}_{\alpha}\rangle\langle\mathcal{J}_{\alpha}|e^{-iH_{D}t}$ (green) at different time are shown in first column. the second, third, and fourth columns 
		show the array plots of the outer products between the eigenvectors
		of $e^{iH_{D}t}\mathcal{P}_{D}e^{-iH_{D}t}$, $e^{iH_{D}t}\mathcal{P}'_{D}e^{-iH_{D}t}$,
		and $e^{iH_{D}t}|\mathcal{J}_{\alpha}\rangle\langle\mathcal{J}_{\alpha}|e^{-iH_{D}t}$.
		From first column, the degeneracies appear in long-time limit for the imaginary part of eigenvalues of
$e^{iH_{D}t}\mathcal{P}_{D}e^{-iH_{D}t}$ and
$e^{iH_{D}t}|\mathcal{J}_{\alpha}\rangle\langle\mathcal{J}_{\alpha}|e^{-iH_{D}t}$,
and their eigenvector outer product exhibit patterns the same with the initial ones.
	}
\end{figure}


Each diagonal element of
$e^{iH_{D}t}\mathcal{P}_{D}e^{-iH_{D}t}$
can be thermalized with time evolution (as a consequence, its eigenvectors becomes mutually orthogonal and self-normalized in long-time limit).
While each eigenvectors of
$e^{iH_{D}t}\mathcal{P}_{D}'e^{-iH_{D}t}$ exhibits insistent 
level repulsion and thus be mutually independent during the evolution,
which leads to the absence of thermalization after the quench (unitary time evolution).
Also,
the persistive level repulsion for
$e^{iH_{D}t}\mathcal{P}_{D}'e^{-iH_{D}t}$
can be seem from the variance evolution,
as it is the only explaination for the enhancement of variance 
(avoided accidental degeneration) after a long-time evolution.
Thus for $e^{iH_{D}t}\mathcal{P}_{D}'e^{-iH_{D}t}$,
the level repulsion exists between its replicas at different time.

As a consequence,
$e^{iH_{D}t}\mathcal{P}_{D}'e^{-iH_{D}t}$ as a (non-extensive) collection of initial states
can be thermalized with respect to the observable 
${\bf J}^{T}_{\alpha}{\bf J}_{\alpha}$.
This is why the level repulsion can be observed only from the begining and end
of the evolution.
In other word, the localization for $e^{iH_{D}t}\mathcal{P}_{D}'e^{-iH_{D}t}$ 
at a single time can be observed only begining and end of the spectrum
(or variance of the mean values).
Meanwhile, the unitary operator $e^{iH_{D}t}$ plays an important role in troducing the 
quenches away from the equilibrium.

Also,
it should be distinguished from the above-discussed case, where the 
${\bf \Psi}_{\alpha_{1}}$ keeps a non-defective degenerate region to guarantees the 
repulsion between degenerated levels,
while the $e^{iH_{D}t}\mathcal{P}_{D}e^{-iH_{D}t}$
does not keeping the level repulsion even in log scale.
In constrast, since
$e^{iH_{D}t}\mathcal{P}_{D}'e^{-iH_{D}t}$ withou the level repulsion and also
without the non-defective character,
there is only one (collective) observable that be able to thermalized with respect
to ${\bf J}^{T}_{\alpha}{\bf J}_{\alpha}$,
unlike the case for the ${\bf \Psi}_{\alpha_{1}}$ where
its degenerated levels can be treated as thermalized with respect to 
itself as guaranteed by the robust diagonal correlation between 
the nondegenerated block (whose dimension tends to be 1 in many-subsystem limit)
and the degenerated block. Note that in this case, the off-diagonal part
for each subsystem does not form symmetry sector to guarantees the level repulsion
and non-Hermicity.

As for $\tilde{\rho}_{res}$,
the local measurement through the observables of the two subsystems
turns to be entangled thermal states (of an unknown parent Hamiltonian),
such unthermalized and extensive final states usually leads to Barren plateaus
which is the difficulty in decoding the initial state information\cite{Rappaport}.
Then in terms of the information theory,
the final states become less entangled (still entangled and whole system becomes thermalized)
for known (unknown) measurement outcomes\cite{Rappaport}.
Such distinction is less significant here due to the low times of measurements.
Another way to obtain unentangled final states is measure using the rank-1 observables
(some rank-invariant reconstructions are also capable to do so,
like ${\bf J}_{\alpha}{\bf J}_{\alpha}^{T}$).
We found that,
for the trivial observable (like ${\bf J}_{\alpha}^{T}{\bf J}_{\alpha}$), 
there is always level repulsion on the imaginary spectrum for $\rho_{res}$;
While for teh nontrivial observables (like the rank-1 measurement),
the long-time average match the Haar measurement better
compares to the trivial ones,

\begin{figure}
	\centering
	\includegraphics[width=1\linewidth]{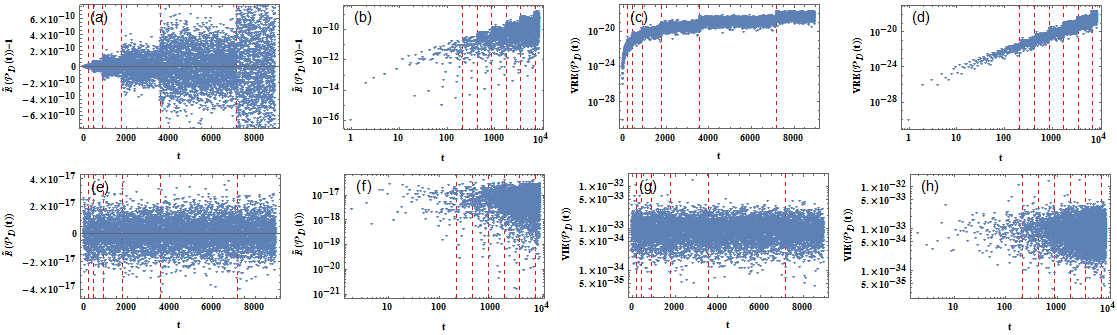}\captionsetup{font=small}
	\caption{
	Time evolutions of the mean value and variance of the imaginary part 
and real part of eigenvalues for $e^{iH_{D}t}\mathcal{P}_{D}e^{-iH_{D}t}$.
The imaginary part of eigenvalues exhibit no signal of thermalization or level repulsion 
From real part of eigenvalues,
we can clearly see the quantization feature,
i.e., the real part of spectrum evolved with time is consist of discontinuously connected segments
where each one of them is governed by a non-Hermitian symmetry (as the edge burst can be seen from the left boundary of each segment; 
The red dashed lines indicate the position of each edge burst
where the interval for each segment does not increases by a fix ratio, but shift to larger ratio with the evolution.
This means that there must be a edge at long-time limit,
in which case the corresponding distincted levels correspond to the intervals of segments in ascending order.
The pseudo-periodic reveals of thermalization for the real part of eigenvalues can be seen in terms of
the mean value and variance for $\mathcal{P}_{D}$, while for $\mathcal{P}_{D}'$ only that in terms of mean value preserve.
As a result of stability of variance,
the correspondence to the above-mentioned edge at long-time limit can be overcomed by the fixed variance
(as a nonlocal symmetry sector whose size is variant).
Consequantly, the edge burst guaranteed by such nonlocal symmetry can be seen in imaginary spectrum 
only for $\mathcal{P}_{D}'$.
In conclusion,
the evolution of initial state of $\mathcal{P}_{D}'$ is able to be non-Hermitian
while that for $\mathcal{P}_{D}$ can only be a piece of spectrum of non-Hermitian system.
In a sharp difference to the 
evolution of
$e^{iH_{D}t}{\bf J}_{\alpha}^{T}{\bf J}_{\alpha}e^{-iH_{D}t}$,
we reproduce the result in a larger scale.
We found that,
the imaginary spectrum of
$e^{iH_{D}t}{\bf J}_{\alpha}^{T}{\bf J}_{\alpha}e^{-iH_{D}t}$
also exhibit quantization character but not as obvious as that for $\mathcal{P}_{D}$
and $\mathcal{P}_{D}'$.
This, again, explain the absence of level repulsion.
Now one may wondering what will happen when it reaches the above-mentioned edge at long-time limit
where the edge burst for $\mathcal{P}_{D}'$.
We can see that in this limit,
$e^{iH_{D}t}\mathcal{P}_{D}e^{-iH_{D}t}$ and keep static untill the edge reaches,
where
and $e^{iH_{D}t}\mathcal{P}_{D}'e^{-iH_{D}t}$ turn to the same value and exponentially increase together.
In terms of the whole evolutionary spectrum,
although thermalization happen for both these three initial states,
we can regard the left part as the degenerated region where
only the evolution of $e^{iH_{D}t}{\bf J}_{\alpha}^{T}{\bf J}_{\alpha}e^{-iH_{D}t}$
keeps the non-defective feature.
	}
\label{44}
\end{figure}

\begin{figure}
	\centering
	\includegraphics[width=1\linewidth]{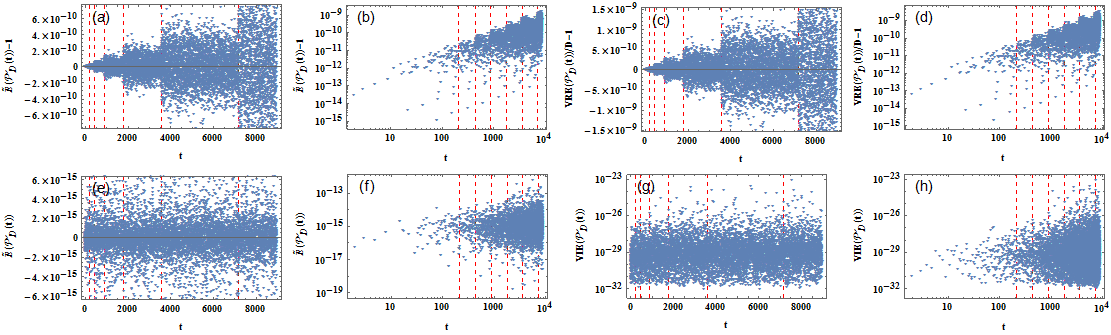}\captionsetup{font=small}
	\caption{	Time evolutions of $e^{iH_{D}t}\mathcal{P}'_{D}e^{-iH_{D}t}$.
	}\label{45}
\end{figure}
\begin{figure}
	\centering
	\includegraphics[width=1\linewidth]{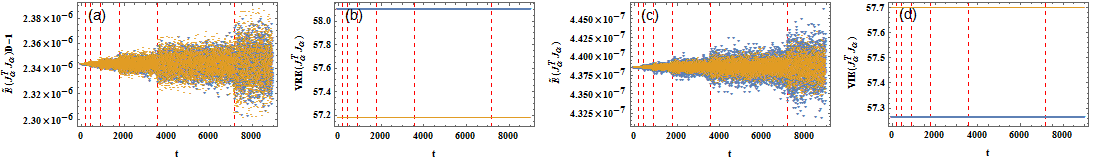}\captionsetup{font=small}
	\caption{Time evolutions of $e^{iH_{D}t}{\bf J}_{\alpha}^{T}{\bf J}_{\alpha}e^{-iH_{D}t}$.
	}
\end{figure}
\begin{figure}
	\centering
	\includegraphics[width=1\linewidth]{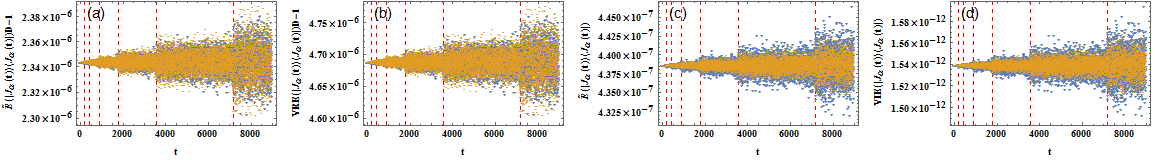}\captionsetup{font=small}
	\caption{Time evolutions of $e^{iH_{D}t}|\mathcal{J}_{\alpha}\rangle \langle \mathcal{J}_{\alpha}|
e^{-iH_{D}t}$.
	}
\end{figure}
\begin{figure}
	\centering
	\includegraphics[width=1\linewidth]{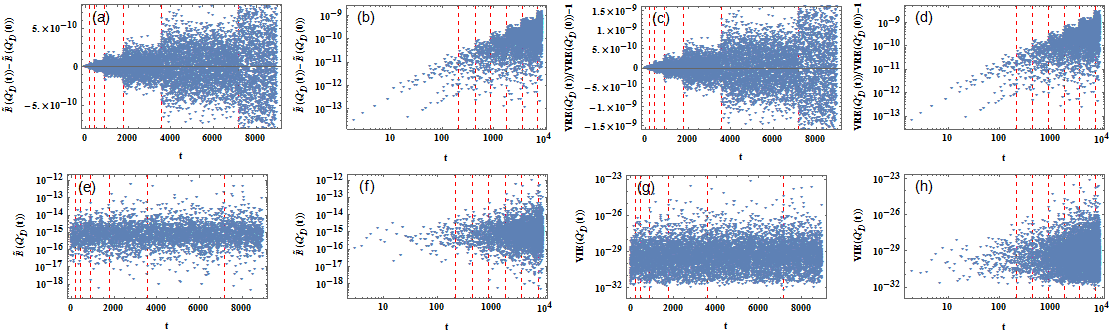}\captionsetup{font=small}
	\caption{Time evolutions of $e^{iH_{D}t}\mathcal{Q}_{nH;1}e^{-iH_{D}t}$.
	}
\end{figure}

\subsection{Effect of multi-edge}

A completely thermalized result can be seem from Eq.(\ref{10191}),
where the time-evolved $e^{-iH_{D}t}|\mathcal{J}_{\alpha}\rangle\langle\mathcal{J}_{\alpha}|
e^{iH_{D}t}$ exhibits a linear increasing without saturation
in terms of integrable state superposition.
As a result, the final expectation for the thermalized state
$ e^{-iH_{D}t}|\mathcal{J}_{\alpha}\rangle\langle\mathcal{J}_{\alpha}|
e^{iH_{D}t}$ with respect to ${\bf J}^{T}_{\alpha_{1}}{\bf J}_{\alpha_{1}}$ and
 ${\bf J}^{T}_{\alpha_{2}}{\bf J}_{\alpha_{2}}$ exhibit a obvious correlation
($3:1$ or $1:3$ as shown in Eq.(\ref{10191})).
In other word,
in a system containing the initial state 
$|\mathcal{J}_{\alpha}\rangle\langle\mathcal{J}_{\alpha}|$,
${\bf J}^{T}_{\alpha}{\bf J}_{\alpha}$ provides the only "edge burst",
i.e., the only source of the level repulsion between
states $e^{-iH_{D}t}|\mathcal{J}_{\alpha}\rangle\langle\mathcal{J}_{\alpha}|e^{iH_{D}t} (t)$.
That is why the traces in Eq.(\ref{10191}) reflect character of isolated system that completely decoupled
with the environment.
In contrast to this,
there is already "edge burst" which can be seem from the evolution of
$e^{-iH_{D}t}\mathcal{P}_{D}'e^{iH_{D}t}$,
where
we numerically verify
\begin{equation} 
	\begin{aligned}
\langle {\rm Tr}[{\bf J}^{T}_{\alpha_{2}}{\bf J}_{\alpha_{2}}]\rangle_{t}
\gtrsim
\langle {\rm Tr}[{\bf J}^{T}_{\alpha_{1}}{\bf J}_{\alpha_{1}}]\rangle_{t},
	\end{aligned}
\end{equation}
where
\begin{equation} 
	\begin{aligned}
\langle {\rm Tr}[{\bf J}^{T}_{\alpha_{i}}{\bf J}_{\alpha_{i}}]\rangle_{t}:=
\lim_{T\rightarrow\infty}\frac{1}{T}
		\int^{T}_{0}dt
	{\rm Tr}[	e^{-iH_{D}t}\mathcal{P}_{D}' e^{iH_{D}t}{\bf J}^{T}_{\alpha_{i}}{\bf J}_{\alpha_{i}}].
	\end{aligned}
\end{equation}
The reason is that ${\bf J}^{T}_{\alpha}{\bf J}_{\alpha}$ does not provide
the only edge burst here,
thus the resulting trace still have finite response to the "environment",
and the level repulsion always exist between $
\langle {\rm Tr}[{\bf J}^{T}_{\alpha_{i}}{\bf J}_{\alpha_{i}}]\rangle_{t}$'s.
Similar phenomenon can be observed from
the evolution of initial states 
\begin{equation} 
	\begin{aligned}
		&
		\lim_{T\rightarrow\infty}\frac{1}{DT}
		\int^{T}_{0}dt
		{\rm Tr}[	e^{-iH_{D}t}
\mathcal{Q}_{\alpha_{1}}
		e^{iH_{D}t}{\bf J}^{T}_{\alpha_{1}}{\bf J}_{\alpha_{1}}]
		\lesssim
		\lim_{T\rightarrow\infty}\frac{1}{DT}
		\int^{T}_{0}dt
		{\rm Tr}[	e^{-iH_{D}t}
\mathcal{Q}_{\alpha_{1}}
		e^{iH_{D}t}{\bf J}^{T}_{\alpha_{2}}{\bf J}_{\alpha_{2}}]\\
		&
		\lesssim
		\lim_{T\rightarrow\infty}\frac{1}{DT}
		\int^{T}_{0}dt
		{\rm Tr}[	e^{-iH_{D}t}
\mathcal{Q}_{\alpha_{2}}
		e^{iH_{D}t}{\bf J}^{T}_{\alpha_{2}}{\bf J}_{\alpha_{2}}]
		\lesssim
		\lim_{T\rightarrow\infty}\frac{1}{DT}
		\int^{T}_{0}dt
		{\rm Tr}[	e^{-iH_{D}t}
\mathcal{Q}_{\alpha_{2}}
		e^{iH_{D}t}{\bf J}^{T}_{\alpha_{1}}{\bf J}_{\alpha_{1}}],
	\end{aligned}
\end{equation}
where 
$\mathcal{Q}_{\alpha}=\sum_{i}|{\bf \Psi}_{\alpha;i}\rangle \sum_{i}\langle {\bf \Psi}_{\alpha;i}^{*}|$.
Numerically the above expression converge to $0.12466,\ 0.12502,\ 0.1272,\ 0.1298$, respectively,
at long time limit, where despite the minor fluctuations,
there is always a fixed difference between them, which indicates the persistent level repulsion.

As the initial state need to be able to evolve according to a time-independent Hamiltonian,
and localization (delocalization) of initial states directly determine
the finial expectation
with respect to the local-type observables ${\bf J}^{T}_{\alpha}{\bf J}_{\alpha}$.
Such local-type observable can be called as non-destructive edge,
as it does not eliminate the non-Hermicity of initial states.
While the observables like ${\bf \Psi}_{\alpha}$, 
${\bf J}^{H}_{\alpha}{\bf J}_{\alpha}$,
and $|\mathcal{J}_{\alpha}\rangle\langle\mathcal{J}_{\alpha}|$
are destructive edges,
which can be evidenced by the absence of level repulsion.
Although such destructive edge can also leads to a relaxation to equilibrium state,
but such relaxation is not completely originates from the level repulsion
from original initial state without the edge.
Thus the non-destructive edge must be:
(1): not able to evolve 
according to the time-independent Hamiltonian $H_{D}$;
(2): Without correlation with other observable.
Here, the destructive observables ${\bf \Psi}_{\alpha}$
and $\mathcal{P}_{D}'|\mathcal{J}_{\alpha}\rangle\langle\mathcal{J}_{\alpha}|$
violate the condition (2), while
 $\mathcal{P}_{D}'{\bf J}_{\alpha}^{T}{\bf J}_{\alpha}$,
${\bf J}^{H}_{\alpha}{\bf J}_{\alpha}$,
$\sum_{i}|{\bf \Psi}_{\alpha;i}\rangle\langle{\bf \Psi}_{\alpha;i}^{*}|$ and
$|\mathcal{J}_{\alpha}\rangle\langle\mathcal{J}_{\alpha}|$ violate condition (1).

For $\mathcal{P}_{D}$ which is Hermitian and cannot be thermalized,
there is not the evolution for
$e^{-iH_{D}t} \mathcal{P}_{D} e^{iH_{D}t} {\bf J}^{T}_{\alpha}{\bf J}_{\alpha}$;
for initial state $\mathcal{P}_{D}'$ which is non-Hermitian but cannot be completely thermalized with respect to 
${\bf J}^{T}_{\alpha}{\bf J}_{\alpha}$, the evolution of 
$e^{-iH_{D}t} \mathcal{P}_{D}' e^{iH_{D}t} {\bf J}^{T}_{\alpha}{\bf J}_{\alpha}$
exhibit level repulsion for $\alpha=1$ and $\alpha=2$;
for initial state $ |\mathcal{J}_{\alpha}\rangle\langle\mathcal{J}_{\alpha}|
$ which is Hermitian and can be completely thermalized with respect to 
${\bf J}^{T}_{\alpha}{\bf J}_{\alpha}$,
the evolution of
$e^{-iH_{D}t} |\mathcal{J}_{\alpha}\rangle\langle\mathcal{J}_{\alpha}|
e^{iH_{D}t} {\bf J}^{T}_{\alpha}{\bf J}_{\alpha}$ exhibit no level repulsion.
Importantly,
the above-mentioned level repulsion only exist for the imaginary part of eigenvalues,
which reflect a crucial dependence on the complex eigenvalue.
It is also important to notice that the
complex eigenfrequencies in a circuit have 
dramatical effects on the output signal resolution as well as the noise (frequencies).
Further, the level repulsion disappear
for initial state $\mathcal{P}_{D}'$ with respect to 
$|\mathcal{J}_{\alpha}\rangle\langle\mathcal{J}_{\alpha}|$,
${\bf J}^{H}_{\alpha}{\bf J}_{\alpha}$,
or ${\bf \Psi}_{\alpha}$,
which means the level repulsion requires the observable does not contain non-defective degeneracies
or other kinds of triggers for the correlations.
Meanwhile this indicates that
the observable does not have to be Hermitian (real eigenvalues),
like the case
where the observable ${\bf J}^{T}_{\alpha}{\bf J}_{\alpha}$ (despite the complex eigenvalue) behaves
more locally than the Hermitian one ${\bf J}^{H}_{\alpha}{\bf J}_{\alpha}$
since there is higher degree of overlap between the eigenvalues of
${\bf J}^{T}_{\alpha_{1}}{\bf J}_{\alpha_{1}}$ and ${\bf J}^{T}_{\alpha_{2}}{\bf J}_{\alpha_{2}}$,
compares to that between
${\bf J}^{H}_{\alpha_{1}}{\bf J}_{\alpha_{1}}$, and ${\bf J}^{H}_{\alpha_{2}}{\bf J}_{\alpha_{2}}$.
Thus
level repulsion only appear between the evolutions (thermalization) 
of $\mathcal{P}_{D}'$ with respect to two uncorrelated observables
where each one of them behaves locally (due to the diagonal contribution),
and such observable in this article should only be ${\bf J}_{\alpha}^{T}{\bf J}_{\alpha}$.

\begin{figure}
	\centering
	\includegraphics[width=0.3\linewidth]{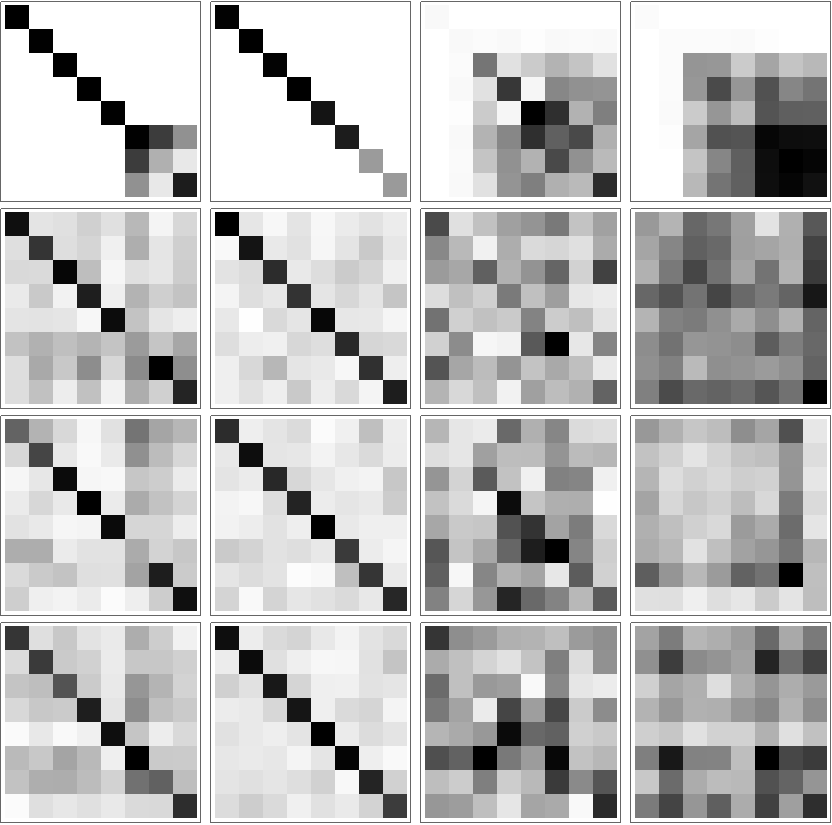}\captionsetup{font=small}
	\caption{
As an evidence for the localization of ${\bf J}^{T}_{\alpha}{\bf J}_{\alpha}$ and the delocalization 
of $|\mathcal{J}_{\alpha}\rangle\langle \mathcal{J}_{\alpha}|$,
we show the eigenvector outer products of ${\bf J}^{T}_{\alpha_{1}}{\bf J}_{\alpha_{1}}$ (first column),
${\bf J}^{T}_{\alpha_{2}}{\bf J}_{\alpha_{2}}$ (second column),
$|\mathcal{J}_{\alpha_{1}}\rangle\langle \mathcal{J}_{\alpha_{1}}|$ (third column),
$|\mathcal{J}_{\alpha_{2}}\rangle\langle \mathcal{J}_{\alpha_{2}}|$ (fourth column),
where the first to last rows correspond to $t=0$, $t=10$, $t=50$, and $t=\infty$, respectively.
	}
\end{figure}

\subsection{Hilbert-Schmidt product}

In terms of projector
$\mathcal{P}_{D}$ (onto the size-dependent sector Hamiltonian $H_{D}$) with normalized basis
($\langle E_{j}|E_{j'}\rangle=\delta_{jj'},\ 
\sum_{j}\langle E_{j}|E_{j}\rangle=D$),
we can rewrite the above trace in the form
of microcanonical ensemble average,
\begin{equation} 
	\begin{aligned}
		{\rm Tr}[{\bf J}_{\alpha}^{T}{\bf J}_{\alpha}]
		&
		={\rm Tr}\left[e^{iH_{D}t} 
		\mathcal{P}_{D}
		{\bf J}_{\alpha}^{T}{\bf J}_{\alpha} e^{-iH_{D}t}\right]
		={\rm Tr}\left[e^{iH_{D}t} 
		{\bf J}_{\alpha}^{T}{\bf J}_{\alpha} \mathcal{P}_{D}
		e^{-iH_{D}t}\right]
		={\rm Tr}\left[e^{iH_{D}t} 
		\mathcal{P}_{D}
		e^{-iH_{D}t}{\bf J}_{\alpha}^{T}{\bf J}_{\alpha}\right]
		={\rm Tr}\left[{\bf J}_{\alpha}^{T}{\bf J}_{\alpha}
		e^{iH_{D}t} 
		\mathcal{P}_{D}
		e^{-iH_{D}t}\right]
	\end{aligned}
\end{equation}
where $[\mathcal{P}_{D},
{\bf J}_{\alpha}^{T}{\bf J}_{\alpha}]=0,\ 
[e^{iH_{D}t} 
\mathcal{P}_{D}
e^{-iH_{D}t},{\bf J}_{\alpha}^{T}{\bf J}_{\alpha}]=0$.
Here
$\intercal_{\alpha;1}:=e^{iH_{D}t} 
{\bf J}_{\alpha}^{T}{\bf J}_{\alpha} \mathcal{P}_{D}
e^{-iH_{D}t}$ and 
$\intercal_{\alpha;2}:={\bf J}_{\alpha}^{T}{\bf J}_{\alpha}
e^{iH_{D}t} 
\mathcal{P}_{D}e^{-iH_{D}t}$ are a pair of  similar matrix with rank larger than one and conserved trace (${\rm Tr}\intercal_{\alpha;1}={\rm Tr}\intercal_{\alpha;2}=1$),
and they have the same singular values which are independent of time.
Indeed, we have verified that $\intercal_{\alpha;2}$ is the Hilbert-Schmidt operator (Endomorphism) while
$\intercal_{\alpha;1}$ is not although it share the same Hilbert-Schmidt norm in some cases:
\begin{equation} 
	\begin{aligned}\label{9101}
		&
		||\intercal_{\alpha;1}||_{HS}
		=||\sqrt{\intercal_{\alpha;1}^{\dag}\intercal_{\alpha;1}}||_{HS}
		=||\sqrt{\intercal_{\alpha;1}\intercal_{\alpha;1}^{\dag}}||_{HS}
		=\sqrt{{\rm Tr}[\intercal_{\alpha;1}^{\dag}\intercal_{\alpha;1}]}
		=\sqrt{{\rm Tr}[\intercal_{\alpha;1}\intercal_{\alpha;1}^{\dag}]}\\
		&
		=||\intercal_{\alpha;2}||_{HS}
		=||{\bf J}^{T}_{\alpha}{\bf J}_{\alpha}||_{HS}
		=||\sqrt{\intercal_{\alpha;2}^{*}\intercal_{\alpha;2}}||_{HS}
		=||\sqrt{\intercal_{\alpha;2}\intercal_{\alpha;2}^{*}}||_{HS}
		=\sqrt{{\rm Tr}[\intercal_{\alpha;2}^{*}\intercal_{\alpha;2}]}
		=\sqrt{{\rm Tr}[\intercal_{\alpha;2}\intercal_{\alpha;2}^{*}]}\\
		&
		={\rm Tr}[\sqrt{\intercal_{\alpha;1}^{\dag}\intercal_{\alpha;1}}]+o(1)
		={\rm Tr}[\sqrt{\intercal_{\alpha;2}^{*}\intercal_{\alpha;2}}]+o(1),\\
		&
		{\rm Tr}[\sqrt{\intercal_{\alpha;1}^{\dag}\intercal_{\alpha;1}}]
		={\rm Tr}[\sqrt{\intercal_{\alpha;1}\intercal_{\alpha;1}^{\dag}}]
		={\rm Tr}[\sqrt{\intercal_{\alpha;2}^{*}\intercal_{\alpha;2}}]
		={\rm Tr}[\sqrt{\intercal_{\alpha;2}\intercal_{\alpha;2}^{*}}]
		\neq 
		{\rm Tr}[\sqrt{\intercal_{\alpha;1}^{*}\intercal_{\alpha;1}}]
		={\rm Tr}[\sqrt{\intercal_{\alpha;1}\intercal_{\alpha;1}^{*}}],
	\end{aligned}
\end{equation}
whose values depends only on the subscript $\alpha$ insteds of time $t$.
$\intercal_{\alpha;1}$ and $\intercal_{\alpha;2}$ share the same (normalized) trace, and 
${\rm Tr}[\intercal_{\alpha;1}^r]={\rm Tr}[\intercal_{\alpha;2}^r]$ be independent of $t$, where the superscript $r$ denotes the matrix power.
Since $\intercal_{\alpha;1}$ and $\intercal_{\alpha;2}$ are similar,
they share the same linear map, 
and in the case that $\intercal_{\alpha;1}$ and $\intercal_{\alpha;2}$ are all full rank, they are bijective.

While for $|\intercal_{\alpha;1}|$ and $|\intercal_{\alpha;2}|$, which means replacing all the entrices by its absolute value,
${\rm Tr}[|\intercal_{\alpha;2}|^r]$ are $t$-independent,
but ${\rm Tr}[|\intercal_{\alpha;1}|^r]$ is $t$-dependent.
The key reason for this difference is that the $\intercal_{\alpha;2}$
is symmetry ($\intercal_{\alpha;2}^{\dag}=\intercal_{\alpha;2}^{*}$) while $\intercal_{\alpha;1}$ is not for $t\neq 0$.
Due to this reason, we also have
\begin{equation} 
	\begin{aligned}
		&
		{\rm Re}[\intercal_{\alpha;2}^2]=	{\rm Re}[(\intercal_{\alpha;2}^{*})^2],
		{\rm Im}[\intercal_{\alpha;2}^2]=-{\rm Im}[(\intercal_{\alpha;2}^{*})^2],
		{\rm Re}[\intercal_{\alpha;2}\intercal_{\alpha;2}^{*}]=	{\rm Re}[\intercal_{\alpha;2}^{*}\intercal_{\alpha;2}],
		{\rm Im}[\intercal_{\alpha;2}\intercal_{\alpha;2}^{*}]=-{\rm Im}[\intercal_{\alpha;2}^{*}\intercal_{\alpha;2}],\\
		&
		{\rm Tr}[|\intercal_{\alpha;2}|^2]={\rm Tr}[\intercal_{\alpha;2}\intercal_{\alpha;2}^{*}]
		={\rm Tr}[\intercal_{\alpha;2}^{*}\intercal_{\alpha;2}]
		={\rm Tr}[\intercal_{\alpha;1}\intercal_{\alpha;1}^{\dag}]
		={\rm Tr}[\intercal_{\alpha;1}^{\dag}\intercal_{\alpha;1}],
	\end{aligned}
\end{equation}
where the matrices in second line are all positive-defined.
Also, the matrices $|\intercal_{\alpha;2}|^2$, $\intercal_{\alpha;2}\intercal_{\alpha;2}^{*}$, and
$\intercal_{\alpha;2}^{*}\intercal_{\alpha;2}$ share the same (real) diagonal elements.
According to the property of Hilbert–Schmidt operator,
the complex and symmetry matrix $\intercal_{\alpha;2}$ corresponds to a linear endomorphism.
From Eq.(\ref{9101}),
the unitary dynamic with local measurements can be seem from
$||\intercal_{\alpha;2}\mho||_{HS}
=\sqrt{{\rm Tr}[\intercal_{\alpha;2}\intercal_{\alpha;2}^{*}]}$,
where $\mho$ is arbitary unitary matrix.
While the nonunitary dynamics can be seem from the time-dependence of the non-positive-defined matrices
$\intercal_{\alpha;1}\intercal_{\alpha;1}^{*}$ or
$\intercal_{\alpha;1}^{*}\intercal_{\alpha;1}$ 
governed by an effective non-Hermitian Hamiltonian.
We have
\begin{equation} 
	\begin{aligned}
		&
		\langle {\rm Tr}[\intercal_{\alpha;1}\intercal_{\alpha;1}^{*}]
		\rangle_{t}:=
		\lim_{T\rightarrow\infty}\frac{1}{T}\int^{T}_{0}dt
		{\rm Tr}[\intercal_{\alpha_{1};1}\intercal_{\alpha_{1};1}^{*}]
		=
		\lim_{T\rightarrow\infty}\frac{1}{T}\int^{T}_{0}dt
		{\rm Tr}[\intercal_{\alpha_{2};1}\intercal_{\alpha_{2};1}^{*}]
		=
		{\rm Tr}[\intercal_{\alpha;2}\intercal_{\alpha;2}^{*}]+o(1),\\
		&
		{\rm Tr}[\intercal_{\alpha_{1};1}\intercal_{\alpha_{2};1}^{*}]
		={\rm Tr}[\intercal_{\alpha_{2};1}\intercal_{\alpha_{1};1}^{*}],\\
		&
		\lim_{T\rightarrow\infty}\frac{1}{T}\int^{T}_{0}dt
		{\rm Tr}[\intercal_{\alpha_{1};1}\intercal_{\alpha_{2};1}^{*}]
		=\frac{1}{D}.
	\end{aligned}
\end{equation}
Here ${\rm Tr}[\intercal_{\alpha;2}\intercal_{\alpha;2}^{*}]$
is independent of time, and has close values for $\alpha_{1}$
and $\alpha_{2}$
\begin{equation} 
	\begin{aligned}
		&
		{\rm Tr}[\intercal_{\alpha;2}\intercal_{\alpha;2}^{*}]
		=	{\rm Tr}[{\bf J}_{\alpha}^{T}{\bf J}_{\alpha}
		({\bf J}_{\alpha}^{T}{\bf J}_{\alpha})^*],\\
		&
		\frac{	{\rm Tr}[\intercal_{\alpha_{1};2}\intercal_{\alpha_{1};2}^{*}]
			+	{\rm Tr}[\intercal_{\alpha_{2};2}\intercal_{\alpha_{2};2}^{*}]}{2}
		\approx
		\frac{2}{D}	{\rm Tr}
		[|\mathcal{J}_{\alpha}\rangle\langle \mathcal{J}_{\alpha}|],\\
		&
		{\rm Tr}
		[|\mathcal{J}_{\alpha_{1}}\rangle\langle \mathcal{J}_{\alpha_{1}}|
		(|\mathcal{J}_{\alpha_{1}}\rangle\langle \mathcal{J}_{\alpha_{1}}|)^*]=
		{\rm Tr}
		[|\mathcal{J}_{\alpha_{2}}\rangle\langle \mathcal{J}_{\alpha_{2}}|
		(|\mathcal{J}_{\alpha_{2}}\rangle\langle \mathcal{J}_{\alpha_{2}}|)^*],\\
		&
		\frac{||\intercal_{\alpha_{1};2}||_{HS}+||\intercal_{\alpha_{2};2}||
			_{HS}}{2}
		\approx
		\frac{1}{2}
		||
		|\mathcal{J}_{\alpha}\rangle\langle\mathcal{J}_{\alpha}|
		||_{HS},
	\end{aligned}
\end{equation}
where the outer product of the initial basis of diagonal ensemble
$\{|\mathcal{J}_{\alpha}\rangle\}$ emphasize 
the ergodicity over the $\alpha_{1}$ and $\alpha_{2}$.

Moreover,
\begin{equation} 
	\begin{aligned}
		&
		D{\rm Tr}[\rho_{\alpha}{\bf J}_{\alpha}^{T}{\bf J}_{\alpha}
		(\rho_{\alpha}
		{\bf J}_{\alpha}^{T}{\bf J}_{\alpha})^*]
		\approx \frac{\langle {\rm Tr}[\intercal_{\alpha;1}\intercal_{\alpha;1}^{*}]
			\rangle_{t}+	{\rm Tr}[\intercal_{\alpha_{2};2}\intercal_{\alpha_{2};2}^{*}]}
		{2}	,\\
		&
		\frac{{\rm Tr}[\rho_{\alpha}{\bf J}_{\alpha}^{T}{\bf J}_{\alpha}
			(\rho_{\alpha}
			{\bf J}_{\alpha}^{T}{\bf J}_{\alpha})^{H}]}
		{\langle {\rm Tr}[\intercal_{\alpha;1}\intercal_{\alpha;1}^{*}]
			\rangle_{t}}
		=	{\rm Tr}[\intercal_{\alpha_{2};2}\intercal_{\alpha_{2};2}^{*}]	,\\
		&
		\frac{{\rm Tr}[\rho_{\alpha}{\bf J}_{\alpha}^{T}{\bf J}_{\alpha}
			(\rho_{\alpha}
			{\bf J}_{\alpha}^{T}{\bf J}_{\alpha})^H]}
		{{\rm Tr}[\rho_{\alpha}{\bf J}_{\alpha}^{T}{\bf J}_{\alpha}
			(\rho_{\alpha}
			{\bf J}_{\alpha}^{T}{\bf J}_{\alpha})^*]}
		\approx \frac{1}{2}D({\rm Tr}[\intercal_{\alpha_{2};2}\intercal_{\alpha_{2};2}^{*}]+
		\langle {\rm Tr}[\intercal_{\alpha;1}\intercal_{\alpha;1}^{*}]
		\rangle_{t}).
	\end{aligned}
\end{equation}

Although the symmetry complex matrix $\intercal_{\alpha;2}$ cannot be a Hermitian or non-Hermitian matrix by itself,
it is closely related to the Hermiticity 
as well as the unitary dynamics of the bipartite system.
This can be seem through the Takagi decomposition,
$\intercal_{\alpha;2}=\mho_{1}\Lambda \mho_{1}^{T}
$ where $\Lambda$ is the diagonal matrix whose diagonal elements are the singular values of $\intercal_{\alpha;2}$,
$\mho_{1}=\mho_{2}\sqrt{\mho_{2}^{H}\nu_{2}^{*}}$
with $\mho_{2}$ and $\nu_{2}$ the matrices whose columns
are the left- and right-singular vectors of $\intercal_{\alpha;2}$,
respectively.

In terms of the normal matrix $(\intercal_{\alpha;2}+i\intercal_{\alpha;2}^{*})$,
\begin{equation} 
	\begin{aligned}
		&
		{\rm Tr}[\intercal_{\alpha;2}\intercal_{\alpha;2}^{*}]
		=\frac{1}{4}
		{\rm Tr}[
		(\intercal_{\alpha;2}+i\intercal_{\alpha;2}^{*})
		(\intercal_{\alpha;2}+i\intercal_{\alpha;2}^{*})^{*}
		+
		(\intercal_{\alpha;2}-i\intercal_{\alpha;2}^{*})
		(\intercal_{\alpha;2}-i\intercal_{\alpha;2}^{*})^{*}],\\
		&
		{\rm Re}[
		{\rm Tr}[
		(\intercal_{\alpha;2}+i\intercal_{\alpha;2}^{*})
		(\intercal_{\alpha;2}-i\intercal_{\alpha;2}^{*})^{*}]]
		={\rm Re}[
		{\rm Tr}[
		(\intercal_{\alpha;2}+i\intercal_{\alpha;2}^{*})^{*}
		(\intercal_{\alpha;2}-i\intercal_{\alpha;2}^{*})]]=0,
	\end{aligned}
\end{equation}
where the last line is due to
$	{\rm Re}[i\intercal_{\alpha;2}\intercal_{\alpha;2}^{*}]=-	{\rm Re}[i\intercal_{\alpha;2}^{*}\intercal_{\alpha;2}],
{\rm Im}[i\intercal_{\alpha;2}\intercal_{\alpha;2}^{*}]={\rm Im}[i\intercal_{\alpha;2}^{*}\intercal_{\alpha;2}]$.
Thus, as far as there is a singule sector with Hilbert space dimension $D$,
the time-dependent projectors (or the orthogonal basis) perform a trace-preserving mapping which indicating a single 
sector.
While the original projector $\mathcal{P}_{D}$ multipled by 
${\bf J}_{\alpha}^{T}{\bf J}_{\alpha}$ on its left or right,
the projector performs nomore the trace conserving mapping,
and this process is equivalents to an estimation of expectation of
${\bf J}_{\alpha}^{T}{\bf J}_{\alpha}$ in the biorthogonal basis
as shown in the last two lines of Eq.(\ref{9121}).

\subsection{$\rho_{\alpha}
	{\bf J}_{\alpha}^{T}{\bf J}_{\alpha}$}

While
${\bf J}_{\alpha}^{T}{\bf J}_{\alpha}
e^{iH_{D}t} \sum_{j}|E_{j}\rangle \sum_{j}\langle E_{j}|
e^{-iH_{D}t}$
and $e^{iH_{D}t} \sum_{j}|E_{j}\rangle \sum_{j}\langle E_{j}|
e^{-iH_{D}t}  {\bf J}_{\alpha}^{T}{\bf J}_{\alpha}$ are a pair of similar matrix that are both rank-1.
Consequently,
their corresponding unique singular values are guaranteed to be the same only at $t=0$.

In Eq.(\ref{8221}),
we further introduce a vector $|\mathcal{J}_{\alpha}\rangle $
which play the same role of ${\bf J}_{\alpha}^{T}$
in diagonal part of the resulting matrix,
\begin{equation} 
	\begin{aligned}
		\langle E_{j}|{\bf J}^{T}_{\alpha}
		{\bf J}_{\alpha}|E_{j}\rangle
		=
		\langle E_{j}|\mathcal{J}_{\alpha}\rangle
		\langle \mathcal{J}_{\alpha}|E_{j}\rangle,\ 
		\forall j,
	\end{aligned}
\end{equation}
which can be normalized as
\begin{equation} 
	\begin{aligned}
		&
		\sum_{j} |\langle E_{j}|\mathcal{J}_{\alpha}\rangle|^2
		=\langle\mathcal{J}_{\alpha}|\mathcal{P}_{D}|\mathcal{J}_{\alpha}\rangle
		=\sum_{j}
		\langle E_{j}|{\bf J}^{T}_{\alpha}
		{\bf J}_{\alpha}|E_{j}\rangle
		=\sum_{j}
		\langle E_{j}|{\bf J}_{\alpha}{\bf J}^{T}_{\alpha}
		|E_{j}\rangle
		=\sum_{j}
		\langle E_{j}|\mathcal{J}_{\alpha}\rangle\langle \mathcal{J}_{\alpha}|E_{j}\rangle\\
		&
		={\rm Tr}
		[
		\pmb{\mathcal{E}}|\mathcal{J}_{\alpha}\rangle
		\langle\mathcal{J}_{\alpha}|\pmb{\mathcal{E}}^{T}]
		={\rm Tr}
		[
		\pmb{\mathcal{E}}|
		{\bf J}_{\alpha}{\bf J}^{T}_{\alpha}
		|\pmb{\mathcal{E}}^{T}]
		=1,\\
		&
		{\rm Rank}[|\mathcal{J}_{\alpha}\rangle\langle \mathcal{J}_{\alpha}|]
		={\rm Tr}[|\mathcal{J}_{\alpha}\rangle\langle \mathcal{J}_{\alpha}|]
		={\rm Tr}[{\bf J}_{\alpha}^{T}{\bf J}_{\alpha}]
		=\langle \mathcal{J}_{\alpha}|\mathcal{J}_{\alpha}\rangle
		=1,\ \forall\alpha,\\
		&
		{\rm Rank}[\sum_{\alpha}^{\alpha_{m}}
		|\mathcal{J}_{\alpha}\rangle\langle \mathcal{J}_{\alpha}|]=
		{\rm Tr}[\sum_{\alpha}^{\alpha_{m}}
		|\mathcal{J}_{\alpha}\rangle\langle \mathcal{J}_{\alpha}|]=\alpha_{m},
	\end{aligned}
\end{equation}
where $\pmb{\mathcal{E}}$ is the matrix whose rows are the 
eigenvectors of $H_{j}$, $\langle E_{j}|$.
$\overline{\langle E_{j}|\mathcal{J}_{\alpha_{1}}\rangle}=
\overline{\langle E_{j}|\mathcal{J}_{\alpha_{2}}\rangle}=\frac{1}{D}$.
Then in terms of the diagonal ensemble for the observable ${\bf J}^{T}_{\alpha}{\bf J}_{\alpha}$,
\begin{equation} 
	\begin{aligned}
		\label{8231}
		&	{\rm Tr}[\rho_{\alpha}
		{\bf J}^{T}_{\alpha}{\bf J}_{\alpha}]
		=
		{\rm Tr}[
		\rho_{\alpha}(t)
		{\bf J}^{T}_{\alpha}{\bf J}_{\alpha}]
		={\rm Tr}[|\mathcal{J}_{\alpha}(t)
		\rangle \langle\mathcal{J}_{\alpha}(t)|\rho_{\alpha}]\\
		&
		=\langle\mathcal{J}_{\alpha}|\rho_{\alpha}|\mathcal{J}_{\alpha}
		\rangle
		=\langle\mathcal{J}_{\alpha}(t)|\rho_{\alpha}|\mathcal{J}_{\alpha}(t)\rangle\ (\forall t)\\
		&=\sum_{j}
		\langle E_{j}| e^{iE_{j}t}\langle \mathcal{J}_{\alpha}|E_{j}\rangle
		|\langle E_{j}|\mathcal{J}_{\alpha}\rangle |^2 
		|E_{j}\rangle \langle E_{j}|
		\langle E_{j}|\mathcal{J}_{\alpha}\rangle e^{-iE_{j}t}|E_{j}\rangle
		\\
		&
		=
		\lim_{T\rightarrow\infty}
		\frac{1}{T}\int^{T}_{0}dt
		\langle\mathcal{J}_{\alpha}(t)|
		{\bf J}^{T}_{\alpha}{\bf J}_{\alpha}
		|\mathcal{J}_{\alpha}(t)\rangle \\
		&
		=
		\lim_{T\rightarrow\infty}
		\frac{1}{T}\int^{T}_{0}dt
		{\rm Tr}[
		|\mathcal{J}_{\alpha}(t)
		\rangle \langle\mathcal{J}_{\alpha}(t)|{\bf J}^{T}_{\alpha}{\bf J}_{\alpha}]\\
		&	
		=	\sum_{j}|\langle E_{j}|\mathcal{J}_{\alpha}\rangle |^2
		\langle {\bf J}^{T}_{\alpha}{\bf J}_{\alpha} \rangle_{mc}
		+o(1),\\
		&
		\rho_{\alpha}=\sum_{j}
		|\langle E_{j}|\mathcal{J}_{\alpha}\rangle |^2 
		|E_{j}\rangle \langle E_{j}|
		=\sum_{j}
		\langle E_{j}|\mathcal{J}_{\alpha}\rangle |E_{j}\rangle
		\langle E_{j}| \langle \mathcal{J}_{\alpha}|E_{j}\rangle,\\
		&
		\left(
		=e^{iH_{D}t}\sum_{j}|\langle E_{j}|\mathcal{J}_{\alpha}\rangle |^2 
		|E_{j}\rangle \langle E_{j}|e^{-iH_{D}t}
		=\sum_{j}
		\langle E_{j}|\mathcal{J}_{\alpha}\rangle e^{iE_{j}t}|E_{j}\rangle
		\langle E_{j}| e^{-iE_{j}t}\langle \mathcal{J}_{\alpha}|E_{j}\rangle
		\right)\\
		&
		\rho_{\alpha}(t)\mathcal{O}
		:=e^{iH_{D}t}\rho_{\alpha}\mathcal{O}e^{-iH_{D}t},\\
		&
		|\mathcal{J}_{\alpha}(t)\rangle=
		\sum_{j}
		\langle E_{j}|\mathcal{J}_{\alpha}\rangle e^{-iE_{j}t}
		|E_{j}\rangle=		
\sum_{j}
		\langle E_{j}|\mathcal{J}_{\alpha}\rangle e^{-iH_{D}t}
		|E_{j}\rangle
,
	\end{aligned}
\end{equation}
where $\rho_{\alpha}$ (${\rm Tr}\rho_{\alpha}=1,\ 
{\rm Rank}\rho_{\alpha}=D$) is the stationary density matrix (microcanonical ensemble) of the diagonal ensemble.

\begin{figure}
	\centering
	\includegraphics[width=0.5\linewidth]{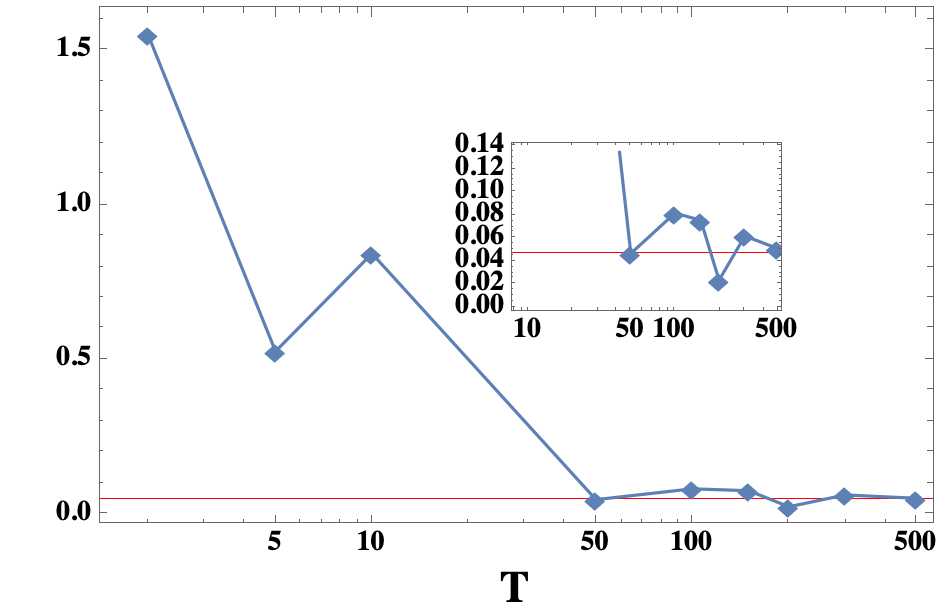}\captionsetup{font=small}
	\caption{
				Time fluctuation of the integral in Eq.(\ref{8231}),
		$	
		\frac{1}{DT}\int^{T}_{0}dt
		{\rm Tr}[
		|\mathcal{J}_{\alpha}(t)
		\rangle \langle\mathcal{J}_{\alpha}(t)|{\bf J}^{T}_{\alpha}{\bf J}_{\alpha}]$
		as a function of $T$.
		The red line indicates the approaching tp $\frac{3}{D^{2}}
		=0.046875$.
	}
\end{figure}

For bipartite system, we found that
\begin{equation} 
	\begin{aligned}
\label{10191}
&	{\rm Tr}[
\sum_{j}
		|\langle E_{j}|\mathcal{J}_{\alpha_{1}}\rangle|^2
 |E_{j}\rangle		\langle E_{j}| \langle 
{\bf J}^{T}_{\alpha_{1}}{\bf J}_{\alpha_{1}}]=
	{\rm Tr}[
\sum_{j}
		|\langle E_{j}|\mathcal{J}_{\alpha_{2}}\rangle|^2
 |E_{j}\rangle		\langle E_{j}| \langle 
{\bf J}^{T}_{\alpha_{2}}{\bf J}_{\alpha_{2}}]\\
&
=		\lim_{T\rightarrow\infty}
		\frac{1}{T}\int^{T}_{0}dt
		{\rm Tr}[
		|\mathcal{J}_{\alpha_{1}}(t)
		\rangle \langle\mathcal{J}_{\alpha_{1}}(t)|{\bf J}^{T}_{\alpha_{1}}{\bf J}_{\alpha_{1}}]\\
&
=		\lim_{T\rightarrow\infty}
		\frac{1}{T}\int^{T}_{0}dt
		{\rm Tr}[
		|\mathcal{J}_{\alpha_{2}}(t)
		\rangle \langle\mathcal{J}_{\alpha_{2}}(t)|{\bf J}^{T}_{\alpha_{2}}{\bf J}_{\alpha_{2}}]
=\frac{3}{D^2},\\
&	{\rm Tr}[
\sum_{j}
		|\langle E_{j}|\mathcal{J}_{\alpha_{1}}\rangle|^2
 |E_{j}\rangle		\langle E_{j}| \langle 
{\bf J}^{T}_{\alpha_{2}}{\bf J}_{\alpha_{2}}]=
	{\rm Tr}[
\sum_{j}
		|\langle E_{j}|\mathcal{J}_{\alpha_{2}}\rangle|^2
 |E_{j}\rangle		\langle E_{j}| \langle 
{\bf J}^{T}_{\alpha_{1}}{\bf J}_{\alpha_{1}}]\\
&
=		\lim_{T\rightarrow\infty}
		\frac{1}{T}\int^{T}_{0}dt
		{\rm Tr}[
		|\mathcal{J}_{\alpha_{1}}(t)
		\rangle \langle\mathcal{J}_{\alpha_{1}}(t)|{\bf J}^{T}_{\alpha_{2}}{\bf J}_{\alpha_{2}}]\\
&
=		\lim_{T\rightarrow\infty}
		\frac{1}{T}\int^{T}_{0}dt
		{\rm Tr}[
		|\mathcal{J}_{\alpha_{2}}(t)
		\rangle \langle\mathcal{J}_{\alpha_{2}}(t)|{\bf J}^{T}_{\alpha_{1}}{\bf J}_{\alpha_{1}}]
=\frac{1}{D^2},
	\end{aligned}
\end{equation}

\begin{equation} 
	\begin{aligned}
		&	{\rm Tr}[|\mathcal{J}_{\alpha}\rangle\langle\mathcal{J}_{\alpha}|]
		=	{\rm Tr}[|\mathcal{J}_{\alpha}(t)\rangle\langle\mathcal{J}_{\alpha}(t)|]
		\\
		&={\rm Tr}
		[\sum_{j}
		\langle E_{j}|\mathcal{J}_{\alpha}\rangle e^{-iE_{j}t}
		|E_{j}\rangle \langle E_{j}|e^{iE_{j}t} \langle  \mathcal{J}_{\alpha}|E_{j}\rangle 
		]\\
		&=\sum_{j}\left(	
		{\rm Tr}[|\mathcal{J}_{\alpha}(t)\rangle
		\langle E_{j}|]\ 
		{\rm Tr}[E_{j}\rangle \langle\mathcal{J}_{\alpha}(t)|]
		\right)
		=\langle\mathcal{J}_{\alpha}(t)|\mathcal{J}_{\alpha}(t)\rangle\\
		&
		=\sum_{j}	e^{iE_{j}t}\langle \mathcal{J}_{\alpha}|E_{j}\rangle
		\langle E_{j}|\mathcal{J}_{\alpha}\rangle e^{-iE_{j}t} 
		=1=
		\langle E_{j}|E_{j}\rangle,\ (\forall j).
	\end{aligned}
\end{equation}
Thus the is $|\mathcal{J}_{\alpha}\rangle$ indeed a Hilbert-Schmidt operator

\begin{equation} 
	\begin{aligned}
		{\rm Tr}[|\mathcal{J}_{\alpha}\rangle\langle\mathcal{J}_{\alpha}|]
		&
		=\sum_{jj'}
		{\rm Tr}
		[\langle E_{j}|\mathcal{J}_{\alpha}\rangle e^{-iE_{j}t}
		|E_{j}\rangle \{\langle {\bf e}_{j'}|\}]\ 
		{\rm Tr}[
		\{|{\bf e}_{j'}\rangle \}
		\langle E_{j}|e^{iE_{j}t} \langle  \mathcal{J}_{\alpha}|E_{j}\rangle]
		\ \delta(\{\langle {\bf e}_{j'}|\},\{\langle E_{j'}|\})\\
		&
		=
		{\rm Tr}
		[\sum_{jj'}
		\langle E_{j}|\mathcal{J}_{\alpha}\rangle e^{-iE_{j}t}
		|E_{j}\rangle \{\langle {\bf e}_{j'}|\}]\ 
		\{|{\bf e}_{j'}\rangle \}
		\langle E_{j}|e^{iE_{j}t} \langle  \mathcal{J}_{\alpha}|E_{j}\rangle]
		\ \delta(\{\langle {\bf e}_{j'}|\},\{\langle E_{j'}|\}),
	\end{aligned}
\end{equation}
where the right-hand-side of first line is independent of $t$
for arbitary basis $\{\langle {\bf e}_{j'}|\}$,
but 
\begin{equation} 
	\begin{aligned}
		&{\rm Tr}[|\mathcal{J}_{\alpha}\rangle\langle\mathcal{J}_{\alpha}|]
		=\sum_{j}
		{\rm Tr}
		[\langle E_{j}|\mathcal{J}_{\alpha}\rangle e^{-iE_{j}t}
		|E_{j}\rangle \{\langle {\bf e}_{j}|\}]\ 
		{\rm Tr}[
		\{|{\bf e}_{j}\rangle \}
		\langle E_{j}|e^{iE_{j}t} \langle  \mathcal{J}_{\alpha}|E_{j}\rangle]\\
		&
		=\sum_{jj'}
		{\rm Tr}
		[\langle E_{j}|\mathcal{J}_{\alpha}\rangle e^{-iE_{j}t}
		|E_{j}\rangle \{\langle {\bf e}_{j'}|\}]\ 
		{\rm Tr}[
		\{|{\bf e}_{j'}\rangle \}
		\langle E_{j}|e^{iE_{j}t} \langle  \mathcal{J}_{\alpha}|E_{j}\rangle]\\
		&
		=\sum_{j\le j'}
		{\rm Tr}
		[\langle E_{j}|\mathcal{J}_{\alpha}\rangle e^{-iE_{j}t}
		|E_{j}\rangle \{\langle {\bf e}_{j'}|\}]\ 
		{\rm Tr}[
		\{|{\bf e}_{j'}\rangle \}
		\langle E_{j}|e^{iE_{j}t} \langle  \mathcal{J}_{\alpha}|E_{j}\rangle]
		\\	&
		=\sum_{j'}
		{\rm Tr}
		\left[
		\sum_{j}
		\langle E_{j}|\mathcal{J}_{\alpha}\rangle e^{-iE_{j}t}
		|E_{j}\rangle \{\langle {\bf e}_{j'}|\}]\ 
		{\rm Tr}[
		\sum_{j}
		\{|{\bf e}_{j'}\rangle \}
		\langle E_{j}|e^{iE_{j}t} \langle  \mathcal{J}_{\alpha}|E_{j}\rangle
		]
		\right]
		\\	&
		=\sum_{j'}
		{\rm Tr}
		\left[
		\sum_{j}
		\langle E_{j}|\mathcal{J}_{\alpha}\rangle e^{-iE_{j}t}
		|E_{j}\rangle \{\langle {\bf e}_{j'}|\}\ 
		{\rm Tr}
		[\{|{\bf e}_{j'}\rangle \}
		\langle E_{j}|e^{iE_{j}t} \langle  \mathcal{J}_{\alpha}|E_{j}\rangle]
		\right]
	\end{aligned}
\end{equation}
is valid iff $\{\langle {\bf e}_{j'}|\}=\{\langle E_{j'}|\}$, and
${\rm Tr}[|\mathcal{J}_{\alpha}(t)\rangle\langle\mathcal{J}_{\alpha}(t)| 
\mathcal{O}]$ be time-independent only in the case of 
$\mathcal{O}={\bf I}$ and $\mathcal{O}=H_{D}$.

In terms of the nonequilibrium fluctuations,
the dynamical form of the density matrix is $\rho_{dia}(t)$,
which can be obtained through the unitary operator $U_{1}:=e^{iH_{j}t}$ (unitary time evolution),
where $H_{D}(iH_{D}t)$ is hermitian (skew hermitian).
Besides, we notice
\begin{equation} 
	\begin{aligned}
		\label{8251}
		{\rm Tr}[\rho_{\alpha}
		{\bf J}^{T}_{\alpha}{\bf J}_{\alpha}]
		&
		=	{\rm Tr}[	{\bf J}^{T}_{\alpha}{\bf J}_{\alpha} e^{iH_{D}t}
		\sum_{j}|\langle E_{j}|\mathcal{J}_{\alpha}\rangle |^2
		|E_{j}\rangle \langle E_{j}|
		e^{-iH_{D}t}]\\
		&=	{\rm Tr}[e^{iH_{D}t}	{\bf J}^{T}_{\alpha}{\bf J}_{\alpha}
		\sum_{j}|\langle E_{j}|\mathcal{J}_{\alpha}\rangle |^2
		|E_{j}\rangle \langle E_{j}|
		e^{-iH_{D}t}]\\
		&=	{\rm Tr}[ e^{iH_{D}t}
		\sum_{j}|\langle E_{j}|\mathcal{J}_{\alpha}\rangle |^2
		{\bf J}^{T}_{\alpha}{\bf J}_{\alpha}
		|E_{j}\rangle \langle E_{j}|
		e^{-iH_{D}t}]\\
		&=	{\rm Tr}[ e^{iH_{D}t}
		\sum_{j}|\langle E_{j}|\mathcal{J}_{\alpha}\rangle |^2
		|E_{j}\rangle \langle E_{j}|	{\bf J}^{T}_{\alpha}{\bf J}_{\alpha}
		e^{-iH_{D}t}],
	\end{aligned}
\end{equation}
where the matrices within the bracket have different entries,
but they have the same trace equals to that of 
$\rho_{dia}{\bf J}_{\alpha}^{T}{\bf J}_{\alpha}$
(and also same rank which equals to ${\rm Rank}[{\bf J}_{\alpha}$]).
This can be explained by the trace class
of the Eq.(\ref{8251}) which is a Hilbert–Schmidt product
or the square of Frobenius norm (Hilbert–Schmidt norm).
We also note that,
both the $U_{1}U_{1}^{\dag}$ and $\rho_{dia}$ are normal matrix,
which satisfy
${\rm Tr}[U_{1}U_{1}^{\dag}]=D$,
${\rm Tr}[\rho_{\alpha}]=1$,
${\rm Rank}[U_{1}U_{1}^{\dag}]=D,\ 
{\rm Rank}[\rho_{\alpha}]=D \ (\forall\alpha)$,
however,
as a product of them,
$\rho_{\alpha}(t)$ is not a normal matrix,
although it still have ${\rm Tr}[\rho_{\alpha}(t)]=1$,
${\rm Rank}[\rho_{\alpha}(t)]
=D\ (\forall\alpha,t)$.
Thus the density operator $\rho_{\alpha}$ (or $\rho_{\alpha}(t)$)
is the singular normal matrix.

Unlike the macroscopic observable ${\bf J}^{T}_{\alpha}{\bf J}_{\alpha}$,
$\rho_{dia}$ is a normal matrix with full rank
and thus can be written as $\rho_{dia}=U_{2}(\lambda_{dia}{\bf I})U_{2}^{\dag}$ whose trace equal one.
Then the second line of Eq.(\ref{8251}) can be written as
\begin{equation} 
	\begin{aligned}\label{8252}
		&	{\rm Tr}[e^{iH_{D}t}	{\bf J}^{T}_{\alpha}{\bf J}_{\alpha}
		\sum_{j}|\langle E_{j}|\mathcal{J}_{\alpha}\rangle |^2
		|E_{j}\rangle \langle E_{j}|
		e^{-iH_{D}t}]\\
		&={\rm Tr}
		[U_{1}{\bf J}^{T}_{\alpha}{\bf J}_{\alpha}
		U_{2}(\lambda_{dia}{\bf I})U_{2}^{\dag}
		U_{1}^{\dag}]\\
		&={\rm Tr}
		[{\bf J}^{T}_{\alpha}{\bf J}_{\alpha}
		U_{2}(\lambda_{dia}{\bf I})U_{2}^{\dag}
		]\\
		&={\rm Tr}
		[
		U_{2}(\lambda_{dia}{\bf I})U_{2}^{\dag}
		{\bf J}^{T}_{\alpha}{\bf J}_{\alpha}
		].
	\end{aligned}
\end{equation}
This trace (Eqs.(\ref{8231},\ref{8251},\ref{8252})) depends on only the $H_{D}$,
and be the same for distince eigenvalues labelled by $\alpha$'s.
Now it is interesting to investigate another trace
\begin{equation} 
	\begin{aligned}
		&{\rm Tr}
		[
		U_{2}(\lambda_{\alpha}{\bf I})
		{\bf J}^{T}_{\alpha}{\bf J}_{\alpha}U_{2}^{\dag}		],
	\end{aligned}
\end{equation}
which is different from Eq.(\ref{8252}),
and this trace will be different for distinct $\alpha$'s.
In fact,
for every $\alpha$'s,
$(\lambda_{\alpha}{\bf I}){\bf J}^{T}_{\alpha}{\bf J}_{\alpha}$ never be a normal matrix,
and the diagonlized matrix $(\lambda_{\alpha}{\bf I})$
can be viewed as a result of the altered observable
${\bf J}^{T}_{\alpha}{\bf J}_{\alpha}$ through the unitary evolution,
while the trace in Eq.(\ref{8252}) representing the autocorrelation between them.
But different to $U_{1}$ which is the time evolution in terms of the skew-Hermitian matrix $H_{j}$,
$U_{2}$ is not a symmetry matrix
but just be an unitary matrix ($U_{2}U_{2}^{\dag}
=U_{2}^{\dag}U_{2}={\bf I}$) and thus be unitarily diagonalizable.
Another direct evidence is that the matrix whose columns are the eigenvectors of $U_{2}$ is not an orthogonal matrix.
While $U_{1}$ is symmetry and thus be orthogonally diagonalizable.

\renewcommand\refname{References}

\end{small}


\begin{thebibliography}{99}
	
\bibitem{Reimann}Reimann, Peter. "Foundation of statistical mechanics under experimentally realistic conditions." Physical review letters 101.19 (2008): 190403.
\bibitem{Sayyad}Sayyad, Sharareh. "Protection of all non-defective twofold degeneracies by antiunitary symmetries in non-Hermitian systems." Physical Review Research 4.4 (2022): 043213.
	
	\bibitem{Yurovsky}Yurovsky, Vladimir A. "Exploring integrability-chaos transition with a sequence of independent perturbations." Physical Review Letters 130.2 (2023): 020404.
	
	\bibitem{Brody}Brody, Dorje C. "Biorthogonal quantum mechanics." Journal of Physics A: Mathematical and Theoretical 47.3 (2013): 035305.
	\bibitem{Hamazaki}R. Hamazaki, T. N. Ikeda, and M. Ueda, Generalized Gibbs ensemble in a nonintegrable system with an extensive number of local symmetries, Phys. Rev. E 93, 032116 (2016).
	
	\bibitem{Yang Z}Yang Z, Schnyder A P, Hu J, et al. Fermion doubling theorems in 2d non-hermitian systems[J]. Preprint at http://arxiv. org/abs/1912.02788, 2019.

\bibitem{Parto}Parto, Midya, et al. "Non-Hermitian and topological photonics: optics at an exceptional point." Nanophotonics 10.1 (2020): 403-423.
	
\bibitem{Wang}Wang, Jiaozi, and Wen-ge Wang. "Characterization of random features of chaotic eigenfunctions in unperturbed basis." Physical Review E 97.6 (2018): 062219.

\bibitem{Helbig}Helbig, Tobias, et al. "Generalized bulk–boundary correspondence in non-Hermitian topolectrical circuits." Nature Physics 16.7 (2020): 747-750.

\bibitem{zx}Xiao, Zhicheng, et al. "Enhanced sensing and nondegraded thermal noise performance based on P T-symmetric electronic circuits with a sixth-order exceptional point." Physical Review Letters 123.21 (2019): 213901.

\bibitem{Choi}Choi, Soonwon, et al. "Emergent SU (2) dynamics and perfect quantum many-body scars." Physical review letters 122.22 (2019): 220603.
\bibitem{Brand}Brandao, Fernando GSL, and Michał Horodecki. "An area law for entanglement from exponential decay of correlations." Nature physics 9.11 (2013): 721-726.
\bibitem{Chung}Chung, Myung-Hoon, and D. P. Landau. "Von Neumann entropy and bipartite number fluctuation in quantum phase transitions." Physical Review B 83.11 (2011): 113104.
\bibitem{rig}Rigol, Marcos. "Fundamental asymmetry in quenches between integrable and nonintegrable systems." Physical review letters 116.10 (2016): 100601.
\bibitem{rig2}M. Rigol, Quantum Quenches in the Thermodynamic Limit,
Phys. Rev. Lett. 112, 170601 (2014).
\bibitem{Shao}Shao, Lubing, et al. "Spinless mirror Chern insulator from projective symmetry algebra." Physical Review B 108.20 (2023): 205126.
\bibitem{Yao}Yao, Shunyu, and Zhong Wang. "Edge states and topological invariants of non-Hermitian systems." Physical review letters 121.8 (2018): 086803.
	\bibitem{Wu}Wu, Zefei, et al. "Intrinsic valley Hall transport in atomically thin MoS2." Nature communications 10.1 (2019): 611.
	\bibitem{Kendirlik}Kendirlik, E. M., et al. "The local nature of incompressibility of quantum Hall effect." Nature communications 8.1 (2017): 14082.
	\bibitem{Cao}Cao, Zan, Zhenyu Xu, and Adolfo del Campo. "Probing quantum chaos in multipartite systems." Physical Review Research 4.3 (2022): 033093.
	\bibitem{Kunst}Kunst, Flore K., et al. "Biorthogonal bulk-boundary correspondence in non-Hermitian systems." Physical review letters 121.2 (2018): 026808.
\bibitem{Shiraishi}Shiraishi, Naoto, and Takashi Mori. "Systematic construction of counterexamples to the eigenstate thermalization hypothesis." Physical review letters 119.3 (2017): 030601.
\bibitem{w}Chen-Huan Wu, and Yida Li. "Non-Hermitian effect to the ballistic transport and quantized Hall conductivity in an operable experimental platform." arXiv preprint arXiv:2311.11276 (2023).
\bibitem{Yao2}Yao, Shunyu, Fei Song, and Zhong Wang. "Non-hermitian chern bands." Physical review letters 121.13 (2018): 136802.


\bibitem{Turner}Turner, Christopher J., et al. "Weak ergodicity breaking from quantum many-body scars." Nature Physics 14.7 (2018): 745-749.
\bibitem{Halder}Halder, Indranil. "Global symmetry and maximal chaos." arXiv preprint arXiv:1908.05281 (2019).
\bibitem{van}van Nieuwenburg, Evert, Yuval Baum, and Gil Refael. "From Bloch oscillations to many-body localization in clean interacting systems." Proceedings of the National Academy of Sciences 116.19 (2019): 9269-9274.
\bibitem{Sousa}Sousa, M. G., et al. "From ergodicity to Stark many-body localization in spin chains with single-ion anisotropy." arXiv preprint arXiv:2401.03111 (2024).
\bibitem{Ganeshan}Ganeshan, Sriram, J. H. Pixley, and S. Das Sarma. "Nearest neighbor tight binding models with an exact mobility edge in one dimension." Physical review letters 114.14 (2015): 146601.

\bibitem{Rappaport}Rappaport, Sonny, et al. "Measurement-induced landscape transitions in hybrid variational quantum circuits." arXiv preprint arXiv:2312.09135 (2023).
\bibitem{Anza}Anza, Fabio, Christian Gogolin, and Marcus Huber. "Eigenstate thermalization for degenerate observables." Physical Review Letters 120.15 (2018): 150603.
\bibitem{Popescu}Popescu, Sandu, Anthony J. Short, and Andreas Winter. "Entanglement and the foundations of statistical mechanics." Nature Physics 2.11 (2006): 754-758.


\bibitem{Rigol7}Rigol, Marcos, Vanja Dunjko, and Maxim Olshanii. "Thermalization and its mechanism for generic isolated quantum systems." Nature 452.7189 (2008): 854-858.

\bibitem{Shirai}Shirai, Tatsuhiko, and Takashi Mori. "Thermalization in open many-body systems based on eigenstate thermalization hypothesis." Physical Review E 101.4 (2020): 042116.

\bibitem{Bocchieri}Bocchieri, P., and A. Loinger. "Ergodic theorem in quantum mechanics." Physical Review 111.3 (1958): 668.
\bibitem{Reimann}Reimann, Peter. "Generalization of von Neumann’s approach to thermalization." Physical review letters 115.1 (2015): 010403.
\bibitem{Steinigeweg}Steinigeweg, Robin, et al. "Pushing the limits of the eigenstate thermalization hypothesis towards mesoscopic quantum systems." Physical review letters 112.13 (2014): 130403.

\bibitem{Hamazaki}Hamazaki, Ryusuke, and Masahito Ueda. "Atypicality of most few-body observables." Physical Review Letters 120.8 (2018): 080603.

\end{thebibliography}
\end{document}